# Novel and Improved Stage Estimation in Parkinson's Disease using Clinical Scales and Machine Learning

R. Prashanth[a,*], Sumantra Dutta Roy[a]

[a] Department of Electrical Engineering, Indian Institute of Technology Delhi, India

## Abstract

The stage and severity of Parkinson's disease (PD) is an important factor to consider for taking effective therapeutic decisions. Although the Movement Disorder Society-Unified Parkinson's Disease Rating Scale (MDS-UPDRS) provides an effective instrument evaluating the most pertinent features of PD, it does not allow PD staging. On the other hand, the Hoehn and Yahr (HY) scale which provides staging, does not evaluate many relevant features of PD. In this paper, we propose a novel and improved staging for PD using the MDS-UPDRS features and the HY scale, and developing prediction models to estimate the stage (normal, early or moderate) and severity of PD using machine learning techniques such as ordinal logistic regression (OLR), support vector machine (SVM), AdaBoost- and RUSBoost-based classifiers. Along with this, feature importance in PD is also estimated using Random forests. We observe that the predictive models of SVM, Adaboost-based ensemble, Random forests and probabilistic generative model performed well with the AdaBoost-based ensemble giving the highest accuracy of 97.46%. Body bradykinesia, tremor, facial expression (hypomimia), constancy of rest tremor and handwriting (micrographia) were observed to be the most important features in PD. It is inferred that MDS-

[*]Corresponding author, Email: prashanth.r.iitd@gmail.com, Present Address: Geetha Bhavan, Near Kowdiar Palace, Trivandrum 695003, Kerala, India

UPDRS combined with classifiers can form effective tools to predict PD staging which can aid clinicians in the diagnostic process.

## 1. Introduction

Parkinson's Disease (PD) is a progressive neurodegenerative disorder characterized by a number of motor and non-motor features resulting in a marked reduction in quality of life [1]. It is characterized by the degeneration of dopaminergic neurons in the substantia nigra pars compacta resulting in a substantial reduction of striatal dopamine. This dopamine deficiency causes a variety of symptoms, among which the four cardinal motor features are: tremor at rest, rigidity, bradykinesia and postural instability [1, 2]. These symptoms begin insidiously, usually affecting one side of the body before spreading to the other, and gradually worsen, and if untreated, the symptoms lead to disability with severe immobility and falling [1, 3]. Presently, PD is an incurable condition, and the therapeutic options mainly focus on symptomatic relief.

Stage and severity of PD is important to estimate for taking effective decisions related to therapy. For instance, Rascol *et al.* [4] observed that early stage PD can be managed successfully by initiating treatment with dopamine agonists such as ropinirole alone, rather than starting with levodopa (which is the most potent and effective medication in PD). Drugs such as Monoamine oxidase B (MAO-B) inhibitors have shown to delay the need for levodopa in early PD and have been approved for use in later stages of PD to boost the effects of levodopa. Other drugs such as amantadine can help in early PD for treating tremor and also in late PD by reducing dyskinesias that may occur with levodopa or dopamine agonists

(http://www.pdf.org/pdf/fs_understanding_medications_12.pdf).

There are many scales to evaluate impairment and disability (stage and severity) in PD. Among these, the Unified Parkinson's Disease Rating Scale (UPDRS) and the Hoehn and Yahr (HY) scale are the most commonly used [2]. UPDRS provides a comprehensive assessment of disability and impairment by evaluating the most pertinent clinical features of PD [5], whereas the HY scale provides a gross assessment of disease progression through a staging which ranges from 0 (no sign of disease) to 5 (severe) [6]. The Movement Disorder Society-Unified Parkinson's Disease Rating Scale (MDS-UPDRS) [5] is the updated and revised version of the original UPDRS with new items devoted to several non-motor elements of PD (thus, making it more comprehensive than the original scale). Other important differences include refined scoring instructions and definitions, and emphasis on impairments and disabilities related to milder symptoms and signs [5]. A brief description of MDS-UPDRS and HY scales are given below:

### 1.1 MDS-UPDRS

The MDS-UPDRS provides a comprehensive assessment of the clinical features (both motor as well as non-motor) of PD [5]. The scale consists of 65 items in 4 parts. Part I assesses the non-motor experiences of daily living, and contains 13 items evaluated partly (the first 6 items) by a specialist and partly (the next 7 items) as a patient questionnaire. Part II concerns motor experiences of daily living, and contains 13 items evaluated completely as a patient questionnaire. Part III is motor examination, containing 33 items and evaluated completely by a specialist. Part IV assesses motor complications, containing 6 items and evaluated by a specialist once a PD subject has started PD medication. Part IV assesses two motor complications:

dyskinesias and motor fluctuations which are strongly related to the duration of disease, levodopa dose and the duration of levodopa treatment [7].

Before MDS-UPDRS, the original UPDRS was used as a gold standard for PD evaluation [8]. UPDRS has shown strong correlations with HY stages and has displayed excellent internal consistency across different HY stages [8]. Longitudinal studies show that UPDRS scores increase over time, and that the scores can indicate crucial clinical decision-making points like the need to introduce symptomatic therapy [8, 9]. However, the original UPDRS had few shortcomings that it did not cover the full spectrum of PD (especially the non-motor aspects of PD), and that the scale favored moderate to late PD, than mild or early PD [8]. Validation studies on the MDS-UPDRS have shown that the updated scale strongly correlated with the original UPDRS, and significantly correlated with other disability measures, quality of life scales and disease duration [5, 10-13]. It is also shown that MDS-UPDRS is more sensitive to changes in PD than the original version [14]. These studies show that the MDS-UPDRS provides a wide spectrum of assessments, and is a reliable and sensitive instrument for estimating the progression and severity in PD.

### 1.2 Hoehn and Yahr (HY) Scale

The HY scale is another widely-used scale to assess PD severity. The scale provides an overall assessment of severity through staging. The progression of PD usually starts from unilateral (Stage 1), to bilateral without balance difficulties (Stage 2), followed by the presence of postural instability (Stage 3), to loss of physical independence (Stage 4), and being wheel-chair or bed-bound unless aided (Stage 5).

The HY scale has also been used to categorize PD as early stage (Stage 1 and 2), moderate stage (Stage 3) and late stage (Stages 4 and 5). Studies have shown that stage has a significant correlation with quality of life measures, and a high correlation with the UPDRS [6]. Neuroimaging studies also indicate that progressively higher stages correlate well with dopaminergic loss [15]. These studies support the utility of HY scale to categorize PD subjects based on the severity.

### 1.3 Motivation, Background and Scope of the study

Although the MDS-UPDRS evaluates the most pertinent features of PD, it does not perform staging. On the other hand, the HY scale provides staging of PD, but does not evaluate many pertinent features. HY staging is usually performed using a *'rate-as-you-see'* approach, taking into account all clinical impairments seen, regardless of their direct relationship to PD [6], and is often biased by the experience of the physician completing the scale. This makes it susceptible to inter-rater variability.

There have been attempts to relate UPDRS and HY [16-18]. Scanlon *et al.* [16, 17] proposed formulas to obtain HY stages from UPDRS Part III scores. Tsanas *et al.* [18] optimized this formula by refining its parameters using genetic algorithm (GA). However, these studies had two shortcomings: 1) the formula was based on intuitive rules, and 2) they used only Part III (motor examination) of the UPDRS, and did not make use of the full spectrum of pertinent PD features.

In this work, for the first time, we propose an improved way of PD staging using MDS-UPDRS, HY scale and machine learning. We use all the features from Parts I, II and III of the MDS-UPDRS, as obtained from the Parkinson's Progression Markers Initiative (PPMI) database [19] to develop diagnostic models to estimate the stage (normal stage, early stage and moderate stage) and severity of PD. Along with this, we also estimate the importance of these features in PD. We did not use Part IV in the study as this part mainly focuses on the complications from long-term treatment for PD, and the PPMI database consisted majority of patients who were in their early stages and they have not undergone long term treatments or were showing these complications.

## 2. Materials and Methods

### 2.1 Database

Data used in the preparation of this article were obtained from the Parkinson's Progression Markers Initiative (PPMI) database. For up-to-date information on the study, please visit http://www.ppmi-info.org. PPMI is a landmark, large-scale, comprehensive, observational, international, and multicenter study to identify PD progression biomarkers. For the study, we considered the features from Parts I, II and III of the MDS-UPDRS, making a total number of features as 59. Part IV was not used as it assesses motor complications of dyskinesias and motor fluctuations that are mostly shown by advanced stage PD patients. As our dataset focused on early and moderate PD, this part was not considered. The study dataset includes observations from the same time periods (visits) for both healthy control (HC) and Parkinson's disease (PD) patients.

### 2.2 Cohort Details

The data used in the study were from the same time instances, rather than taking the complete set of observations for the sake of removing any bias, and accordingly it contained MDS-UPDRS scores collected for Month 0 (Baseline or BL), Month 12 (Visit 04 or V04), Month 24 (Visit 06 or V06), Month 36 (Visit 08 or V08), Month 48 (Visit 10 or V10), Month 60 (Visit 12 or V12) and Month 84 (Visit 14 or V14) from age and gender matched PD and HC subjects.

The data contains 1025 observations from 197 HC subjects and 1995 observations from 434 PD subjects, making a total of 3020 observations from 631 subjects. Table 1 shows the number of subjects and their mean age at enrollment based on the gender for the healthy control (HC) and Parkinson' disease (PD) groups. The PPMI has collected data from age and gender matched healthy control and de novo PD subjects. We can observe in the below table the same pattern that both the groups are age and gender matched.

**Table 1: Number of subjects and age at enrollment for HC and PD**

| | **No. of Subjects** | | **Age at Enrollment*** | |
|---|---|---|---|---|
| ***Gender*** | ***HC*** | ***PD*** | ***HC (in yrs)*** | ***PD (in yrs)*** |
| ***Female*** | 71 | 149 | 59.37 ± 11.70 | 60.64 ± 9.61 |
| ***Male*** | 126 | 285 | 61.78 ± 10.83 | 62.19 ± 9.69 |
| ***Total*** | 197 | 434 | | |

* HC and PD represent healthy control and Parkinson's disease, respectively. The age at enrollment details for 3 HC and 7 PD subjects were not available.

Table 2 shows the distribution of subjects with history of PD. Here, family history includes first degree relatives (biological parents, full siblings, and children) and second-degree relatives (maternal and paternal grandparents, maternal and paternal aunts and uncles, and half siblings). It is observed that 10 PD subjects showed a positive history of PD. But it is to be noted that for all these 10 subjects, the history corresponded to the second-degree relatives and none corresponded to the first-degree relatives. For the PD case, the history involved both first and second-order relatives.

Table 3 shows the number of subjects on medication and total number of subjects for different visits. There were no subjects who were on medication at BL visit. The medications used were Levodopa or Dopamine agonist or others. For subjects taking medication, as per the PPMI protocols, they have full MDS-UPDRS assessed off medication, followed by repeat Part III motor exam one hour after dosing in clinic. For predictive modelling, we have used the data taken before medication only. Fig.1 shown below shows the average of Part III scores pre and post medication.

**Table 2: Details of family history of the subjects**

| | **No. of Subjects** | |
|---|---|---|
| ***History*** | ***HC*** | ***PD*** |
| ***0*** | 187 | 326 |
| ***1*** | 10 | 107 |

* History of 0 and 1 represent not having history and having history, respectively. Family history of one PD subject was not available.

**Table 3: Number of subjects on medication and total number of subjects for different visits**

| | **No. of Subjects** | | |
|---|---|---|---|
| ***Visit*** | ***On med*** | ***Total*** | ***%age*** |
| ***V04*** | 212 | 348 | 60.92% |
| ***V06*** | 283 | 332 | 85.24% |
| ***V08*** | 305 | 329 | 92.71% |
| ***V10*** | 290 | 304 | 95.39% |
| ***V12*** | 222 | 234 | 94.87% |
| ***V14*** | 15 | 15 | 100.00% |

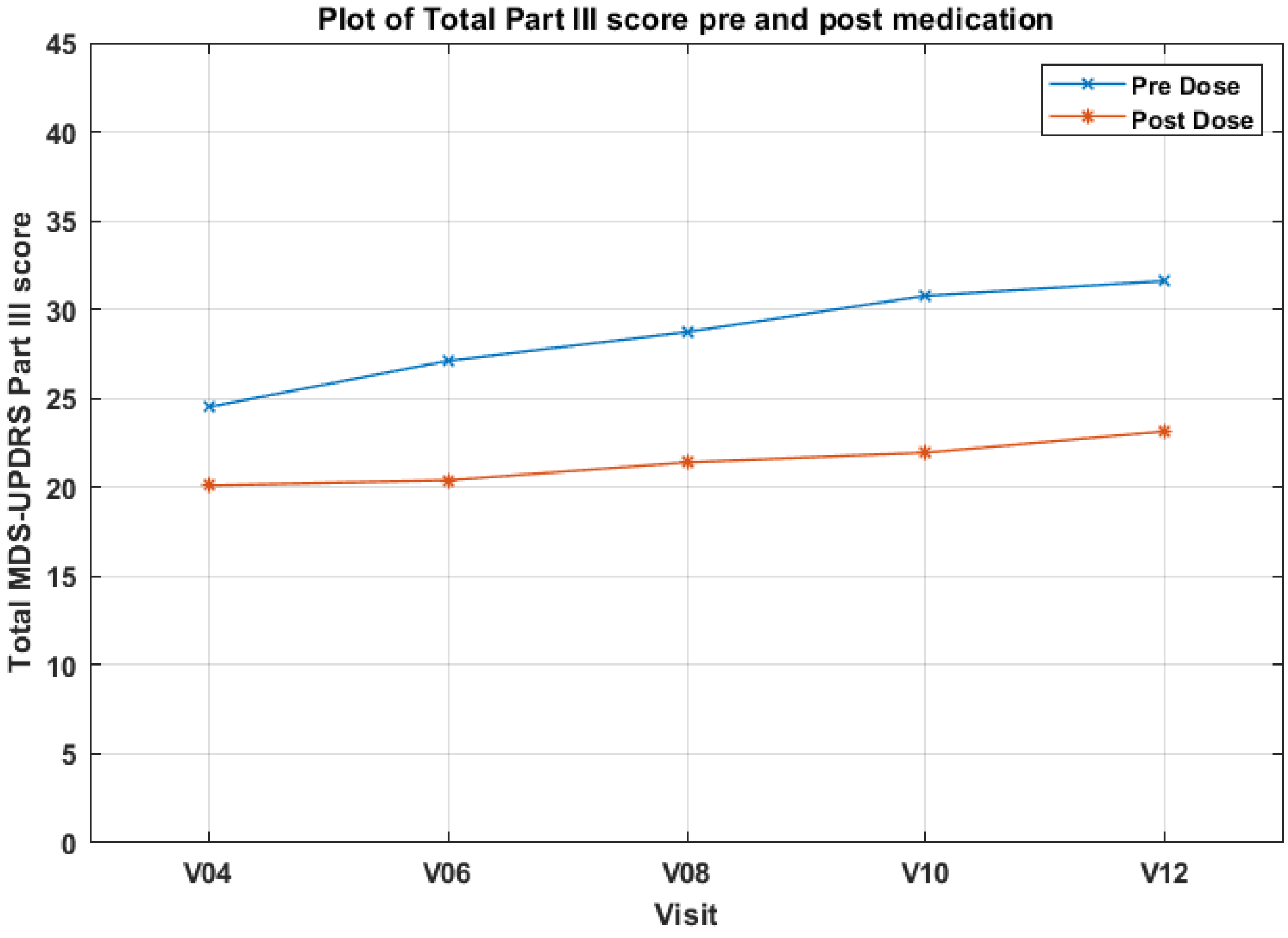

Fig. 1. Mean of all total scores of MDS-UPDRS Part III for different visits for pre and post medication. The visit at month 84 (V14) was not included in the plot as the number of observations for V14 was very low.

Table 4 shows the duration of PD at enrollment for both HC and PD subjects. At enrollment, all subjects were in their early and untreated stage or without medication. Table 5 shows the distribution of subjects for HC and PD based on a particular visit and severity as indicated by the Hoehn and Yahr (HY) stage. And for PD, the data used corresponded to the one that was collected prior to the medication. It can be observed that 5 PD observations (from a total of 1995 PD observations) showed severity of HY = 0. These observations although had a total of Parts I, II and III of MDS-UPDRS in the range 21 to 68 with mean as 39.40 ± 18.09 which is a value shown by PD subjects. It is to be noted that HY staging is usually performed using a *'rate-as-you-see'* approach making it susceptible to bias based on the clinician rating the subject. Along with this, it is to be noted that the number of subjects in V12 was much lower than the previous visits. This is because PPMI is a longitudinal study and the enrollment date of the subjects varied. A total of 111 subjects (32 Healthy control and 79 PD) had an enrollment date during 2013 and their data for the V12 visit are yet to be updated in the PPMI database.

Along with this, 16 HC observations (from a total of 1025 observations) showed severity of HY ≠ 0. These observations had a total of Parts I, II and III of MDS-UPDRS in the range 3 to 35 with mean as 12.88 ± 8.99. These observations might also be a possible case of bias from clinician, which is almost inevitable in large clinical trial study like PPMI, rating the subject.

The change in severity is considered in the study. The severity of a considerable number of PD subjects changed during the course of the data collection. A total of 180 subjects showed a change in severity from HY stage 1 to 2, 35 subjects showed a change in severity from HY stage 2 to 3, 3 subjects showed a severity change from HY stage 1 to stage 3, and 19 subjects showed a change in severity from 1 to 2 then 3. A low number of subjects whose severity changed from HY stage 1 to 3 indicate that during early stages of the disease, the progression might be mostly linear.

**Table 4: Duration of PD at enrollment for HC and PD**

| *Gender* | *Duration of PD at Enrollment (in months)* |
|---|---|
| ***Female*** | 18.24 ± 19.16 |
| ***Male*** | 16.29 ± 15.26 |

* Duration of PD at Enrollment for 8 PD subjects was not available.

**Table 5: Distribution of subjects based on severity and visit**

| *Visit* | *Group* | *Hoehn & Yahr Stage* | | | | *Total* |
|---|---|---|---|---|---|---|
| | | **0** | **1** | **2** | **3** | |
| ***BL*** | **HC** | 195 | 2 | | | 197 |
| | **PD** | | 191 | 240 | 2 | 433 |
| ***V04*** | **HC** | 177 | 3 | 4 | | 184 |
| | **PD** | 1 | 101 | 234 | 12 | 348 |
| ***V06*** | **HC** | 170 | 2 | 1 | | 173 |
| | **PD** | 2 | 83 | 235 | 12 | 332 |
| ***V08*** | **HC** | 164 | 1 | 1 | | 166 |
| | **PD** | 1 | 62 | 244 | 22 | 329 |
| ***V10*** | **HC** | 159 | 1 | 1 | | 161 |
| | **PD** | 1 | 47 | 227 | 29 | 304 |
| ***V12*** | **HC** | 132 | | | | 132 |
| | **PD** | | 25 | 195 | 14 | 234 |
| ***V14*** | **HC** | 12 | | | | 12 |
| | **PD** | | | 13 | 2 | 15 |

* BL data for one PD subject was not available

Fig. 2 shows the stacked bar chart for all the 59 features for normal, early and moderate stage observations. Along with this, we have also shown plots that depicts the time development of all 59 features in PD as well as HC in Fig. C1 (in Appendix). Fig. 3 shows the mean of all features for HC and PD groups as a stacked bar chart. It is clear from the graph that all features showed greater severity during PD. Among the features, the features 7 to 13 comparatively showed lower severity. This must be due to the fact that they represent the non-motor aspects of experiences in daily life and they were filled by the patient themselves. Studies have shown that age is a crucial factor for influencing non-motor features [20] and as the dataset involves mostly aged cohort, non-motor symptoms were more prevalent in the HC group.

As observed from Table 5 (shown before) 191 subjects and 240 subjects were in HY stage 1 and 2 respectively, combinedly making 431 subjects in early stage PD. Only 2 subjects were in moderate stage (HY stage 3) at baseline and this is because PPMI mainly focus on recruiting early PD subjects.

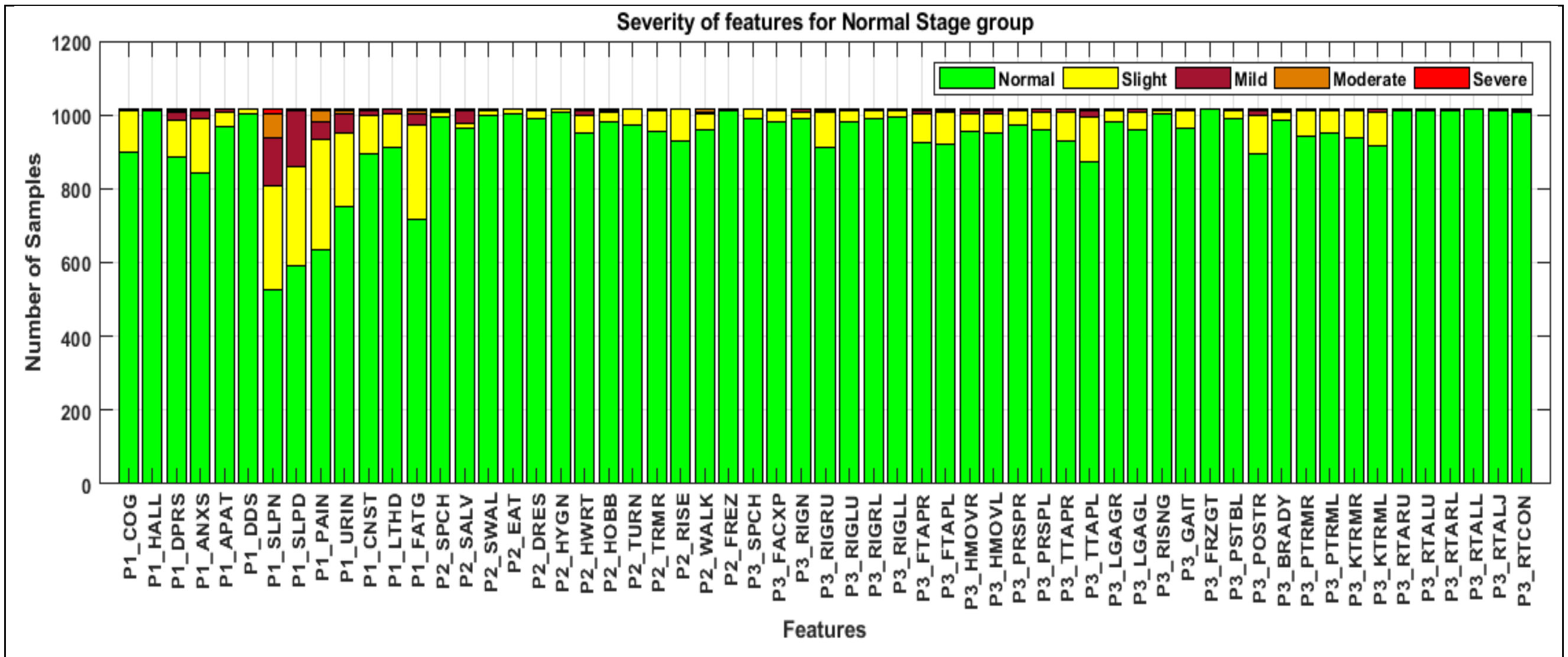

(a)

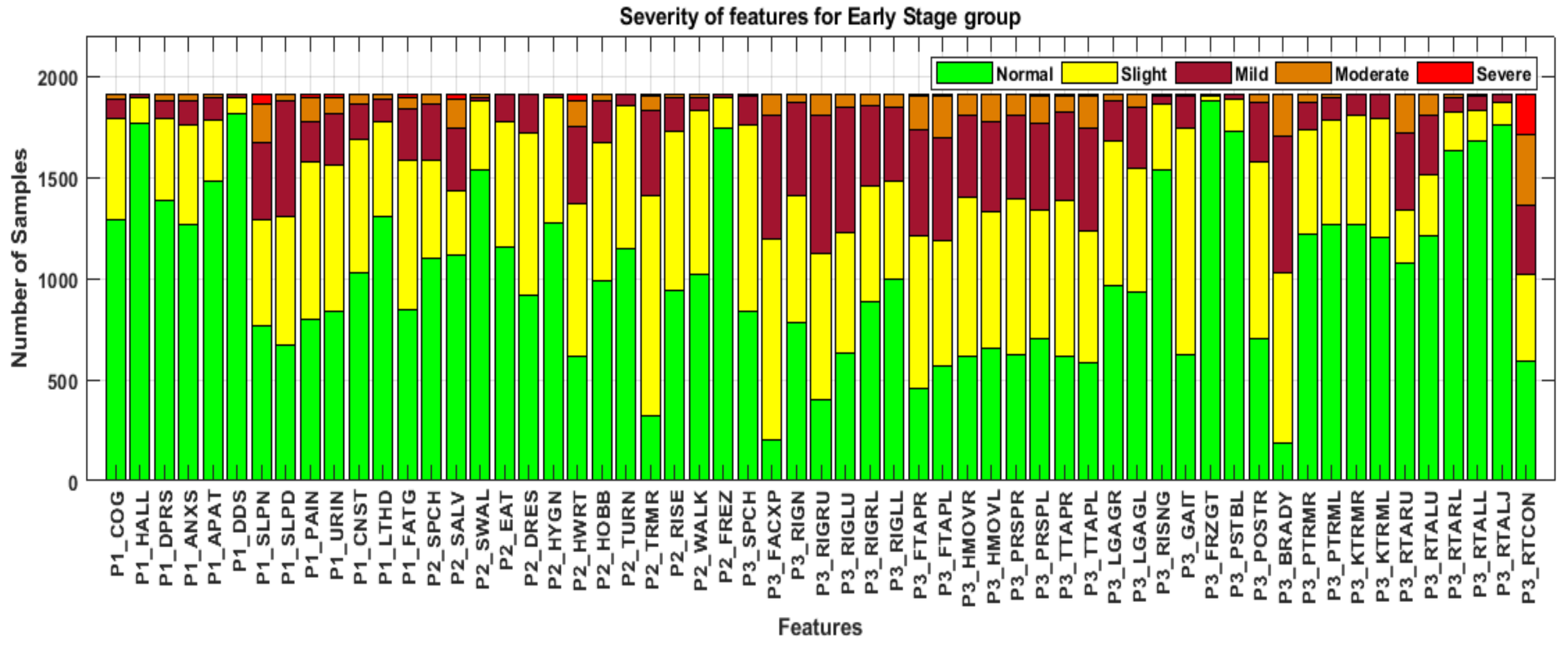

(b)

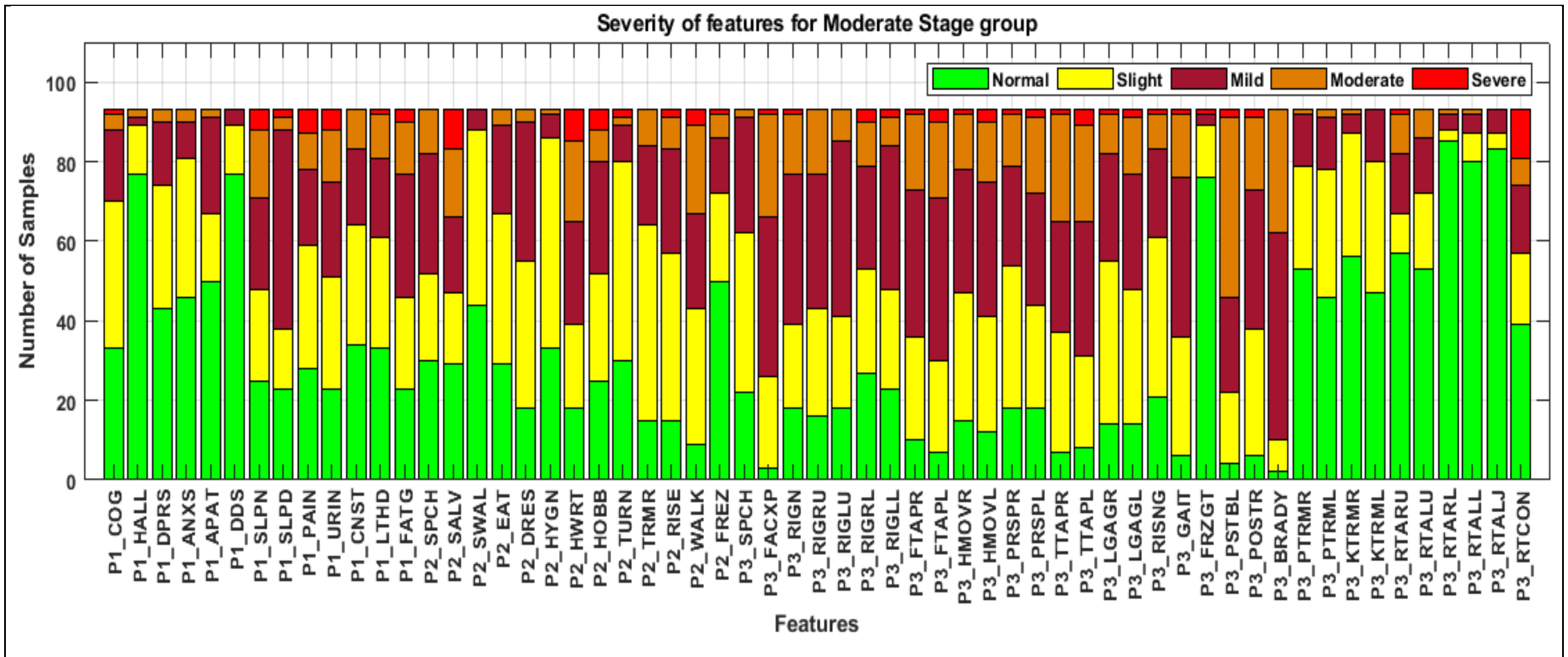

(c)

Fig. 2. The plot of severity of the features in MDS-UPDRS Parts I, II and III (a) Normal Stage (b) Early Stage, (c) Moderate stage

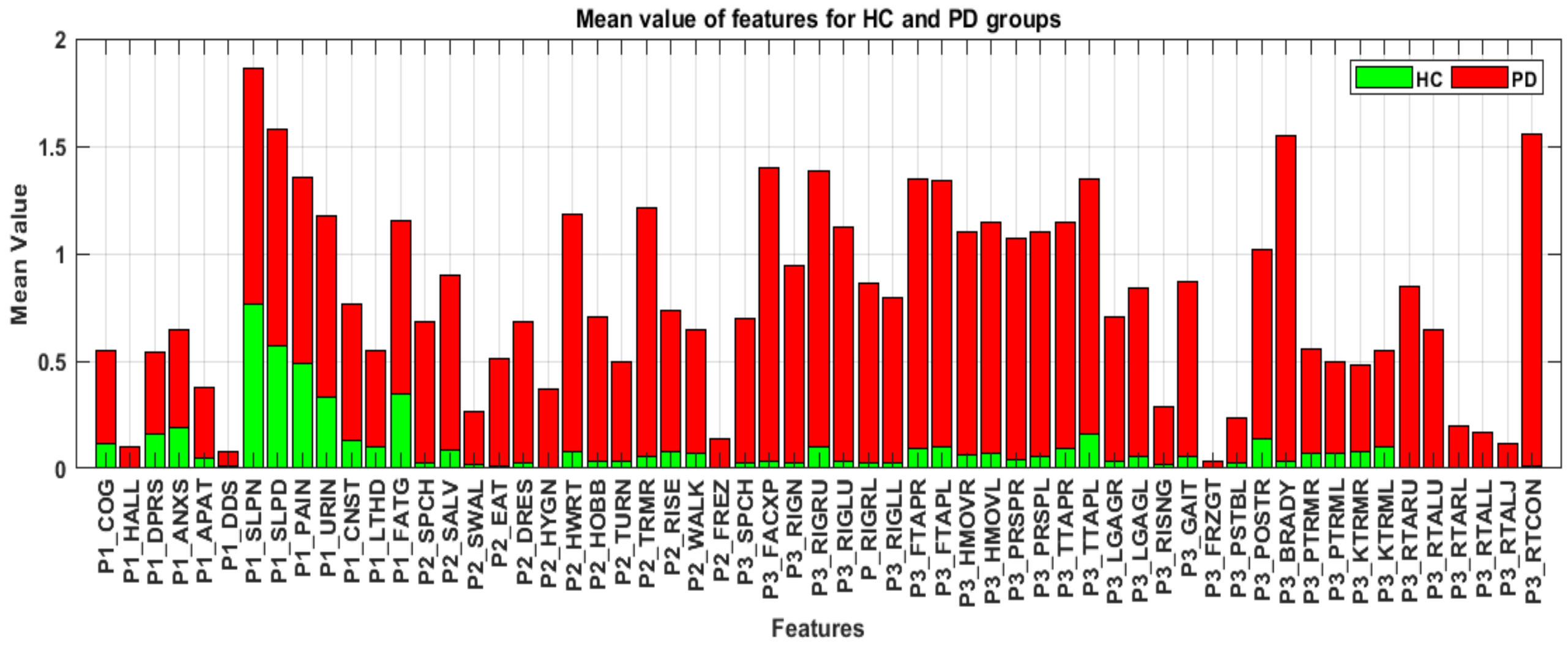

Fig. 3. The plot of mean value of all features for HC and PD groups

## 2.3 Statistical Analysis of features

A flowchart of the proposed analysis is shown in Fig. 4. All the 59 features are analyzed for statistical significance. Features were selected based on filter methods. Statistical techniques of Kruskal-Wallis test for the 3-class classification problem (which was estimating the stage of a subject) and Wilcoxon rank sum test for the 2-class problem (which was for estimating the feature importance). All the statistically significant features ($p$-value<0.05) are used for subsequent stage classification, severity and feature importance estimation.

### 2.4 A note on imbalance in the dataset used

The dataset used for the study was imbalanced with respect to normal stage (16.62%), and majorly imbalanced with respect to moderate stage PD (3.48%) as compared to early stage PD (79.90%). This imbalance is of *extrinsic* type where the imbalance is not directly related to the nature of data space [21]. The sample size of moderate stage PD subjects was small as PPMI (which is one of the large-scale studies in PD) had limited number of subjects who were at this stage.

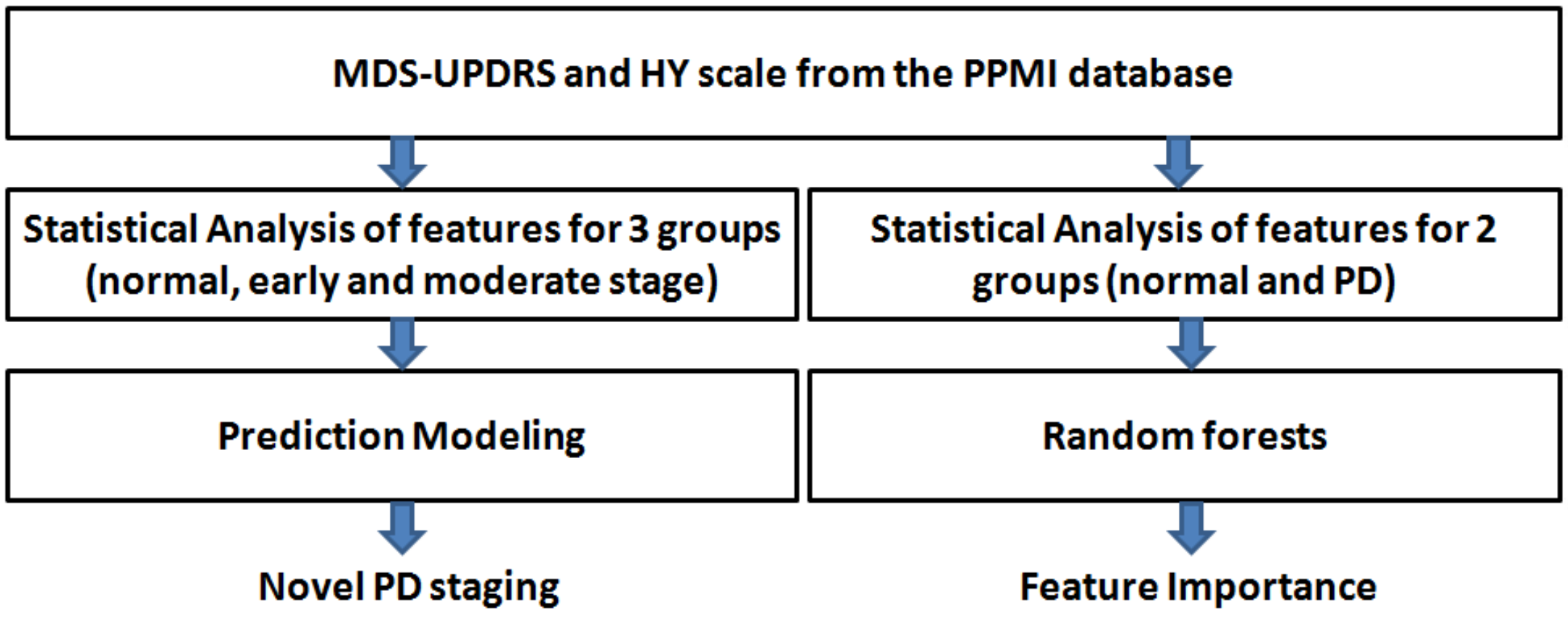

Fig. 4. Flowchart for predictive modeling for novel and improved PD staging, and feature importance estimation.

Fig. 2 shows that there is higher amount of class overlap between early and moderate stage PD, than between normal and early stage PD. In other words, there is higher class separability between the normal and early stage PD (which is more crucial), than between the early and moderate stages. The data complexity arises due to small sample size of the moderate PD class, and the overlapping distributions between early PD and moderate PD classes. Therefore, it is evident that for this domain, we require a classifier that will provide a high overall accuracy, along with a high detection of the minority class (the moderate stage PD), without severely jeopardizing the accuracy of other classes (normal and early stage PD classes).

### 2.5 Prediction modelling for PD severity and stage estimation

The aim of this study is to estimate stage and severity in PD from the MDS-UPDRS ratings using HY scale for validation using prediction models. All the features that are statistically significant are used to develop prediction models. For our study, we use traditional as well as algorithms that are specific for multi-class imbalance problem.

Cost-sensitive learning and sampling based methods dominate the current research in imbalanced data learning [21]. Sampling methods use resampling approaches to rebalance the class distribution, and cost-sensitive learning approaches involve biasing the minority class (by assigning higher misclassification costs for the minority class).

For our experiments, we use ordinal logistic regression (as the response variable is ordinal), cost-sensitive SVM [22], cost-sensitive AdaBoost-based ensemble [23, 24] and sampling based RUSBoost-based ensemble [25], k-nearest neighbor classifier [26], Random forests [27], probabilistic generative model [28] and neural networks [28] to perform prediction modelling. To estimate the optimal parameters for cost-sensitive and sampling based methods, we use the same approach as followed by Sun, Kamel, Wong and Wang [23, 29], where they use genetic algorithm (GA). We use LIBSVM library [30] for classification using SVM, and statistics and machine learning toolbox in MATLAB for classification using AdaBoost-based ensemble,

RUSBoost-based ensemble, Random forests, Probabilistic generative model, *k*-nearest neighbor and neural networks.

The classification performances were evaluated based on a 10-fold cross validation setting that is repeated 100 times. The evaluation measures used are accuracy, precision, recall and F-measure (as described in Appendix A) which is computed for each class. Along with these, we also compute the Brier score for the logistic model, which is a well-known evaluation measure for probabilistic classifiers.

Let the input data (or training data) be represented as $S = \{(\boldsymbol{x}_i, y_i)\}$ where $\boldsymbol{x}_i$ is an observation given by $\boldsymbol{x}_i = \left\{x_i^{(j)}\right\}; j = 1, \dots, m; i = 1, \dots, n$; *m* is the number of features and *n* is the number of observations. The response variable (or class label) be $y_i; y_i \in \{1, 2, 3\}$ where '1', '2' and '3' represent the 3 categories: normal, early and moderate stage PD, respectively.

### 2.5.1 Predictive modeling using Ordinal Logistic Regression

The ordinal logistic regression (OLR) model (also called cumulative logit model or proportional odds model) is among the most popular methods for ordinal data analysis [31, 32]. As the number of categories is 3 in our study, there are 2 equations for the OLR model which are as follows:

$$\log\left(\frac{P(y_i \leq 1)}{P(y_i > 1)}\right) = \log\left(\frac{\pi_1}{\pi_2 + \pi_3}\right) = \alpha_1 + \beta_1 x_i^{(1)} + \beta_2 x_i^{(2)} + \cdots + \beta_m x_i^{(m)} \quad (1)$$

$$\log\left(\frac{P(y_i \leq 2)}{P(y_i > 2)}\right) = \log\left(\frac{\pi_1 + \pi_2}{\pi_3}\right) = \alpha_2 + \beta_1 x_i^{(1)} + \beta_2 x_i^{(2)} + \cdots + \beta_m x_i^{(m)} \quad (2)$$

where $\alpha_1$ and $\alpha_2$ are intercept parameters and satisfy $\alpha_1 \leq \alpha_2$; and $\beta = [\beta_1, \beta_2, \dots, \beta_m]$ are the regression coefficients.

### 2.5.2 Cost-sensitive Support Vector Machine (SVM)

SVM is a powerful classification method that finds an optimal hyperplane (decision boundary) with the largest margin [33]. We use *one-vs-one* approach which is a competitive approach for multi-class classification [34]. This approach constructs $k(k-1)/2$ classifiers, where each one is trained on two classes, for a $k$ (in our case, $k$=3) class problem. For training data from classes $i$ and $j$, the optimization problem is given by

$$\min_{w^{ij}, b^{ij}, \xi^{ij}} \frac{1}{2} \left(\boldsymbol{w}^{ij}\right)^T \boldsymbol{w}^{ij} + C \sum_t \left(\xi^{ij}\right)_t \tag{3}$$

subject to

$$\left(w^{ij}\right)^T \varphi(\boldsymbol{x}_t) + b^{ij} \geq 1 - \left(\xi^{ij}\right)_t; \text{ if } y_t = i \tag{4}$$

$$\left(w^{ij}\right)^T \varphi(\boldsymbol{x}_t) + b^{ij} \leq -1 + \left(\xi^{ij}\right)_t; \text{ if } y_t = j \tag{5}$$

$$\left(\xi^{ij}\right)_t \geq 0 \tag{6}$$

For classification, a voting strategy is used: if $sign\left(\left(w^{ij}\right)^T \varphi(\boldsymbol{x}) + b^{ij}\right)$ indicates $\boldsymbol{x}$ is in class $i$, then the vote for class $i$ is increased by 1, else the vote for class $j$ is increased by 1. $\boldsymbol{x}$ is classified to the class which gets the largest vote. We use the radial basis function (RBF) kernel, given by $K(\boldsymbol{x}, \boldsymbol{x}_i) = \exp(-\gamma \|\boldsymbol{x} - \boldsymbol{x}_i\|^2); \gamma > 0$, for the SVM.

SVMs are believed to be less prone to the class imbalance problem, since boundaries between classes are calculated with respect to only a few support vectors and the class sizes may not affect the class boundary too much [29]. Nevertheless, we use a cost-sensitive approach with SVM where different penalty parameters are used for different classes, making errors on the minority instances costlier than the majority class instances.

In our case, we employ GA for searching an optimal misclassification cost setup. Let $C_1, C_2$ and $C_3$ denote the misclassification costs for the classes 1 (normal stage), 2 (early stage) and 3 (moderate stage), respectively. The cost vector $[C_1, C_2, C_3]$ which is encoded into the fitness function (which the GA optimizes). We have used the average F-measure (the mean of F-measures of all the classes) as the fitness function for the GA. The final output of the GA is a vector which yields the most satisfactory performance among all tests. A brief description of the working of GA is given in Appendix B.

### 2.5.3 Cost-sensitive AdaBoost-based classifier ensemble

The AdaBoost (Adaptive Boosting) algorithm is an effective boosting algorithm which not only improves the classification accuracies of 'weak' learners, but is also an attractive technique for tackling the class imbalance problem. We use the multi-class extension of AdaBoost which is the *AdaBoost.M2* algorithm as proposed by Freund and Schapire [35], for learning the weak learners (in our case, classification trees). The optimal cost (in our case, $[C_{ab1}, C_{ab2}, C_{ab3}]$ where $C_{abi}$ is the misclassification cost for class *i*) for each class is obtained using GA. The steps to create AdaBoost-based classifier ensemble is as follows:

*Input and Initialization*

1. Given the training set $S = \{(\boldsymbol{x}_i, y_i)\}$.

2. Let the number of iterations (which is the number of weak learners in the ensemble) be $T_{AB}$.

3. Let $B = \{(i, y);\ i \in \{1,2, \dots, n\}, y \in Y - \{y_i\}$

*Algorithm*

1. Initialize the weight $D_1(i, y) = \frac{1}{|B|}$ for $(i, y) \in B$

2. for *t* = *1* to $T_{AB}$

   a) Choose a weak learner, $h_t: X \times Y \rightarrow [0, 1]$, such that it minimizes pseudo-loss error given by

$$e_t = \frac{1}{2} \sum_{i=1}^{n} \sum_{y \neq y_i} D_t(i, y)(1 - h_t(\boldsymbol{x}_i, y_i) + h_t(\boldsymbol{x}_i, y)) \tag{7}$$

where $D_t(i, y)$ contains observation weights at step $t$ for class $y$, $h_t(\boldsymbol{x}_i, y) \in [0,1]$ is the confidence of prediction of sample $\boldsymbol{x}_i$ into class $y$ by the weak learner at step *t*, and $y_i$ is the true class.

   b) Compute the weight update parameter, $\beta_t$ given by

$$\beta_t = \frac{e_t}{1 - e_t} \tag{8}$$

   c) Update the weight distribution using

$$D_{t+1}(i, y) = \frac{D_t(i, y)}{Z_t} \beta_t^{\frac{1}{2}(1 + h_t(\boldsymbol{x}_i, y_i) - h(\boldsymbol{x}_i, y))} \tag{9}$$

where $Z_t = \sum_i D_t(i)$ is the normalization constant.

3. Output the final hypothesis as

$$h_{AB}(\boldsymbol{x}) = \arg\max_{y \in Y} \sum_{t=1}^{T_{AB}} \left(\log \frac{1}{\beta_t}\right) h_t(\boldsymbol{x}, y) \tag{10}$$

The final hypothesis, $h_{AB}$ outputs the label $y$, for an input instance $\boldsymbol{x}$**,** that maximizes a weighted average of the weak hypothesis values $h_t(\boldsymbol{x}, y)$.

### 2.5.4 RUSBoost-based classifier ensemble

RUSBoost is a hybrid technique combining data sampling and boosting algorithm designed to improve performance of models on skewed data [25]. The data sampling used is *random undersampling* (RUS) technique in which it removes examples from the majority class at random until a desired class distribution is achieved. The sampling proportion which is required for RUSBoost algorithm is, often, unknown. We use GA to find a near optimal sampling proportion for each class (in our case $[p_1, p_2, p_3]$ where $p_i$ is the sampling proportion for class *i*). If the sampling proportion leads to a value that is larger than the number of samples in a class, then RUSBoost samples the members with replacement, otherwise it samples the members without replacement. The steps to create a RUSBoost-based classifier ensemble are as follows:

*Input and Initialization*

1. Given the training set $S = \{(\boldsymbol{x}_i, y_i)\}$.

2. Let the number of iterations (which is the number of weak learners in the ensemble) be $T_{RUS}$

3. Let the proportion of samples in a training set after random undersampling be $[p_1, p_2, \ldots, p_k]$ by which the number of samples in class $k$ is $p_k$ times number of samples in the minority class.

*Algorithm*

1. Initialize the weights $D_1(i) = \frac{1}{n}$ for all $i$.

2. For $t = 1$ to $T_{RUS}$

   a) Create temporary training set $S_t^{'}$ with weight distribution $D_t^{'}$ using *random undersampling*.

   b) Choose a (any) weak learner which creates the hypothesis, $h_t: X \times Y \rightarrow [0, 1]$ when $S_t^{'}$ and $D_t^{'}$ are passed to it.

   c) Compute the pseudo-loss error (based on original training set $S$ and weight distribution $D_t$) of $h_t$ using (7).

   d) Compute the weight update parameter $\beta_t$ using (8).

   e) Update the weight distribution $D_{t+1}$ using (9).

3. Output the final hypothesis $h_{RUS}$ given by

$$h_{RUS}(\boldsymbol{x}) = \arg\max_{y \in Y} \sum_{t=1}^{T_{RUS}} \left(\log \frac{1}{\beta_t}\right) h_t(\boldsymbol{x}, y) \tag{11}$$

### 2.5.5 *k*-Nearest Neighbor classifier

From the input data, the *k*-nearest neighbor technique finds the *k* closest points for a test point and assigns the point to that class which is in majority [26]. The distance function used to estimate closeness is Euclidean.

### 2.5.6 Probabilistic Generative Model

In this approach, the conditional probability distribution $P(k|x)$ is estimated in an inference stage, and then subsequently uses this distribution to make optimal predictions [28]. The optimization function is given by

$$\hat{y} = \underset{y=1,2,3}{\operatorname{argmin}} \sum_{k=1}^{3} P(k|x)C(y|k) \tag{12}$$

where $\hat{y}$ is the predicted class, $P(k|x)$ is the posterior probability of class *k* for observation *x*, and $C(y|k)$ is the cost of classifying an observation as *y* when its true class is *k*.

The posterior probability $P(k|x)$ is the product of the prior probability $P(k)$ and the class-conditional densities $P(x|k)$. Taking multivariate normal density as the conditional density, the density function of the multivariate normal with mean ($\mu_k$) and covariance ($\Sigma_k$) at a point *x* is given by

$$P(x|k) = \frac{1}{(2\pi|\Sigma_k|)^{0.5}} \exp\left(-0.5 * (x - \mu_k)^T \Sigma_k^{-1} (x - \mu_k))\right) \tag{13}$$

and the posterior probability $P(k|x)$ is obtained by Bayes' theorem given by

$$P(k|x) = \frac{P(x|k)P(k)}{P(x)} \tag{14}$$

The cost (or prior) could be optimally selected for imbalance class problems. For finding the optimal parameters, genetic algorithm was used.

### 2.5.7 Neural Networks

For the study, feedforward neural networks were used. For training is carried gradient descent algorithm with gradients computed via backpropagation [28]. The weight and bias update equations are given by

$$w_{new} = w_{old} - \eta \frac{\delta L(w,b)}{\delta w} \tag{15}$$

$$b_{new} = b_{old} - \eta \frac{\delta L(w,b)}{\delta b} \tag{16}$$

Along with this, we also try with multilayer feedforward neural networks with 2 hidden layers which is a deep learning technique. The optimal number of neurons in the hidden layer is found using genetic algorithm.

## 2.6 Estimation of feature importance in PD

In our study, we estimate the importance of the features in a 2-class (PD/Normal) classification framework, by computing an importance score for these features. The technique used is the Random forests algorithm [27].

The Random forests algorithm involves drawing *n* out of *n* observations with replacement which omits on average 37% of the observations for each decision tree. These constitute the *'out-of-bag'* observations, and they can be used to estimate feature importance. An out-of-bag estimate of feature importance is obtained by randomly permuting out-of-bag data across one feature at a time, and estimating the increase in the out-of-bag error due to this permutation. The larger the increase, higher the importance score of the feature. The out-of-bag error is estimated by comparing out-of-bag predicted responses and observed responses for all observations used for training. The Random forests algorithm is briefly described below:

*Input and Initialization*

1. Given the training set $S = \{(\boldsymbol{x}_i, y_i)\}$.
2. Let the number of weak learners be $T_{RF}$.

*Algorithm*

1. for *f* = *1* to $T_{RF}$

a) Select a bootstrap sample $[X_f, Y_f]$ by randomly selecting *n* out of *n* observations with replacement from the training set *S*.

b) Construct a decision tree $T_f$ from $[X_f, Y_f]$ by choosing a random subset of *m'* (where $m' = \sqrt{m}$, in our case) from *m* features for every decision split, and recursively repeat this splitting for each terminal node of the tree until a minimum node size is reached.

3. Output the ensemble of trees given by $R_f = \{T_f\}, f = 1, .. T_{RF}$.

## 3. Results and Discussion

### 3.1 Statistical testing of features

The results of the statistical tests on the features are shown in Table 6 as shown below. It was observed that all 59 features are highly statistically significant, and thus are used for subsequent prediction modelling.

**Table 6: Results of Kruskal-Wallis test and Wilcoxon rank sum test**

| *Sno* | *Feature* | *Abbrv.* | *Chi-sq* | *z-stat* | *p-val** |
|---|---|---|---|---|---|
| 1 | Cognitive impairment | P1_COG | 245.45 | 14.07 | 0.00 |
| 2 | Hallucination & Psychosis | P1_HALL | 95.10 | 8.96 | 0.00 |
| 3 | Depressed mood | P1_DPRS | 136.67 | 10.12 | 0.00 |
| 4 | Anxious mood | P1_ANXS | 118.56 | 10.56 | 0.00 |
| 5 | Apathy | P1_APAT | 218.67 | 13.53 | 0.00 |
| 6 | Features of DDS | P1_DDS | 65.09 | 5.65 | 0.00 |
| 7 | Sleep problems | P1_SLPN | 73.17 | 7.96 | 0.00 |
| 8 | Daytime sleepiness | P1_SLPD | 195.45 | 13.51 | 0.00 |
| 9 | Pain & other sensations | P1_PAIN | 147.11 | 11.86 | 0.00 |
| 10 | Urinary problems | P1_URIN | 296.58 | 16.51 | 0.00 |
| 11 | Constipation | P1_CNST | 391.59 | 19.54 | 0.00 |
| 12 | Light headedness | P1_LTHD | 267.03 | 14.62 | 0.00 |
| 13 | Fatigue | P1_FATG | 262.23 | 15.04 | 0.00 |
| 14 | Speech | P2_SPCH | 581.13 | 23.57 | 0.00 |
| 15 | Saliva & Drooling | P2_SALV | 488.27 | 21.27 | 0.00 |
| 16 | Chewing & Swallowing | P2_SWAL | 281.53 | 14.57 | 0.00 |
| 17 | Eating tasks | P2_EAT | 580.48 | 23.50 | 0.00 |
| 18 | Dressing | P2_DRES | 789.94 | 27.61 | 0.00 |
| 19 | Hygiene | P2_HYGN | 497.61 | 21.48 | 0.00 |
| 20 | Handwriting | P2_HWRT | 988.24 | 31.36 | 0.00 |
| 21 | Doing hobbies | P2_HOBB | 671.46 | 25.50 | 0.00 |
| 22 | Turning in bed | P2_TURN | 499.88 | 21.85 | 0.00 |
| 23 | Tremor | P2_TRMR | 1467.90 | 38.80 | 0.00 |
| 24 | Getting out of bed | P2_RISE | 628.25 | 24.23 | 0.00 |
| 25 | Walking & Balance | P2_WALK | 677.54 | 23.61 | 0.00 |
| 26 | Freezing | P2_FREZ | 308.70 | 10.43 | 0.00 |
| 27 | Speech | P3_SPCH | 859.31 | 28.93 | 0.00 |
| 28 | Facial expression | P3_FACXP | 1768.92 | 41.97 | 0.00 |
| 29 | Rigidity neck | P3_RIGN | 910.68 | 29.98 | 0.00 |
| 30 | Rigidity Right Upper Extremity (RUE) | P3_RIGRU | 1220.12 | 35.25 | 0.00 |
| 31 | Rigidity Left Upper Extremity (LUE) | P3_RIGLU | 1063.47 | 32.75 | 0.00 |
| 32 | Rigidity Right Lower Extremity (RLE) | P3_RIGRL | 780.96 | 27.85 | 0.00 |
| 33 | Rigidity Left Lower Extremity (LLE) | P3_RIGLL | 695.42 | 25.70 | 0.00 |
| 34 | Finger tapping Right Hand (RH) | P3_FTAPR | 1179.13 | 34.50 | 0.00 |
| 35 | Finger tapping Left Hand (LH) | P3_FTAPL | 1026.36 | 31.81 | 0.00 |
| 36 | Hand movements RH | P3_HMOVR | 1017.20 | 31.93 | 0.00 |
| 37 | Hand movements LH | P3_HMOVL | 968.97 | 30.77 | 0.00 |
| 38 | PS movements RH | P3_PRSPR | 1044.02 | 32.62 | 0.00 |
| 39 | PS movements LH | P3_PRSPL | 913.79 | 30.14 | 0.00 |
| 40 | Toe tapping Right Foot (RF) | P3_TTAPR | 987.34 | 30.83 | 0.00 |
| 41 | Toe tapping Left Foot (LF) | P3_TTAPL | 897.83 | 29.43 | 0.00 |
| 42 | Leg Agility Right Leg (RL) | P3_LGAGR | 734.22 | 25.84 | 0.00 |
| 43 | Leg Agility Left Leg (LL) | P3_LGAGL | 704.55 | 25.65 | 0.00 |
| 44 | Arising from chair | P3_RISNG | 485.79 | 15.11 | 0.00 |
| 45 | Gait | P3_GAIT | 1124.80 | 32.40 | 0.00 |
| 46 | Freezing of gait | P3_FRZGT | 160.15 | 5.36 | 0.00 |
| 47 | Postural stability | P3_PSTBL | 922.82 | 9.84 | 0.00 |
| 48 | Posture | P3_POSTR | 807.78 | 27.28 | 0.00 |
| 49 | Global spontaneity of movement | P3_BRADY | 1793.20 | 42.25 | 0.00 |
| 50 | Postural tremor RH | P3_PTRMR | 313.77 | 17.52 | 0.00 |
| 51 | Postural tremor LH | P3_PTRML | 304.04 | 16.76 | 0.00 |
| 52 | Kinetic tremor RH | P3_KTRMR | 252.58 | 15.78 | 0.00 |
| 53 | Kinetic tremor LH | P3_KTRML | 274.24 | 16.12 | 0.00 |
| 54 | Rest Tremor Amplitude (RTA) − RUE | P3_RTARU | 589.76 | 24.46 | 0.00 |
| 55 | RTA − LUE | P3_RTALU | 479.45 | 21.98 | 0.00 |
| 56 | RTA − RLE | P3_RTARL | 160.67 | 12.77 | 0.00 |
| 57 | RTA − LLE | P3_RTALL | 134.98 | 11.69 | 0.00 |
| 58 | RTA − Lip/jaw | P3_RTALJ | 80.15 | 9.12 | 0.00 |
| 59 | Constancy of rest tremor | P3_RTCON | 1127.34 | 33.80 | 0.00 |

* All *p*-values were $\ll 0.01$; Abbrv. is abbreviation. *Chi-sq, z-stat* and *p*-val stand for the Chi-square statistic, value of the *z*-statistic and *p*-value. Here, as both Kruskal-Wallis test and Wilcoxon gave very low *p*-values for all features, only one column for *p*-value is shown.

### 3.2 Logistic Model

Table 7 shows the logistic model obtained from the complete data. This model shows no numerical problems, such as multicollinearity among the variables, as observed from the low standard errors (all SE's<2) for the regression coefficients.

The mean Brier score is obtained as 0.0242 which is very low indicating the closeness of probabilistic predictions with actual outcomes. The expected value of the class is computed which is used as an estimate of severity of PD in this study (formula is given in Appendix A). To provide a visual demonstration of the predictions from the logistic model, we plot the expected value of the class with the MDS-UPDRS total score (total from Parts I, II and III) and is shown in Fig. 5. We observe that the lower expected values correctly correspond to normal cases and the higher probabilities correctly correspond to PD. It is interesting to observe that 0.5 nicely seperates normal from early stage PD and 1.5 seperates moderate stage PD from early stage PD, as some clinicians use the modified version of the HY scale, sometimes known under the designation "modified HY" which includes 0.5 increments.

**Table 7: The logistic model**

| No. | Features | $\beta$ | SE | No. | Features | $\beta$ | SE |
|---|---|---|---|---|---|---|---|
| 1. | Cognitive impairment | -0.2586 | 0.21 | 31. | Rigidity LUE | -0.6447 | 0.22 |
| 2. | Hallucination & Psychosis | 0.5336 | 0.34 | 32. | Rigidity RLE | -0.0697 | 0.21 |
| 3. | Depressed mood | -0.0031 | 0.21 | 33. | Rigidity LLE | 0.0716 | 0.23 |
| 4. | Anxious mood | 0.4439 | 0.20 | 34. | Finger tapping RH | -0.3170 | 0.23 |
| 5 | Apathy | 0.1716 | 0.25 | 35. | Finger tapping LH | -0.0513 | 0.24 |
| 6. | Features of DDS* | -0.0466 | 0.38 | 36. | Hand movements RH | 0.3249 | 0.25 |
| 7. | Sleep problems | 0.1595 | 0.11 | 37. | Hand movements LH | 0.0811 | 0.27 |
| 8. | Daytime sleepiness | -0.0970 | 0.13 | 38. | PS movements RH | -0.5399 | 0.23 |
| 9. | Pain & other sensations | 0.0236 | 0.13 | 39. | PS movements LH | -0.4888 | 0.23 |
| 10. | Urinary problems | -0.2205 | 0.14 | 40. | Toe tapping- RF | -0.4750 | 0.22 |
| 11. | Constipation | -0.3703 | 0.17 | 41. | Toe tapping- LF | 0.0258 | 0.20 |
| 12. | Light headedness | -0.5029 | 0.18 | 42. | Leg Agility- RL | -0.0769 | 0.27 |
| 13. | Fatigue | 0.1749 | 0.16 | 43. | Leg Agility- LL | 0.2351 | 0.25 |
| 14. | Speech | -0.2895 | 0.19 | 44. | Arising from chair | 0.0540 | 0.27 |
| 15. | Saliva & Drooling | -0.2749 | 0.13 | 45. | Gait | -1.0285 | 0.23 |
| 16. | Chewing & Swallowing | -0.2342 | 0.25 | 46. | Freezing of gait | 1.2618 | 0.54 |
| 17. | Eating tasks | 0.2484 | 0.25 | 47. | Postural stability | -3.4213 | 0.25 |
| 18. | Dressing | -0.0440 | 0.28 | 48. | Posture | 0.5660 | 0.19 |
| 19. | Hygiene | -0.3696 | 0.33 | 49. | Global spontaneity of movement | -1.3592 | 0.23 |
| 20. | Handwriting | -0.6869 | 0.15 | 50. | Postural tremor RH | -0.0044 | 0.24 |
| 21. | Doing hobbies | -0.1651 | 0.21 | 51. | Postural tremor LH | -0.1570 | 0.28 |
| 22. | Turning in bed | 0.1648 | 0.27 | 52. | Kinetic tremor RH | 0.2299 | 0.27 |
| 23. | Tremor | -1.0636 | 0.19 | 53. | Kinetic tremor LH | -0.3255 | 0.26 |
| 24. | Getting out of bed | 0.7465 | 0.24 | 54. | RTA − RUE | 0.1879 | 0.21 |
| 25. | Walking & Balance | -0.4961 | 0.20 | 55. | RTA − LUE | 0.0575 | 0.23 |
| 26. | Freezing | 0.0769 | 0.32 | 56. | RTA − RLE | 0.3132 | 0.29 |
| 27. | Speech | 0.1759 | 0.25 | 57. | RTA − LLE | -0.3221 | 0.40 |
| 28. | Facial expression | -1.4895 | 0.21 | 58. | RTA − Lip/jaw | 0.2853 | 0.38 |
| 29. | Rigidity neck | -0.0251 | 0.19 | 59. | Constancy of rest tremor | -0.9627 | 0.20 |
| 30. | Rigidity RUE | -0.4709 | 0.19 | | | | |

DDS−Dopamine Dysregulation Syndrome, PS− Pronation-Supination, RTA− Rest Tremor Amplitude, RUE−Right Upper Extremity, LUE−Left Upper Extremity, RLE−Right Lower Extremity, LLE−Left Lower Extremity, RH−Right Hand, LH−Left Hand, RL−Right Leg,

LL−Left Leg, RF−Right Foot, and LF−Left Foot. β is the regression coefficient for a predictor in the logistic model, SE is its standard error. The intercepts, $\alpha_1$ and $\alpha_2$, are obtained as 3.5843 and 21.1390, respectively. The logistic model does not show any numerical instability as observed from the very low standard errors of the regression coefficients.

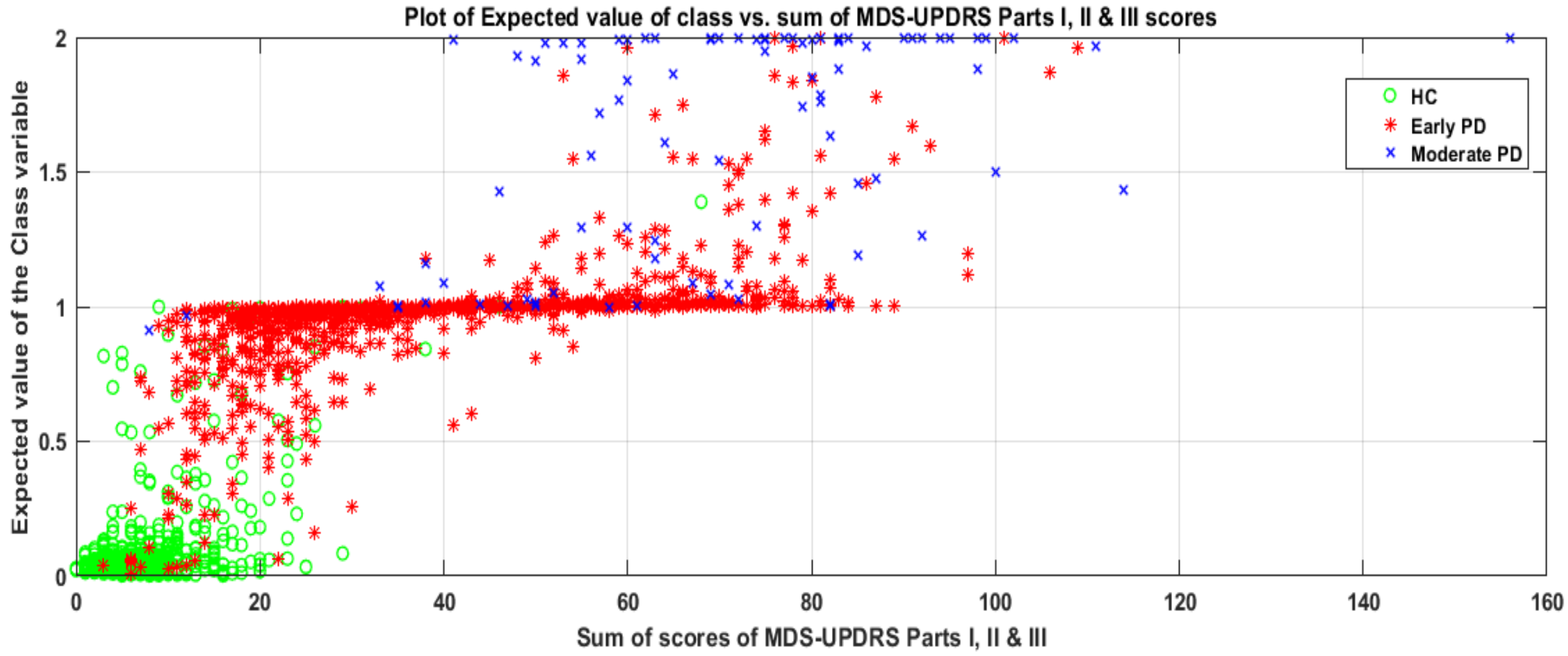

Fig. 5. The plot of expected value of class with the sum of scores from MDS-UPDRS Parts I, II and III. It is observed that the higher expected values (which is the severity of the disease) rightly corresponded to the diseased cases and the lower values to the normal ones.

We also estimate Spearman's rank correlation coefficients between a feature and the true stage (represented as $R$), and between a feature and the expected value of the class which is the predicted stage (represented as $R'$). Fig. 6 shows the values of $R$ and $R'$ for each feature. It is observed that $R'$ for all features, except for the feature corresponding to sleep problems, is higher than $R$. The slightly lower $R'$ for the sleep feature may be because, sleep disorders are very common among the elderly irrespective of PD. A large-scale study by Foley *et al.* [36], involving over 9000 elderly subjects, observed that around 57% reported at least one sleep problem (trouble falling asleep, waking up, awaking too early, needing to nap and not feeling rested).

This is also reflected in the feature severity plot as shown in Fig. 2, where we observe that the two most prominent features showing a sort of severity in normal stage population are the ones that are related to sleep (sleep problems and daytime sleepiness).

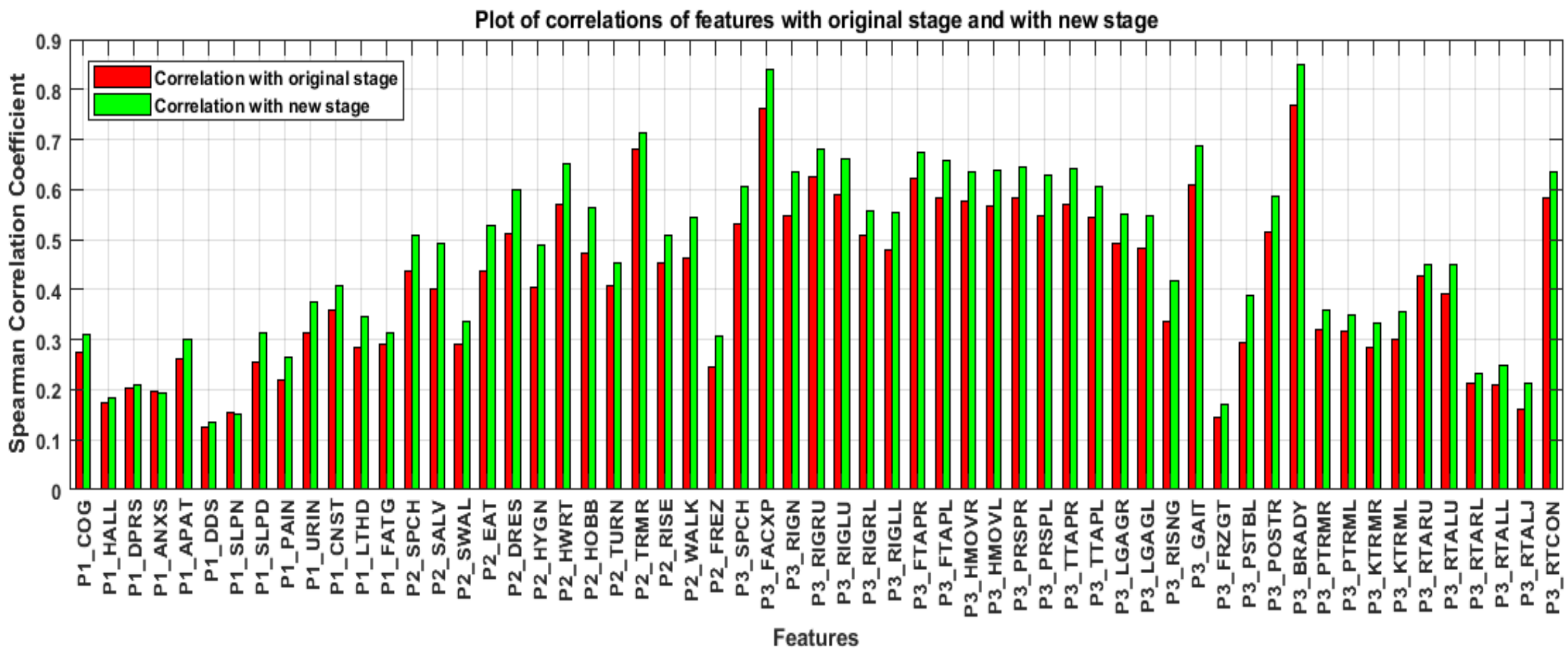

Fig. 6. Plot of correlation of features with original stage and with the predicted stage which is the expected value of the class variable. Almost all the features showed higher correlation with the expected value than with the original stage.

Tremor (P2_TRMR, feature no. 23), loss of facial expression or hypomimia (P3_FACXP, feature no. 28) and global spontaneity of movement or body bradykinesia (P3_BRADY, feature no. 49) were among the most correlated features with the stage, which is consistent with the clinical perspective of the disease [2]. When we observe the most correlated features with the expected value of the class, gait (P3_GAIT, feature no. 45) also shows higher importance along with these 3 features. This higher correlation of gait feature may be because, to evaluate this item in the MDS-UPDRS (testing of gait), it involves walking of the patient away from and towards the examiner, and involves measurement of multiple behaviors: stride amplitude, stride speed, height of foot lift, heel strike during walking, turning, and arm swing, but not freezing, which may have correlation with other features in the MDS-UPDRS. Gait disturbances such as freezing of gait (P3_FRZGT, feature no. 46) showed a very low correlation with both the stage and the expected value of the class (both $R$ and $R'$) which is consistent with previous studies which suggest that these conditions usually occur in advanced PD [37].

Overall, the logistic model performed well as observed from the very low Brier score indicating that lower probabilities correctly corresponded to normal observations and higher probabilities correctly corresponded to PD cases, and higher Spearmen's rank correlation of features with the expected value of the class than with the true category.

### 3.3 Cost and sampling proportion estimation using GA

Genetic algorithm (GA) is used to estimate near optimal costs for SVM and AdaBoost-based ensemble, RUSBoost-based ensemble, Random forests, probabilistic generative model, neural networks (also deep learning). The fitness function used for the GA is the average F-measure of the 3 classes. The other options used for GA are as follows: *Crossover fraction* as 0.8, *elite count* as 2 for both SVM and AdaBoost-based ensemble, and 8 for RUSBoost-based ensemble, *mutation function* is such that it randomly generates directions that are adaptive with respect to the last successful or unsuccessful generation, and the *number of generations* is set as 50. The initial population range was carefully chosen based on empirical experiments.

Fig. 7 shows the plots for the genetic algorithm based parameter selection for different classifiers. We can observe that generally mean errors and least error in a population decreases with iteration, except for RUSBoost, Random forests, Neural Networks and Deep learning methods. For RUSBoost and Random forests, it is due to the inherent sampling involved in these methods and for neural networks and deep learning, it is due to the non-convex optimization which leads to different convergences for different initializations. Fig. 7 also shows the mean error from the population along with the error for the best individual. The difference between the error from the best individual and other members in the population are the following: for Adaboost it is 0.0119, for RUSBoost it is 0.0308, for Random forests it is 0.0196, for SVM it is 0, for probabilistic generative model, it is 0, for neural networks it is 0.024, and for deep learning it is 0.032. The error of 0 means all the members in the present generation are same.

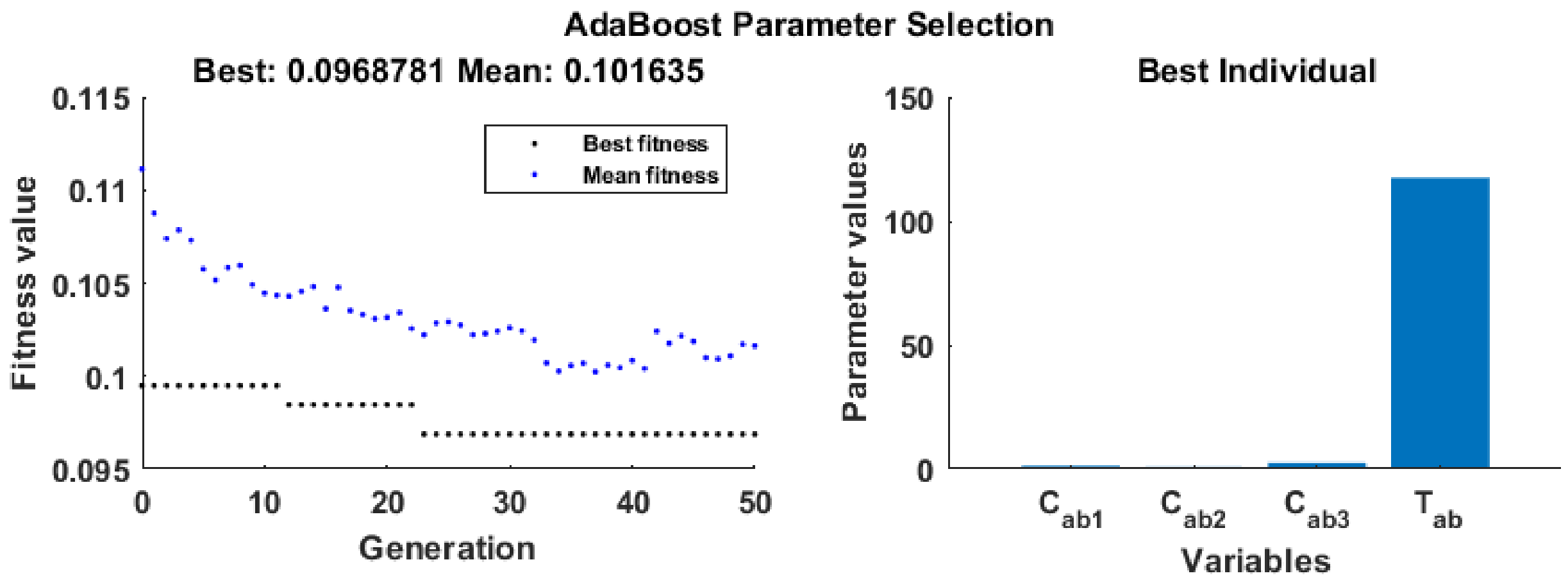

(a) AdaBoost

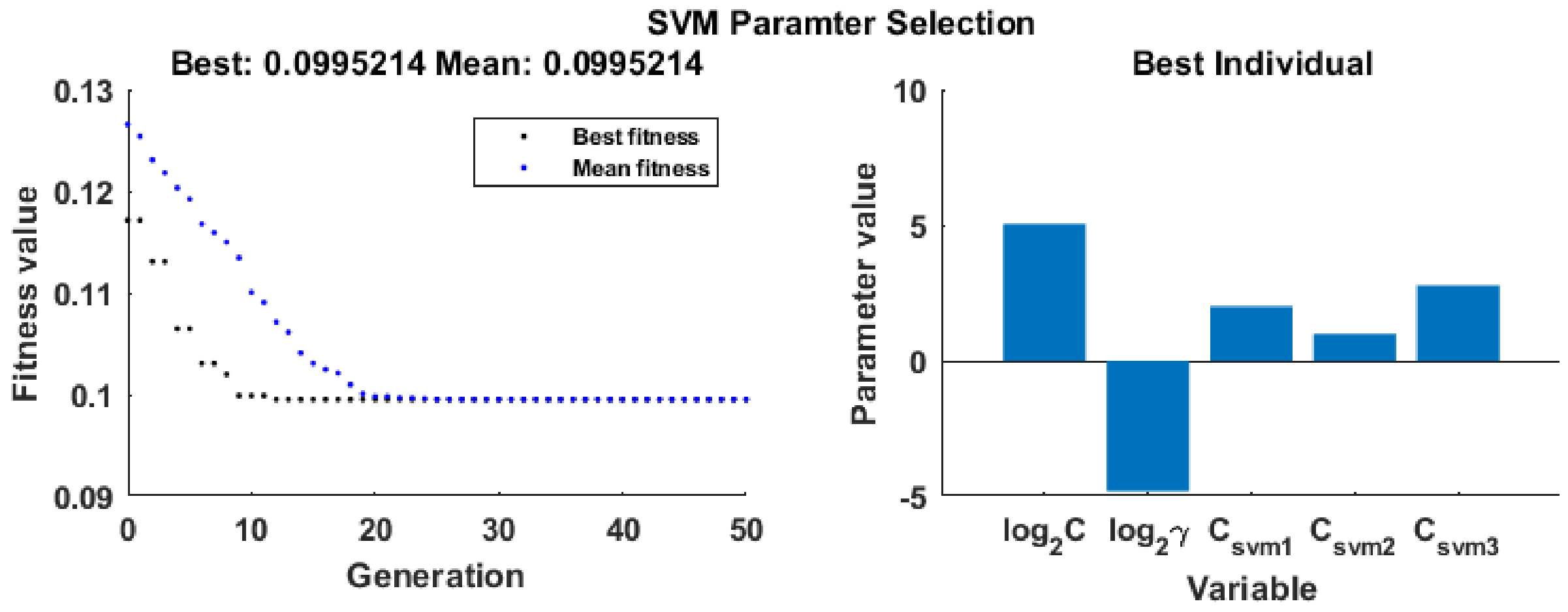

(b) SVM

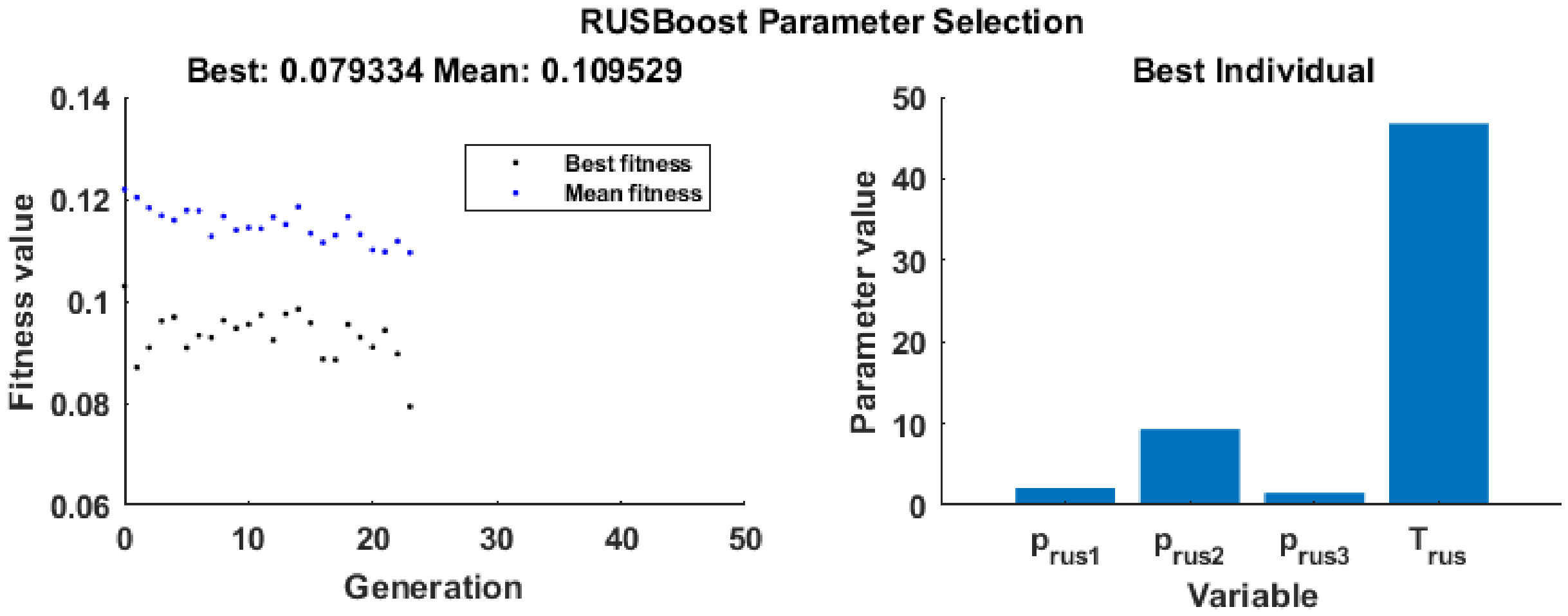

(c) RUSBoost

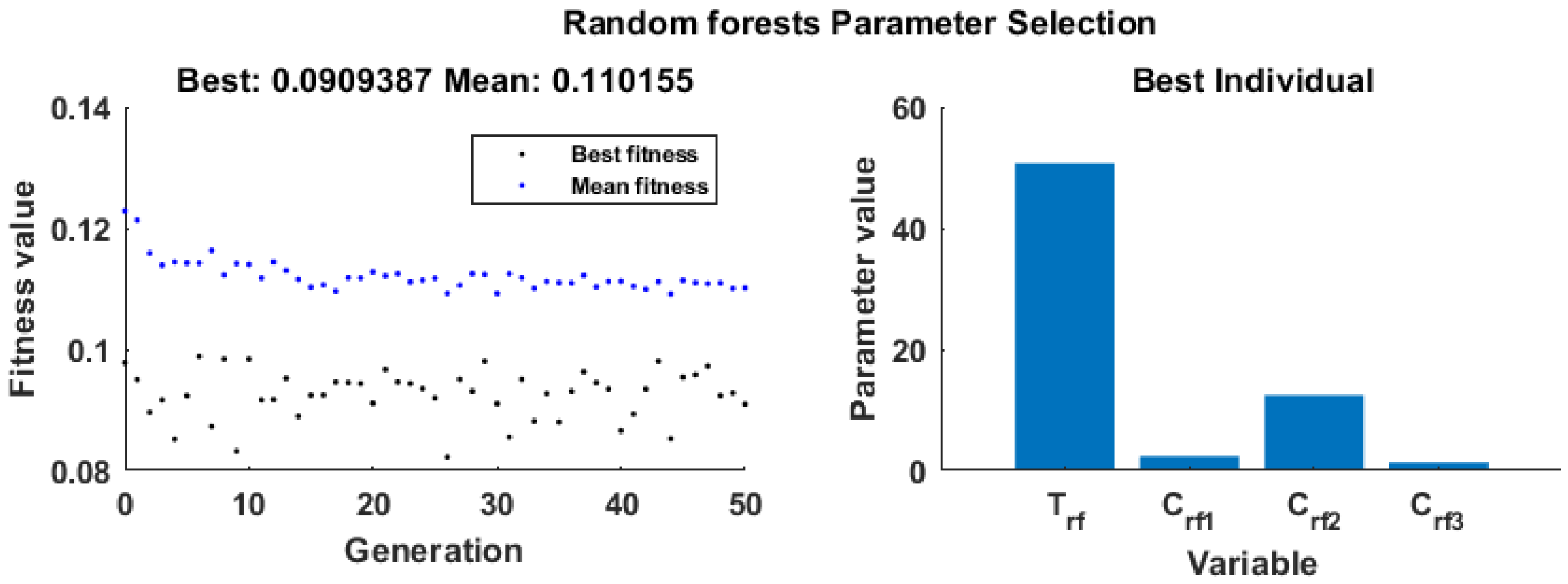

(d) Random forests

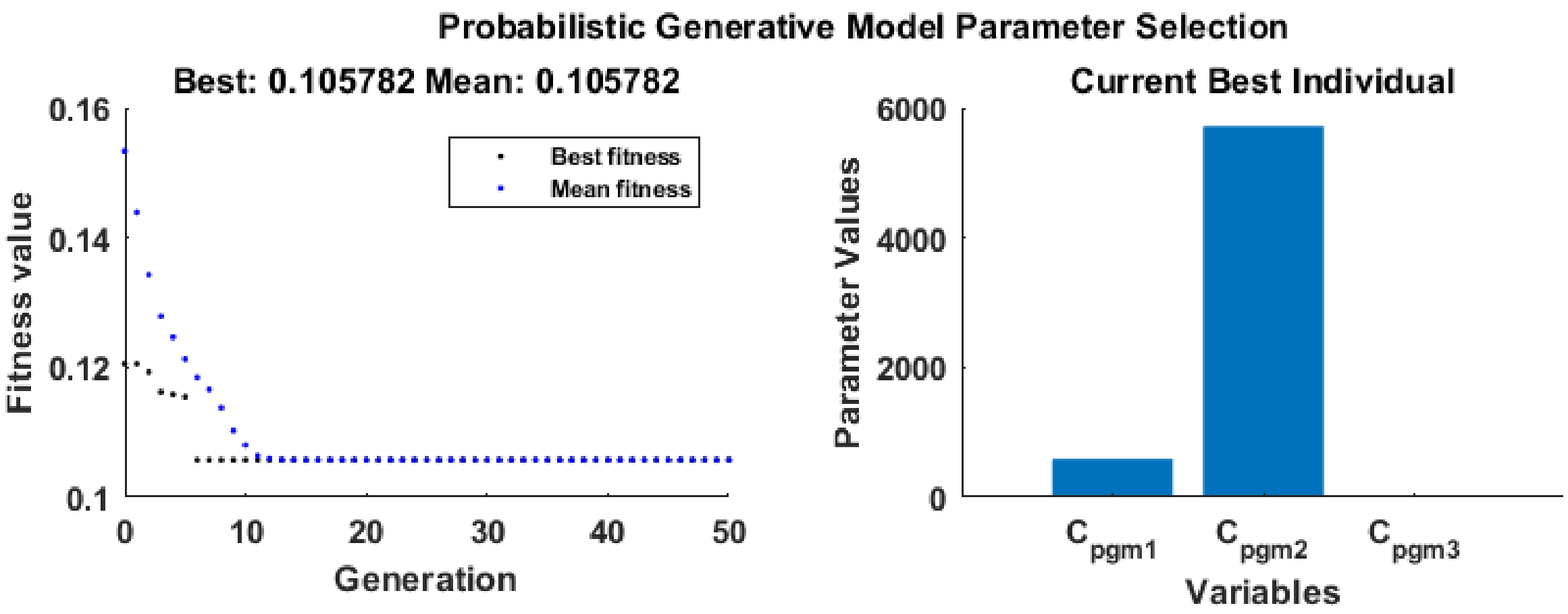

(e) Probabilistic generative model

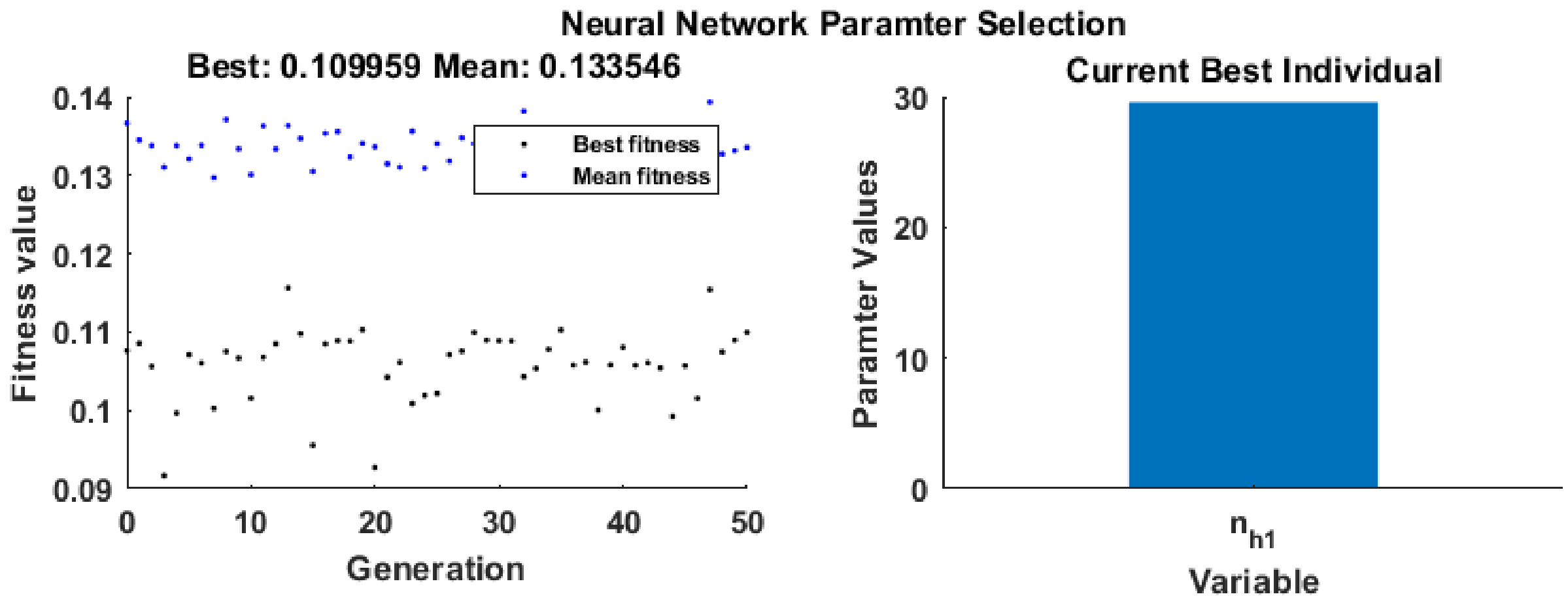

(f) Neural Networks

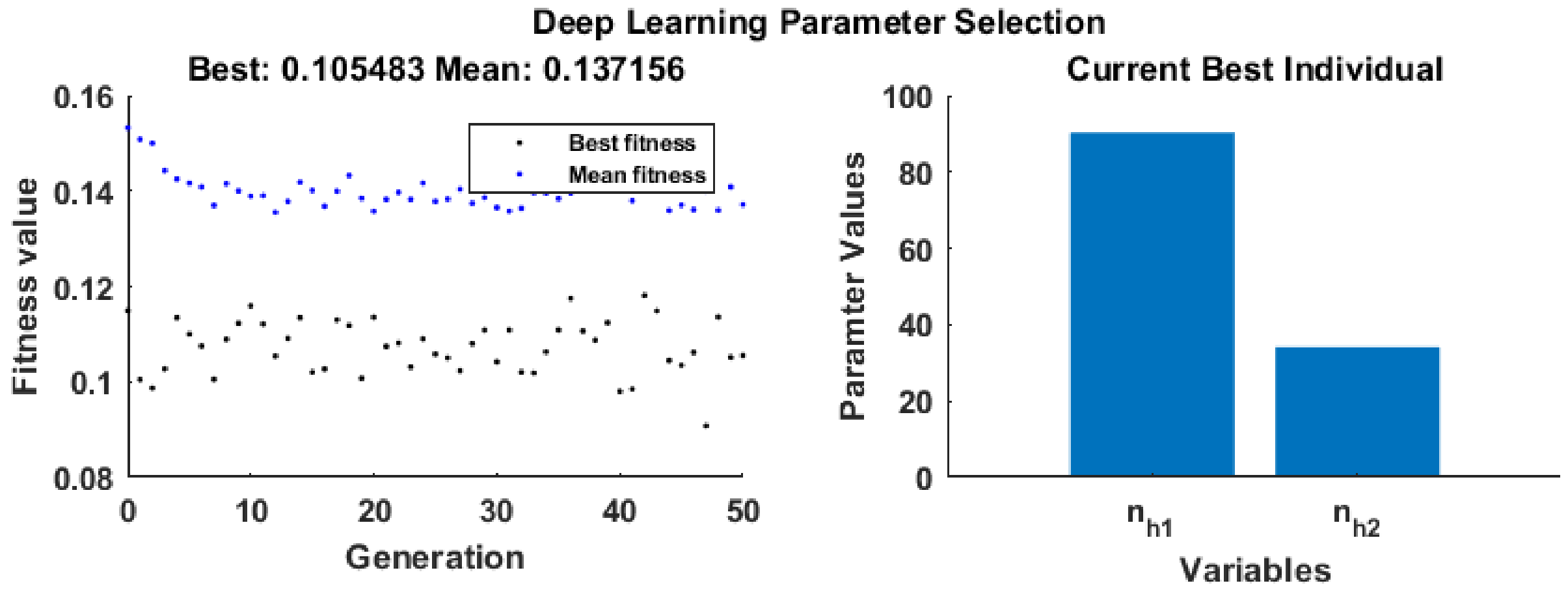

(g) Deep Learning

Fig. 7. Plots for the genetic algorithm based parameter selection for different classifiers. (a) Adaboost (b) SVM (c) RUSBoost (d)Random forests (e) Probabilistic generative model (f) Neural networks (g) Deep Learning

We used LIBSVM [30] library for implementing cost-sensitive SVM which had 5 parameters to be estimated: the cost $C$ of SVM, $\gamma$ for the RBF kernel, parameters $C_1, C_2, C_3$ for the three classes which are multiplied with $C$ which forms the misclassification costs for the 3 classes. The values obtained for $C, \gamma, C_{svm1}, C_{svm2}$ and $C_{svm3}$ are 33.0416, 0.0355, 2.0, 1.0 and 2.7733, respectively.

The AdaBoost-based ensemble had 4 parameters to be estimated: the misclassification costs $C_{ab\,1}, C_{ab\,2}, C_{ab\,3}$ for the 3 classes, and the number of weak learners $T_{ab}$. The optimal values for $C_{ab\,1}, C_{ab\,2}, C_{ab\,3}$ and $T_{ab}$ obtained using GA are 1.3651, 1.0003, 2.6583and 117, respectively. Likewise in SVM, the cost for the moderate stage PD class is higher than the other 2 classes.

The RUSBoost-based ensemble had 4 parameters to be estimated: the sampling proportions $p_{rus\,1}, p_{rus\,2}, p_{rus\,3}$ for the 3 classes, and the number of weak learners $T_{rus}$. The values for $p_{rus\,1}, p_{rus\,2}, p_{rus\,3}, T_{rus}$ were obtained as 2.0781, 9.3262, 1.5190 and 47 , respectively. The sampled training set contains highest number of examples from the early PD category.

The Random forests had 4 parameters to be estimated: the misclassification costs $C_{rf\,1}, C_{rf\,2}, C_{rf\,3}$ for the 3 classes, and the number of weak learners $T_{rf}$. The near optimal values for $C_{rf\,1}, C_{rf\,2}, C_{rf\,3}$ and $T_{rf}$ obtained using GA are 2.3732, 12.4722, 1.3256 and 51, respectively.

The probabilistic generative model had 3 parameters to be estimated: the misclassification costs $C_{pgm\,1}, C_{pgm\,2}, C_{pgm\,3}$ for the 3 classes which were obtained as 598.72, 5724.99 and 28.88, respectively

Neural networks had one parameter to be estimated which was the number of hidden neurons ($n_{h1}$) in the hidden layer which was obtained as 30. And for the multi-layered neural network (deep learning), there were two hidden layers and the number of neurons in these layers $[n_{h1}, n_{h2}]$ were estimated as [100, 38].

### 3.4 Predictive modeling and performance comparison of models

*'Classification trees'* is used as the *'weak learners'* for developing AdaBoost-based ensemble, RUSBoost-based ensemble, and Random forests. The minimum size of the leaf node for a tree is taken as 1. Table 8 shows the average performance measures (overall accuracy, precision, recall, and F-measure) for the classifiers. The AdaBoost-based ensemble, Random forests, SVM and Probabilistic generative model gave very high performances as indicated by the accuracies and F-scores. Logistic regression gave relatively a lower accuracy and F-score indicating the non-linearity involved in the data.

**Table 8: Performance metrics of the classifiers used**

| | | Normal | | | Early | | | Moderate | | |
|---|---|---|---|---|---|---|---|---|---|---|
| ***Classifier*** | ***Acc*** | ***Prec*** | ***Recall*** | ***F-scr*** | ***Prec*** | ***Recall*** | ***F-scr*** | ***Prec*** | ***Recall*** | ***F-scr*** |
| ***SVM*** | 96.84 | 96.66 | 98.66 | 97.65 | 98.04 | 97.07 | 97.55 | 74.08 | 72.37 | 73.21 |
| ***AdaBoost*** | 97.46 | 98.32 | 97.98 | 98.15 | 97.14 | 98.96 | 98.04 | 93.40 | 61.06 | 73.85 |
| ***RUSBoost*** | 96.29 | 96.51 | 98.61 | 97.55 | 98.59 | 95.65 | 97.10 | 61.13 | 84.02 | 70.77 |
| ***RF*** | 97.13 | 98.63 | 97.06 | 97.84 | 96.80 | 98.79 | 97.78 | 85.05 | 63.76 | 72.89 |
| ***LR*** | 95.63 | 96.15 | 96.63 | 96.39 | 96.47 | 96.64 | 96.56 | 70.42 | 63.92 | 67.02 |
| ***k-NN*** | 95.56 | 95.22 | 98.69 | 96.93 | 96.87 | 96.15 | 96.51 | 65.21 | 49.27 | 56.13 |
| ***PGM*** | 96.38 | 96.06 | 97.54 | 96.80 | 97.86 | 96.60 | 97.23 | 72.05 | 79.06 | 75.40 |
| ***NB*** | 90.98 | 84.72 | 98.74 | 91.19 | 96.91 | 88.70 | 92.62 | 56.93 | 53.43 | 55.13 |
| ***NN*** | 96.73 | 96.81 | 98.13 | 97.46 | 97.26 | 97.70 | 97.48 | 81.20 | 61.54 | 70.01 |
| ***DL*** | 96.69 | 97.11 | 97.96 | 97.53 | 97.09 | 97.81 | 97.45 | 79.52 | 59.92 | 68.35 |

* All values in percentage. Acc, Prec and F-scr stands for accuracy, precision and F-score. SVM, AdaBoost, RUSBoost, RF, LR, *k*-NN, PGM, NN and DL stand for Support vector machine,

AdaBoost based ensemble. RUSBoost based ensemble, Random forests, Logistic regression, *k*-nearest neighbor, Probabilistic generative model, Neural network and Deep learning, respectively.

### 3.5 Error Analysis

To visualize the misclassified instances, we use the total of the scores from the Parts I, II and III of the MDS-UPDRS (sum of all 59 features) as the variable for plotting. Fig. 8 shows the box plot of the full data used in the study and the box plots of the observations which were misclassified by the high performing classifiers. From these box plots, we can observe that only those observations which did not follow a pattern were misclassified. Looking at the box plot of full data, it can be observed that the median total score of the normal group is much lower than the early stage group.This pattern is violated for the misclassified instances as observed from the box plots of the misclassified instances.

For instance, when we look at the misclassified instances from SVM, AdaBoost ensemble and Random forests, the median value of the total score of the misclassified early observations were much lower than the median value of the misclassified normal observations.

Normal stage observations with higher MDS-UPDRS scores were misclassified as early stage cases (but not as moderate stage) by the classifiers. Early stage observations with lower MDS-UPDRS scores were misclassified as normal, and with higher MDS-UPDRS scores were misclassified as moderate stage (this was very low in number). And misclassifications of moderate stage as normal or early stage also occurred in a similar manner. The spread of the misclassified moderate stage instances is similar for all the classifiers.

Most of the misclassifications occurred between early stage and moderate stage PD classes. This is because the severity of symptoms (or the features) in moderate PD is not severe, unlike in advanced PD, and due to which a good amount of overlap exists between the features for these two classes. This is also reflected in Fig. 2.

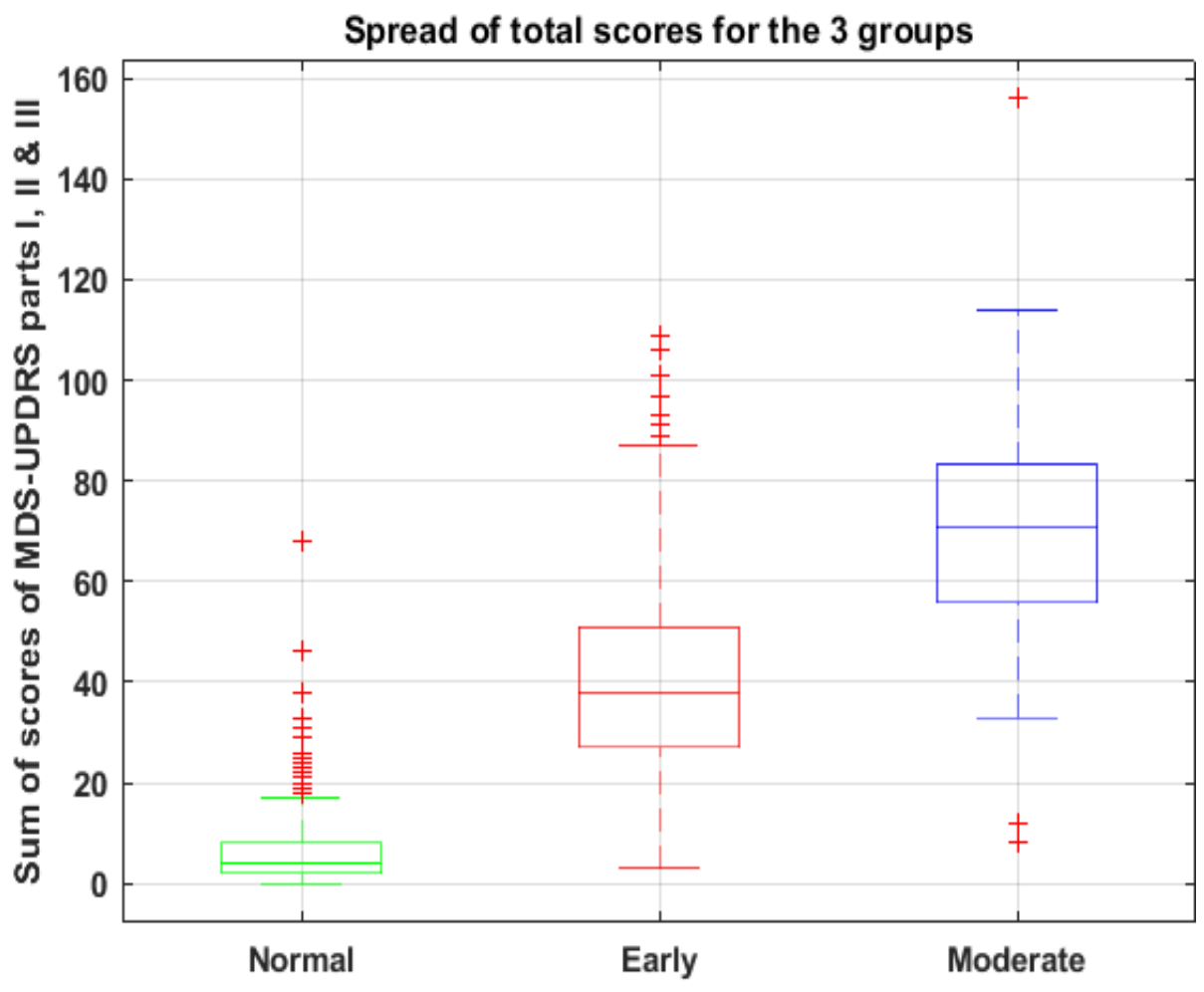

(a) Full data

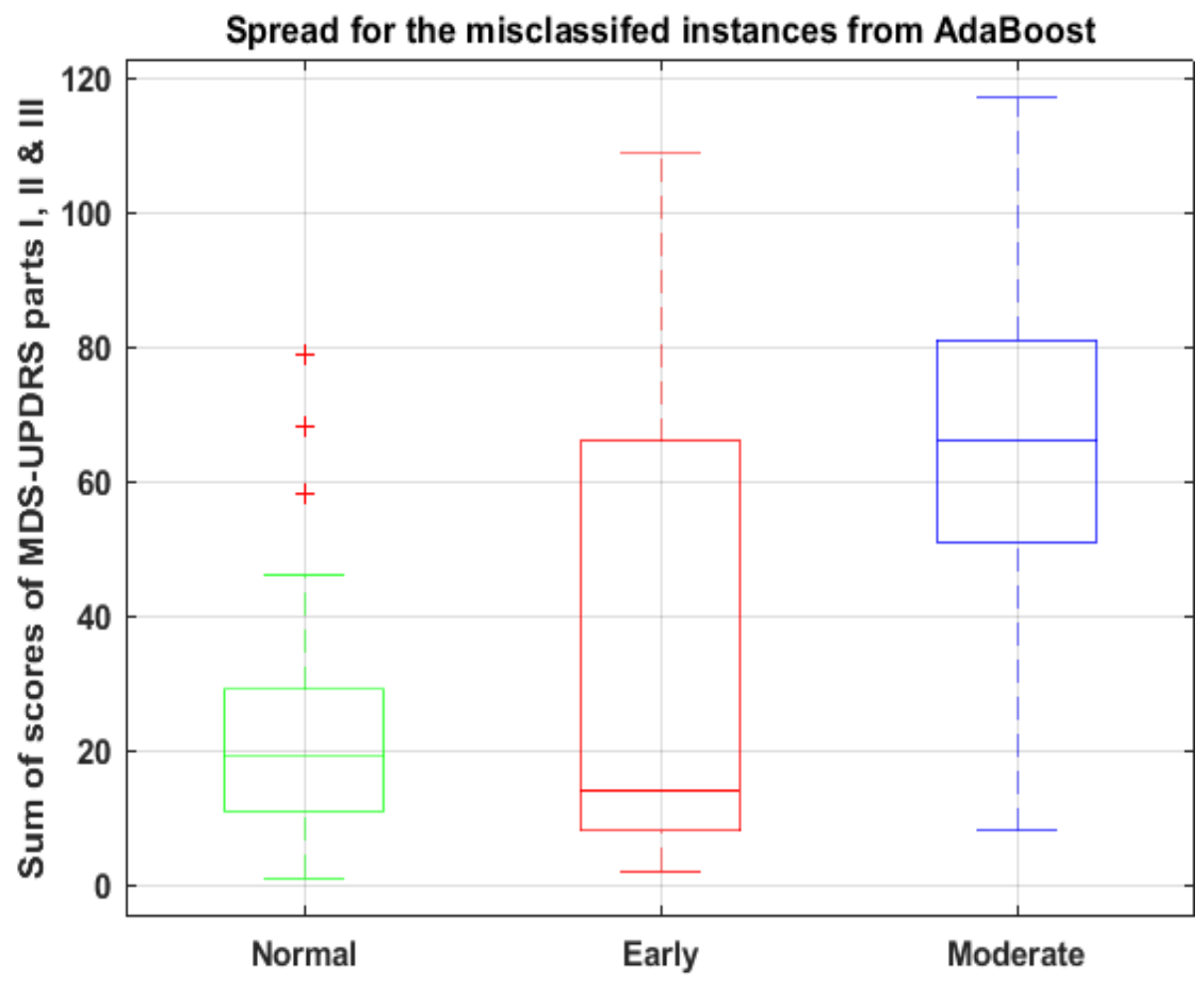

(b) AdaBoost

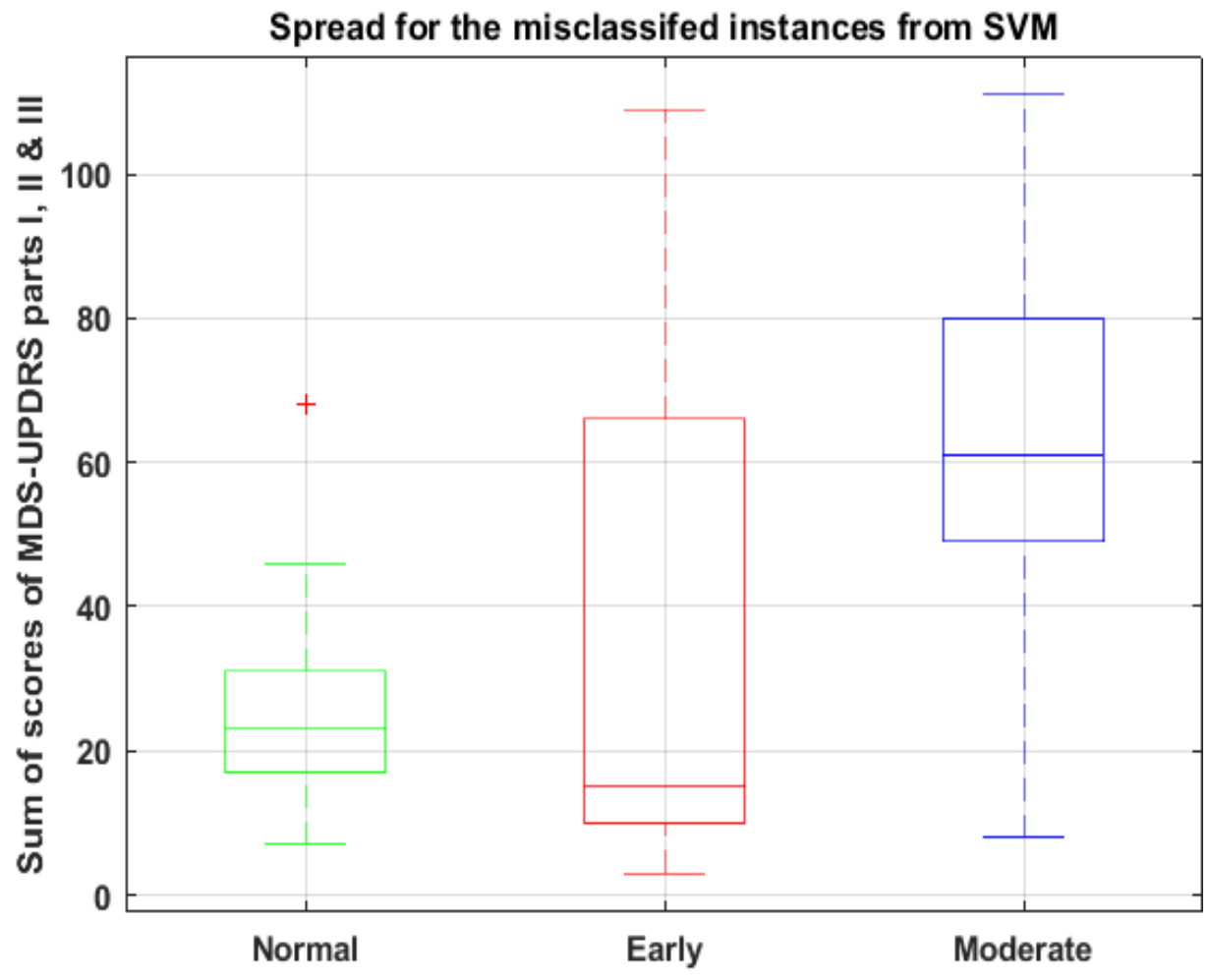

(c) SVM

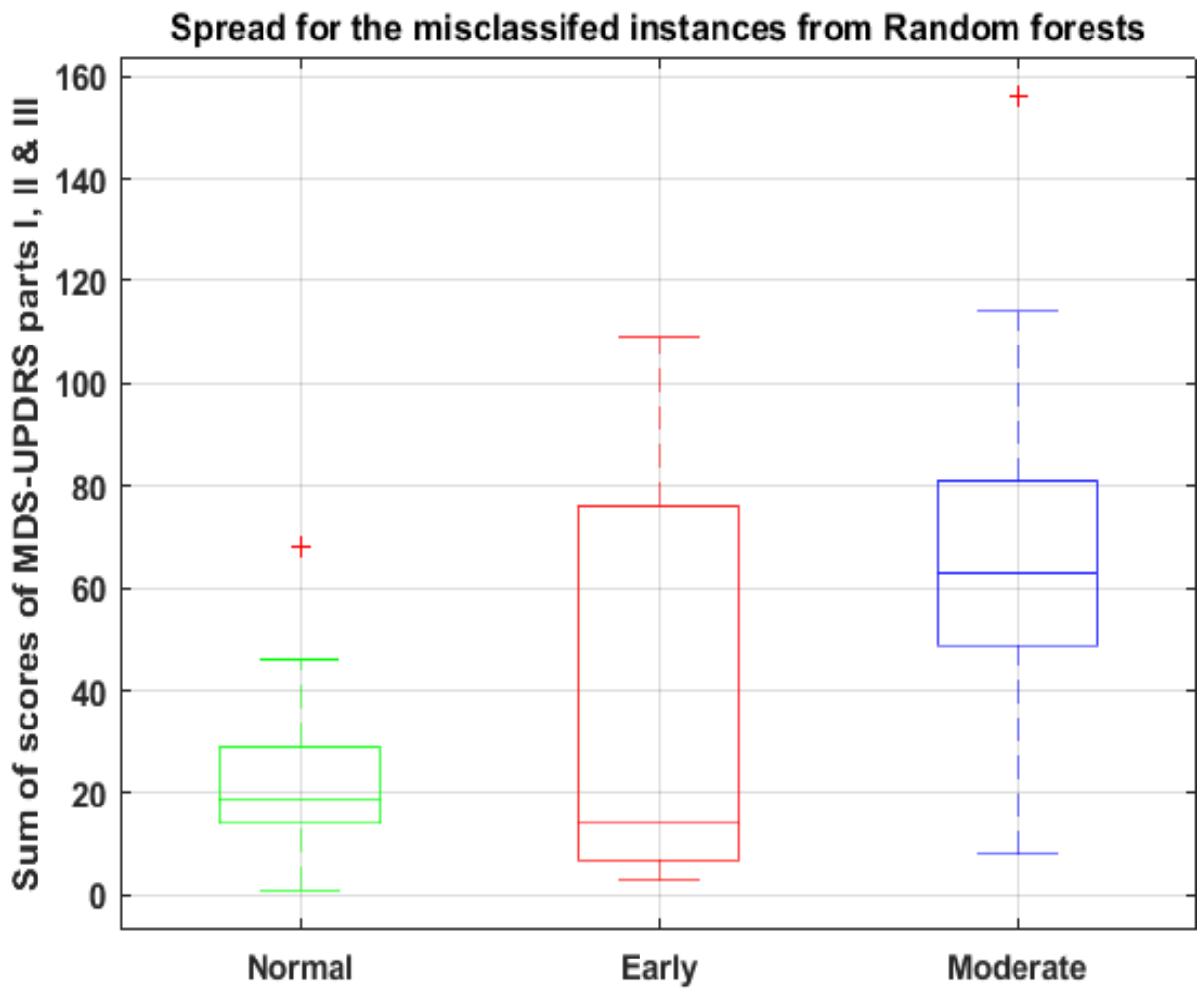

(d) Random forests

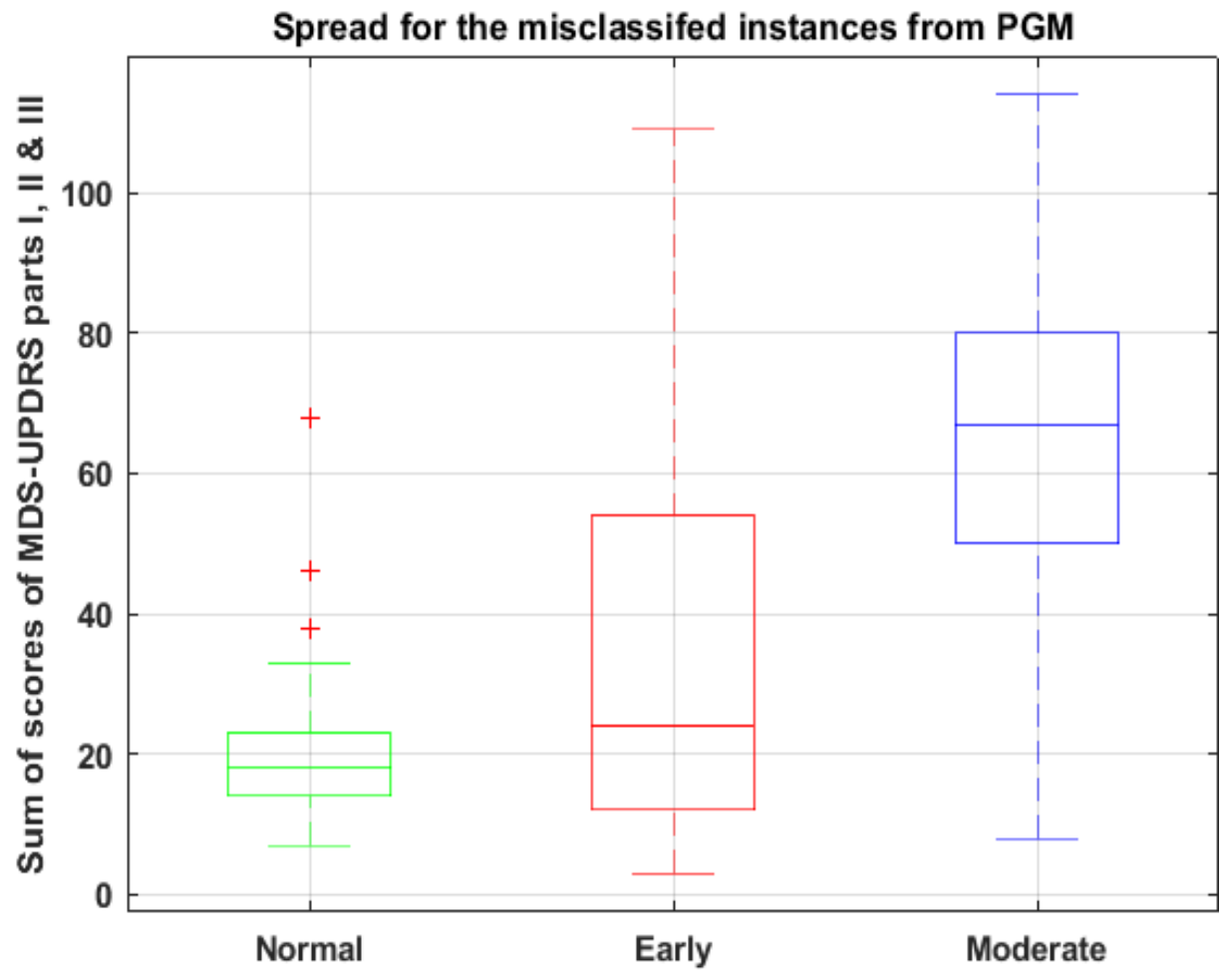

(e) Probabilistic generative model

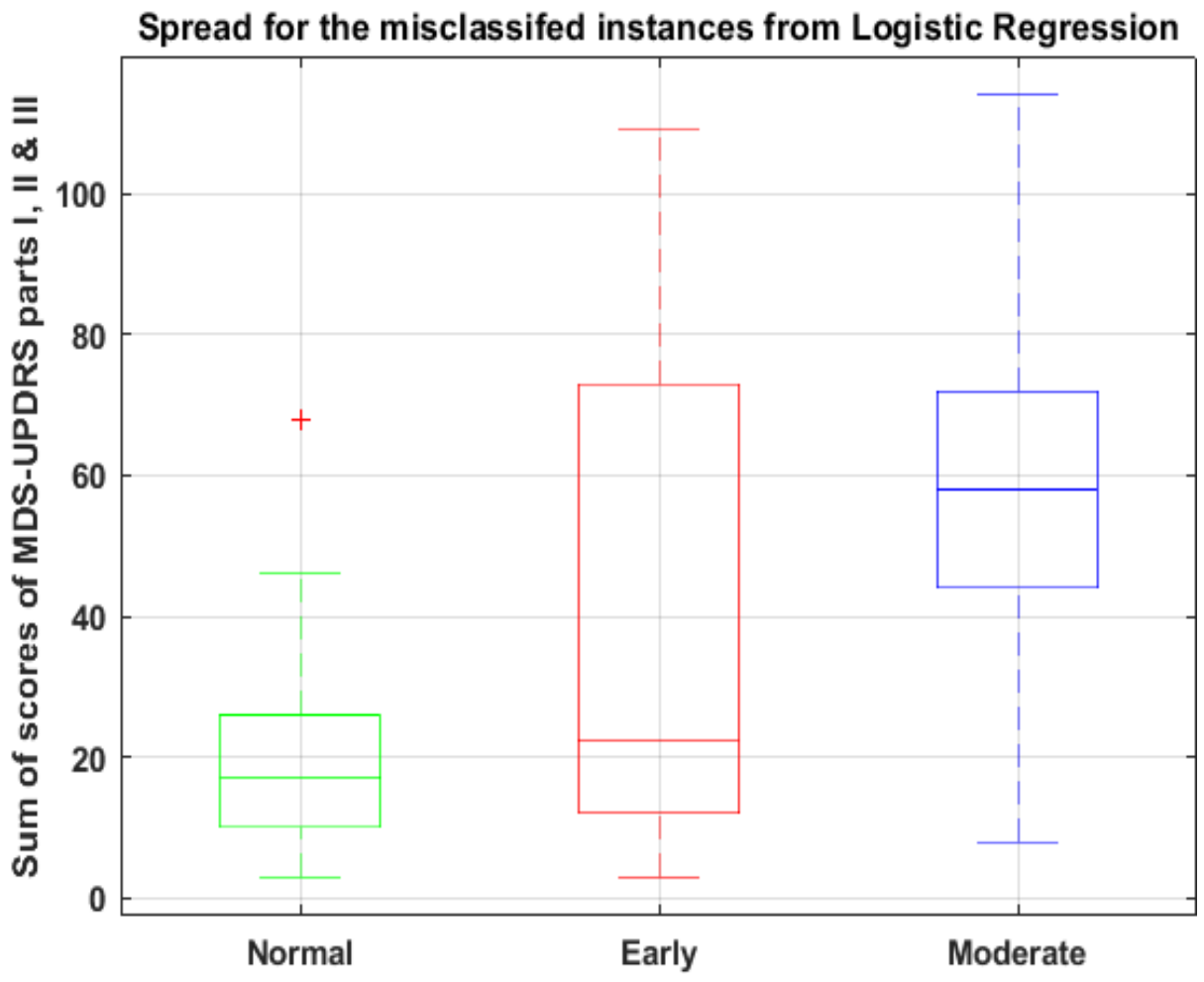

(f) Logistic regression

Fig. 8. Box plots of full data and of misclassified instances from few classifiers used in the study (a) box plot of full data (b) AdaBoost-based ensemble (c) SVM (d) Random forests (e) probabilistic generative model and (f) Logistic regression.

### 3.6 Feature importance estimation in PD

In case of 2-class classification, all 59 features are observed to be statistically significant (*p*-value<0.05) based on Wilcoxon rank sum test. Hence, all 59 features are used for subsequent feature importance estimation using Random forests.

The plot of feature importance is shown below in Fig. 9. Global spontaneity of movement (P3_BRADY), tremor (P2_TRMR), loss of facial expression (P3_FACXP), constancy of rest tremor (P3_RTCON) and handwriting or micrographia (P2_HWRT) were observed to be the most important features which is consistent with the clinical perspective of the disease [2]. Global spontaneity of movement and rest tremor are clinically, the most common and easily recognized symptoms of PD [2]. During the evaluation of global spontaneity of movement rating in the MDS-UPDRS, all observations on slowness, hesitancy, and small amplitude and poverty of movement in general, including a reduction of gesturing and of crossing the legs, is taken into account. This assessment is based on the examiner's global impression after observing spontaneous gestures while sitting, and the nature of arising and walking [5]. Loss of facial expression is also a form of bradykinesia that involves facial muscles. Constancy of rest tremor is evaluated based on one rating for all tremors (RUE−Right Upper Extremity, LUE−Left Upper Extremity, RLE−Right Lower Extremity, LLE−Left Lower Extremity and Lip/Jaw) during the examination period, when different body parts are variously at rest [5]. In the early stages of PD, gait is characterized mainly by reduced speed and decreased amplitude of leg movements, accompanied by reduced arm swing. From Fig. 9., gait (P3_GAIT) is observed as a feature of moderate importance. Falls and complex gait disturbances such as freezing and start hesitations are usually confined to the later stages of PD [37]. Our result (that the feature freezing of gait as the least important feature) is consistent with this as we do not involve samples from late stage PD.

Along with this, a plot showing the difference between percentages of observations showing severity in HC and PD groups is also given below in Fig. 10. Comparing these graphs, we can observe that the highest raked features, from the Random forests technique, such as P2_TRMR (Tremor), P3_FACXP (Facial expression), P3_BRADY (Bradykinesia or Global spontaneity of movement), P3_RTCON (Constancy of rest tremor), also had higher difference of percentages of observations showing severity in HC and PD groups. In Fig. 10, we can observe that P3_FACXP and P3_BRADY have similar difference of percentages. It is to be noted that both these features had strong correlation of 0.76 as compared to other correlations between features, and because of this correlation, one of the features, which in our case P3_BRADY had lower importance estimate as compared to the P3_FACXP from the Random forests technique.

For features from 7-13, they evaluate the non-motor aspects of experiences in daily life and they were filled by the patient themselves. Studies have shown that age is a crucial factor for influencing non-motor features and as the dataset involves mostly aged cohort, the age factor might have played crucial role in increasing the score for these features in the HC group. And these features had lower importance estimates from the Random forests model as observed from the Fig. 9. Although these features showed lower importance, they are certainly important and are highly statistically significant.

The Random forests model which was used to estimate the feature importance gave an accuracy of 99.03%, sensitivity of 99.51% and specificity of 98.10% indicating its high performance in distinguishing PD from HC. The average confusion matrix is as shown below in Table 9. It can be observed that the number of PD observations which were not detected was very low (which is 10 or 0.5%) indicating that the machine learning model might was able to learn the complex patterns in the data which helped in the detection.

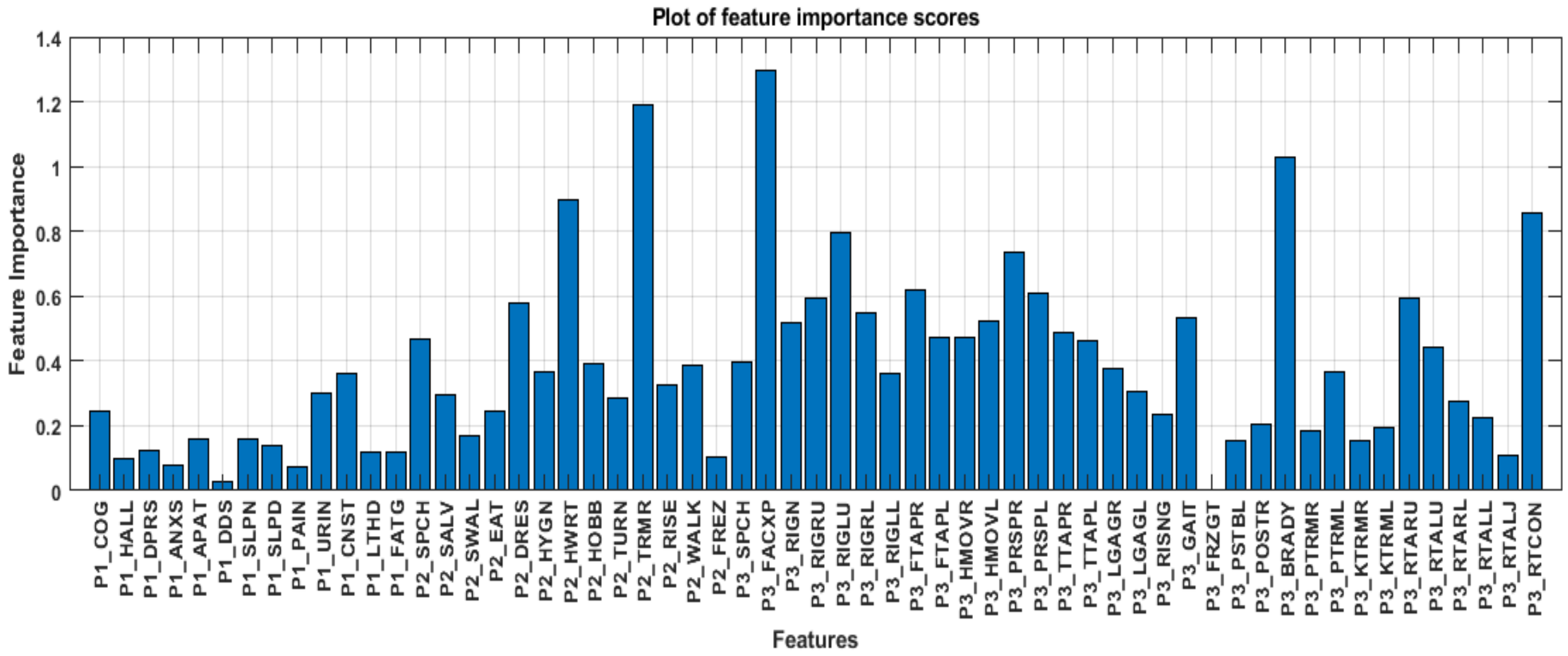

Fig. 9. Plot of feature importance scores using Random forests

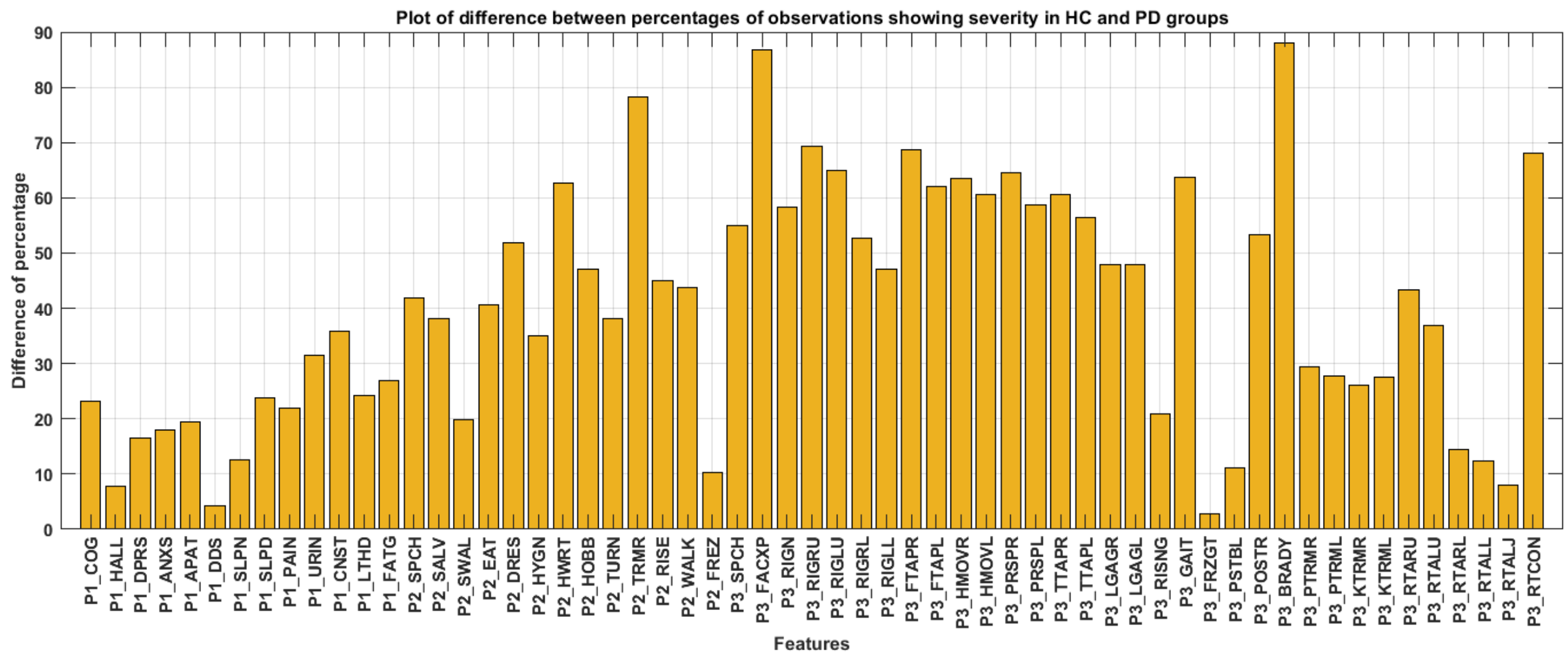

Fig. 10. The plot of difference of percentages of observations showing severity in PD and HC groups

**Table 9: Confusion matrix obtained from the Random forests model for the 2-class case**

| | | ***Predicted*** | |
|---|---|---|---|
| | | **PD** | **HC** |
| ***True*** | **PD** | 1985 | 10 |
| | **HC** | 19 | 1006 |

### 3.7 A note on cost-sensitive and sampling based imbalanced data learning used in the study

The imbalance of data used is of *extrinsic* type. The imbalance is not directly related to the nature of the data space. We tried with the default costs and sampling proportions (i.e., assigning misclassification costs and sampling proportions as 1 for all classes) for SVM, AdaBoost-based ensemble and RUSBoost-based ensemble. It is observed that the SVM and AdaBoost ensemble gave almost similar performance as compared to the cost-sensitive versions. This finding is consistent with the study by Japkowicz and Stephen [38], where they observe that SVMs are less sensitive to the class imbalance problem. On the other hand, the AdaBoost algorithm is an approach which inherently performs a weighting strategy that is equivalent to re-sampling the data space, combining both up-sampling and down-sampling. When the learning is proper, this re-sampling can avoid extra cost estimation for exploring the optimal class distribution [23].

### 3.8 Note on using a combined data from different patients

The course or progression of PD is not linear and it varies between subjects. And this fact is established by many research studies. In this way, the combined dataset might become complex with different kinds of patterns in it. And this was one of the main motivation for using machine learning techniques on this data as it is established that these techniques have the potential to learn the latent (or hidden) information from the data and can effectively use it in prediction. And from our study, we observe that we obtain high performance in estimating the stage of a subject as indicated by the accuracy and other performance measures. It is also observed that the learned stage or the expected value of the class variable from the logistic models showed higher correlation with the features, as shown on Fig. 6 shown below, indicating its usefulness. Furthermore, rather than using a generalized model for prediction, patient specific models can be developed which might have higher performance and usefulness. But for this, a large amount of data for each subject has to be collected which is a limitation for patient specific models.

### 3.9 Limitations and future work

Our study used only samples corresponding to normal, early and moderate stage PD, and did not consider advanced PD due to unavailability of data. However, the approach used in the study is applicable while learning data involving advanced (or late) stage PD also. This is because there is severe deterioration of symptoms in advanced stages of PD leading to significant separation between class distributions. Testing these approaches in an extended dataset involving advanced PD observations is a possible future goal. Along with this, another interesting future study would be consider data from a single visit such as visit at month 60 (V12), to provide further restrictions on the timelines of data collections, for the analysis. However such an analysis could make the data further imbalanced which needs to be addressed appropriately.

Our study also had limited number of moderate stage PD samples. This is an example of *extrinsic* imbalance [21] where the imbalance is not a direct result of the nature of the data space. (The PPMI database has limited observations from moderate stage PD). The Random forests technique, due to its unique *'out of bag'* property, can be an ideal tool to estimate feature importance. We estimated the feature importance in PD, and not corresponding to the stages of PD because moderate stage PD observations were low in number, and the feature importance scores from Random forests is based on the out-of-bag error estimates which is biased towards the majority classes. Estimation of feature importance between stages is another possible future work.

## 4. Conclusion

Although there is no cure for PD yet, there are a variety of treatments that can help a patient lead a fulfilling and productive life for many years. The therapeutic options depend on the stage and severity of PD. MDS-UPDRS is a widely used rating scale evaluating the severity of most pertinent features of PD. In this study, we propose a novel staging for PD using MDS-UPDRS features, HY scale, combined with classifiers that can be effective tools to estimate the stage and severity of PD. Such tools can aid in detection and diagnosis, and thereby, helping in taking effective therapeutic decisions in PD.

**Acknowledgments**

PPMI – a public-private partnership – is funded by the Michael J. Fox Foundation for Parkinson's Research and funding partners, including Abbvie, Avid Radiopharmaceuticals, Biogen Idec, Bristol-Myers Squibb, Covance, GE Healthcare, Genentech, GlaxoSmithKline, Eli Lilly and Company, Lundbeck, Merck & Co., Meso Scale Discovery, Pfizer, Piramal, Hoffmann-La Roche, and UCB (Union Chimique Belge). The authors also sincerely thank Dr. Michael Fox, MD, PhD, at the Beth Israel Deaconnes Medical Center, a teaching affiliate of HMS for his valuable suggestions that helped in the study.

## Appendices

### *A.* ***Performances measures***

Performance measures play a crucial role in both evaluating the classification performance and guiding the classifier modelling. We have used four measures to evaluate the performance of classifiers: *Accuracy, Precision, Recall* and *F-measure.*

$$Accuracy = \frac{TP + TN}{TP + TN + FP + FN}$$

$$Precision = \frac{TP}{TP + FP}$$

$$Recall = \frac{TP}{TP + FN}$$

$$F - measure = \frac{2 \times Precision \times Recall}{Precision + Recall}$$

where TP is true positive, TN is true negative, FP is false positive and FN is false negative. Accuracy is measured as an overall estimate, whereas, Precision, Recall and F-measure are computed for each class. The average F-measure for the 3 classes is chosen as the fitness function, for GA, as it gives equal importance to all classes.

Along with these, we also compute the Brier score for evaluating the logistic model, given by

$$Brier\ score = \frac{1}{n \times k} \sum_{i=1}^{n} \sum_{j=1}^{k} (p(i,j) - \hat{p}(i,j))^2$$

where $p(i,j)$ and $\hat{p}(i,j)$ are true and estimated probability of observation $i$ to belong to class $j$; $k$ is the number of classes (in our case 3), and $n$ is the number of observations. The expected value of the class is used as a measure of severity of PD in the study. It is given by

$$E = \sum_{j=1}^{k} (j-1) \times \hat{p}(i,j)$$

$(j-1)$ is used instead of $j$ to ensure that probabilities of early and moderate stage PD are used in the severity estimation.

### B. ***Genetic Algorithm for optimal cost and sampling proportion estimation***

Genetic algorithm (GA) is a method for solving optimization problems and is based on the theory of natural selection. The two main concepts in GA are 1) Fitness function, and 2) Individuals. Fitness function is the function that is to be optimized, and an individual is any vector (which can also be the solution to the fitness function) that contains the values of variables of the fitness function. The fitness function used for GA is the average F-measure of the 3 classes.

A brief outline of the working of GA is given below:

1) The algorithm begins with an initial population (which can be a random population or a population created from a specified range). A population is a set of individuals.

2) It then creates a sequence of new populations based on the current population. To create a new population from the current population, the algorithm performs the following:

    a) A fitness score is computed for each member in the current population.

b) Few of the individuals that have lower fitness values, as compared to others, are chosen as *elite*, and are passed to the next generation.

c) Selects members (called parents) based on their fitness values. Parents are expected to have better fitness values.

d) It, then produces children from the parents either by *crossover* (by combining two parents) or *mutation* (by making random changes to a single parent).

e) Replaces the current population with the new population consisting of elite children, crossover children and mutation children.

The process of creating a new population is repeated till the stopping criteria is met.

### *C. Progression of features with time*

Fig. C1 shows the progression of all 59 features with time. All the features showed significant variation in PD as compared to healthy control.

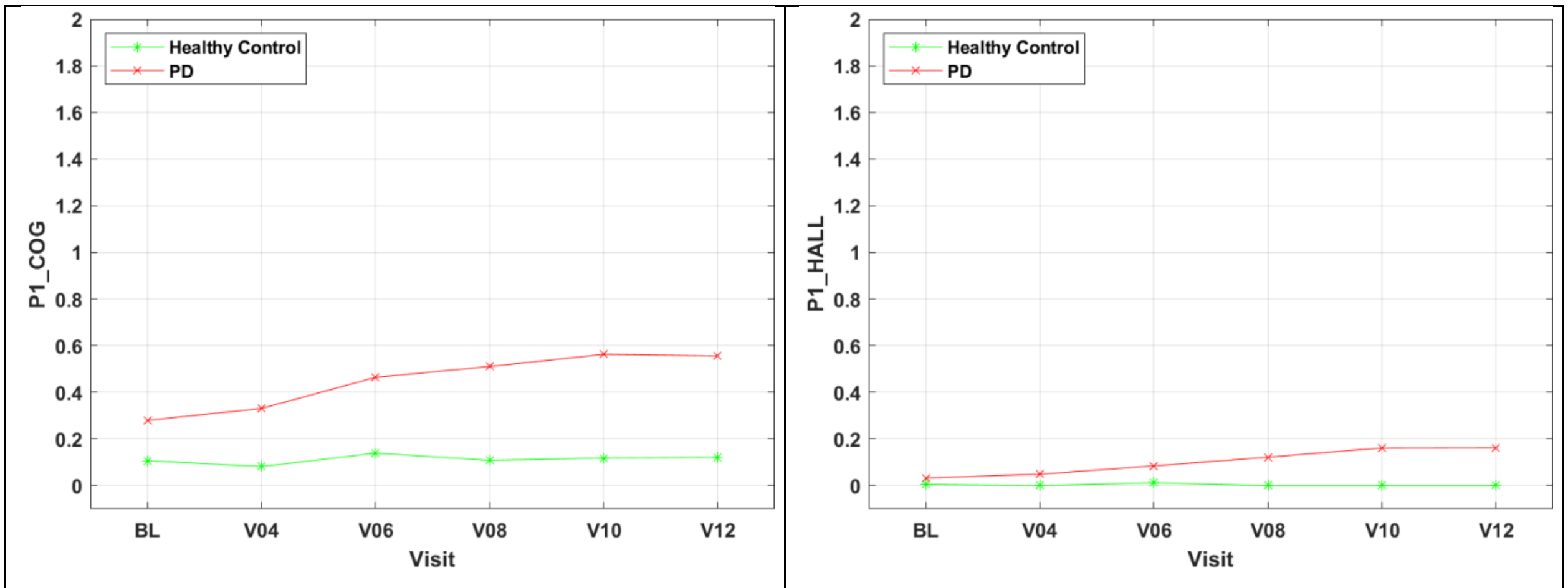

(a) Cognitive impairment (b) Hallucination & Psychosis

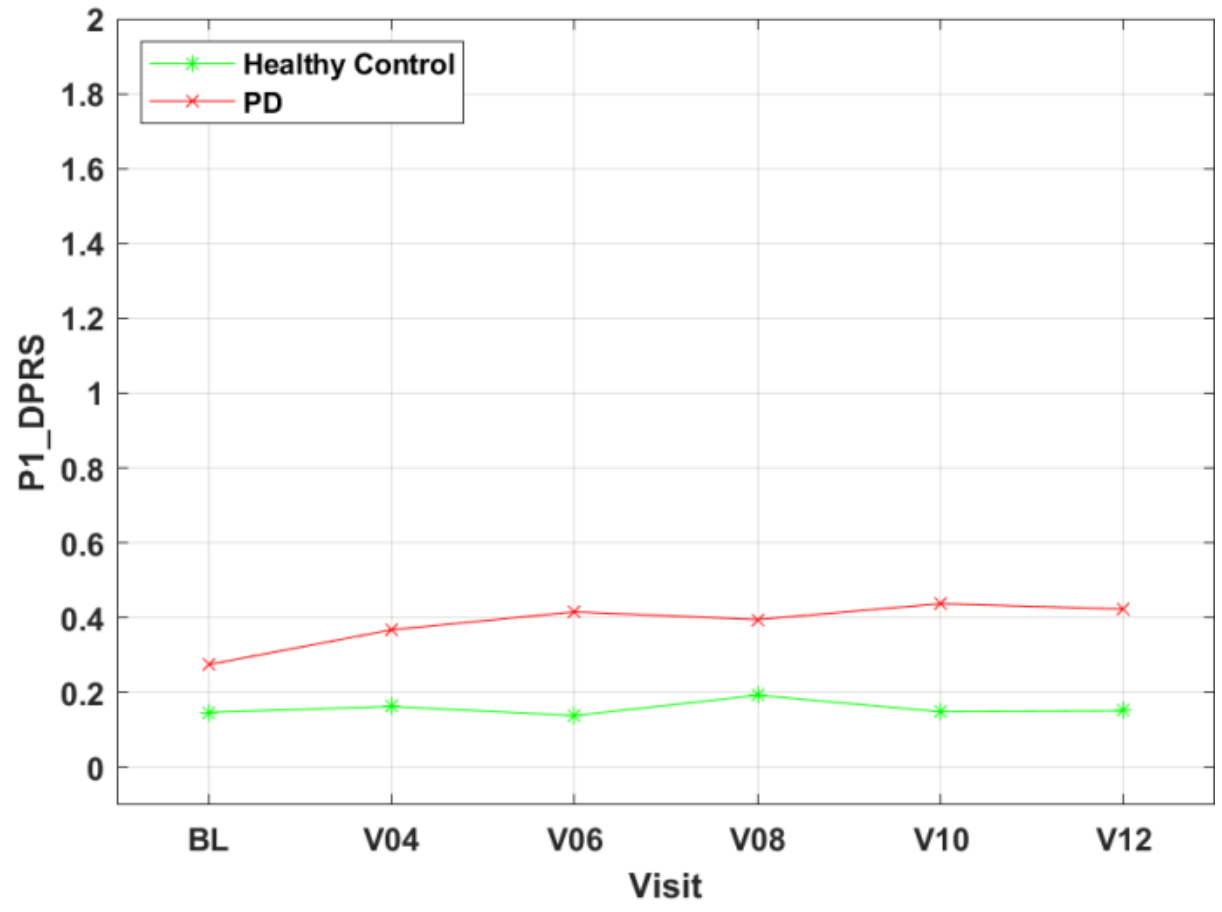

(c) Depressed mood

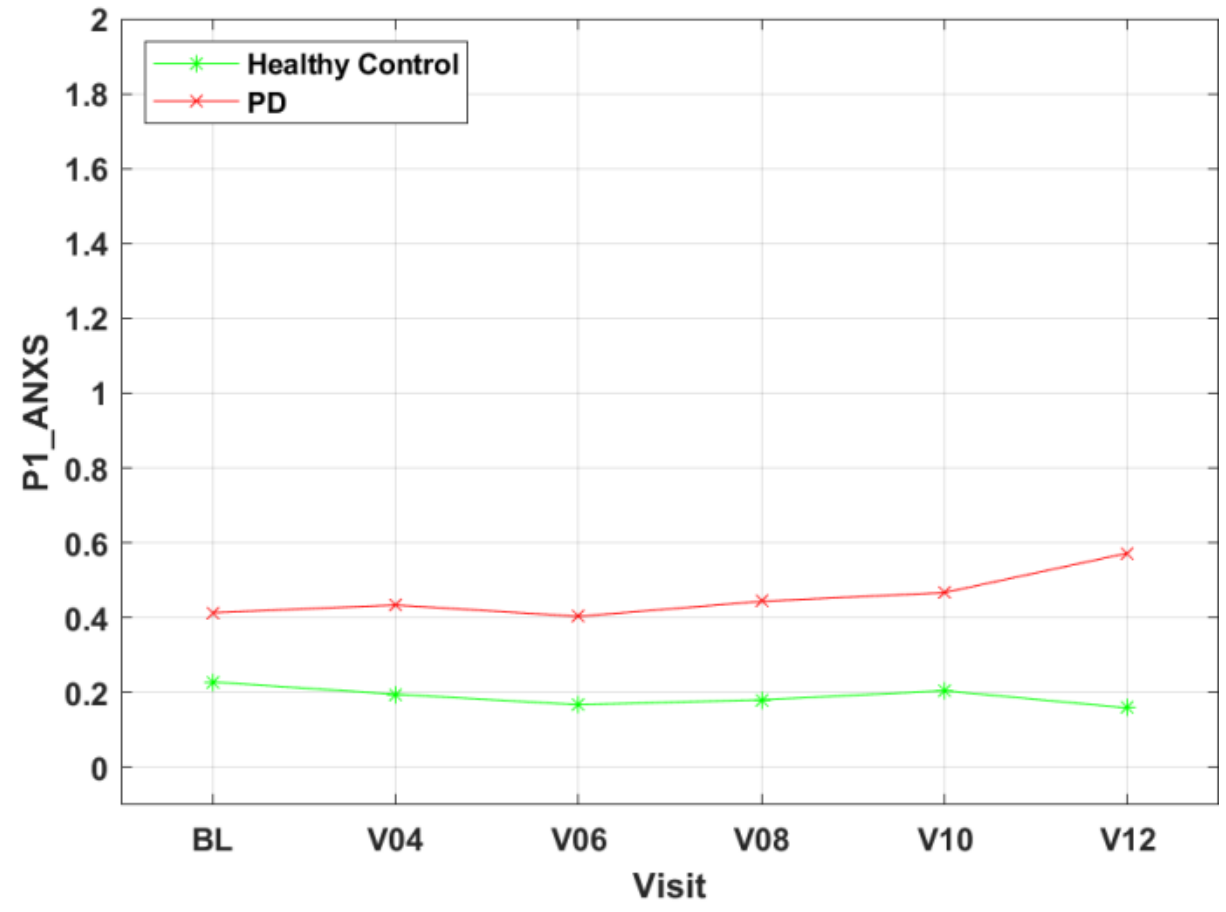

(d) Anxious mood

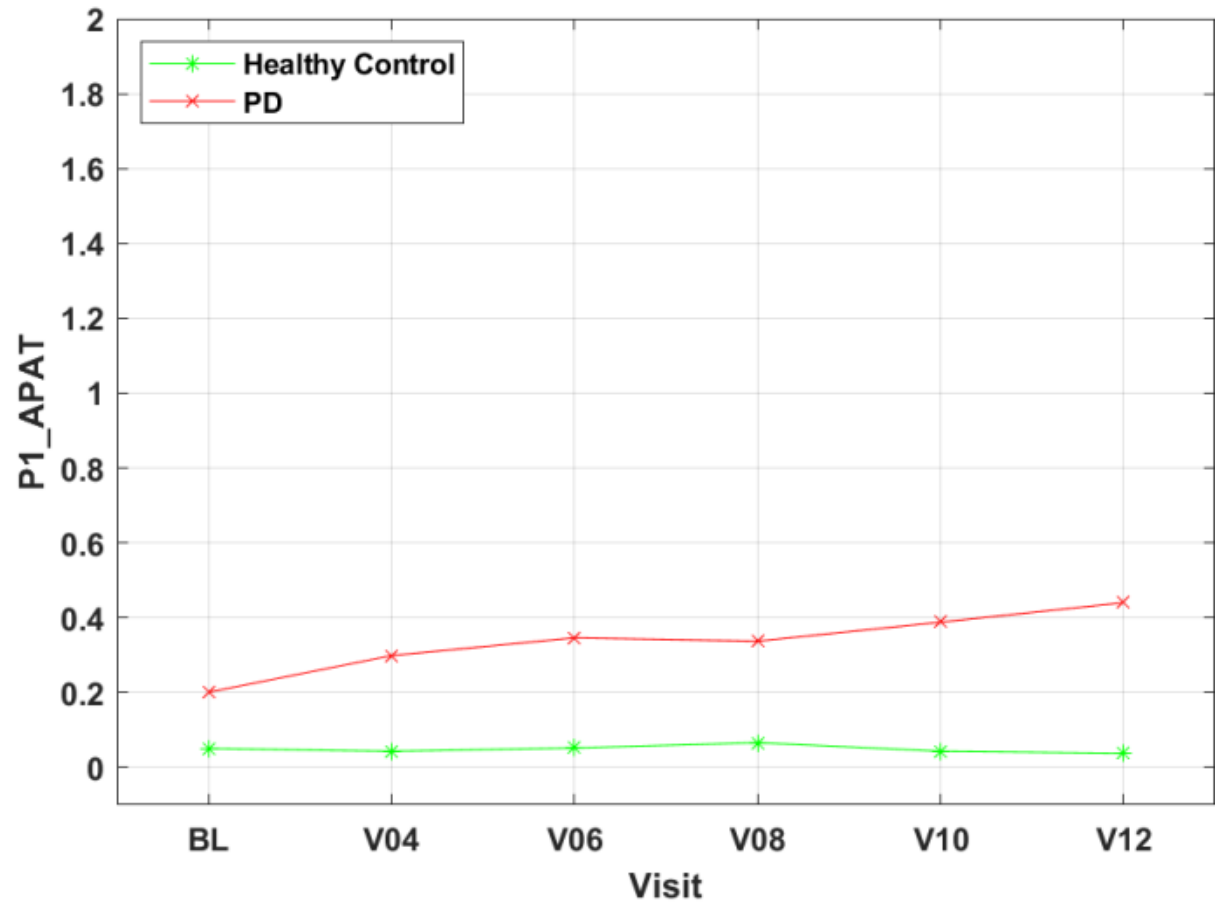

(e) Apathy

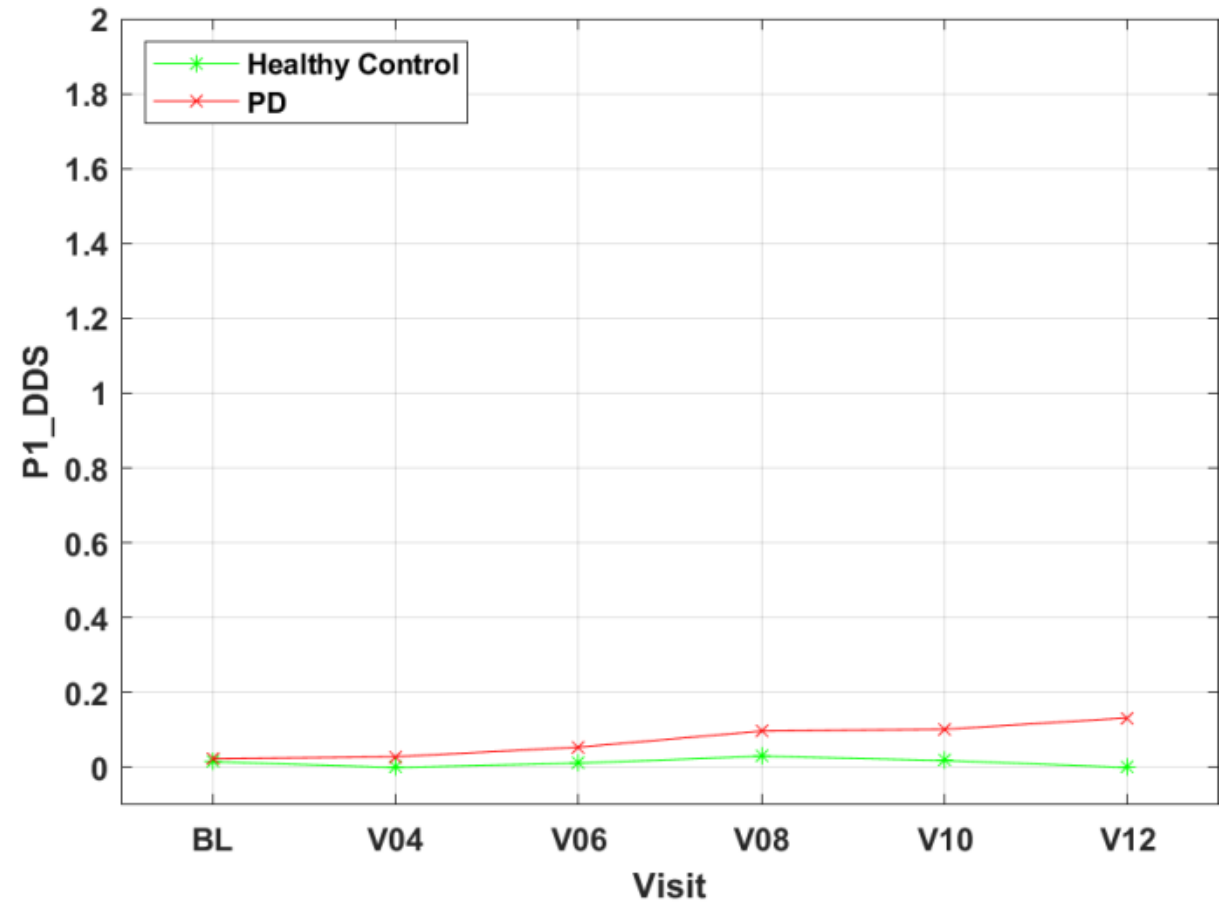

(f) Features of DDS

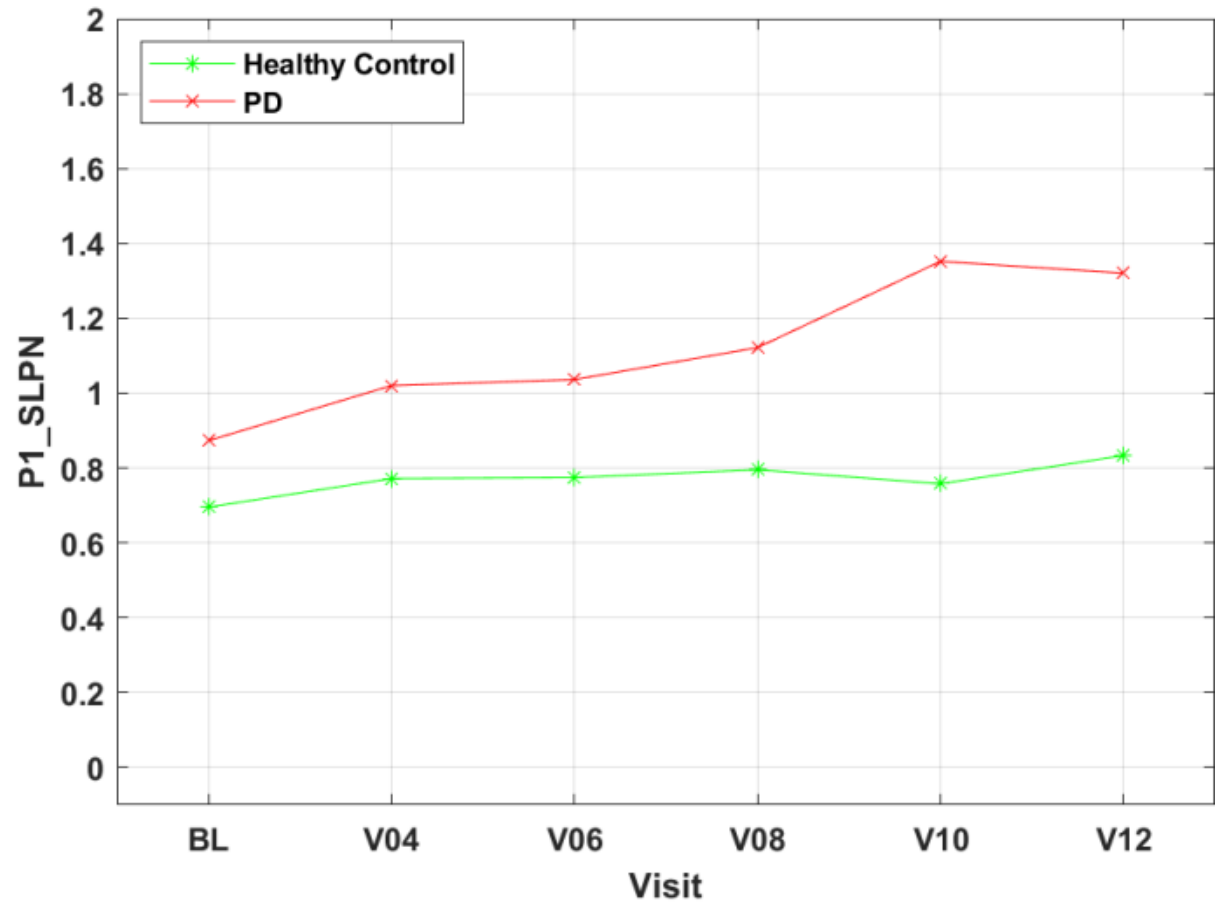

(g) Sleep problems

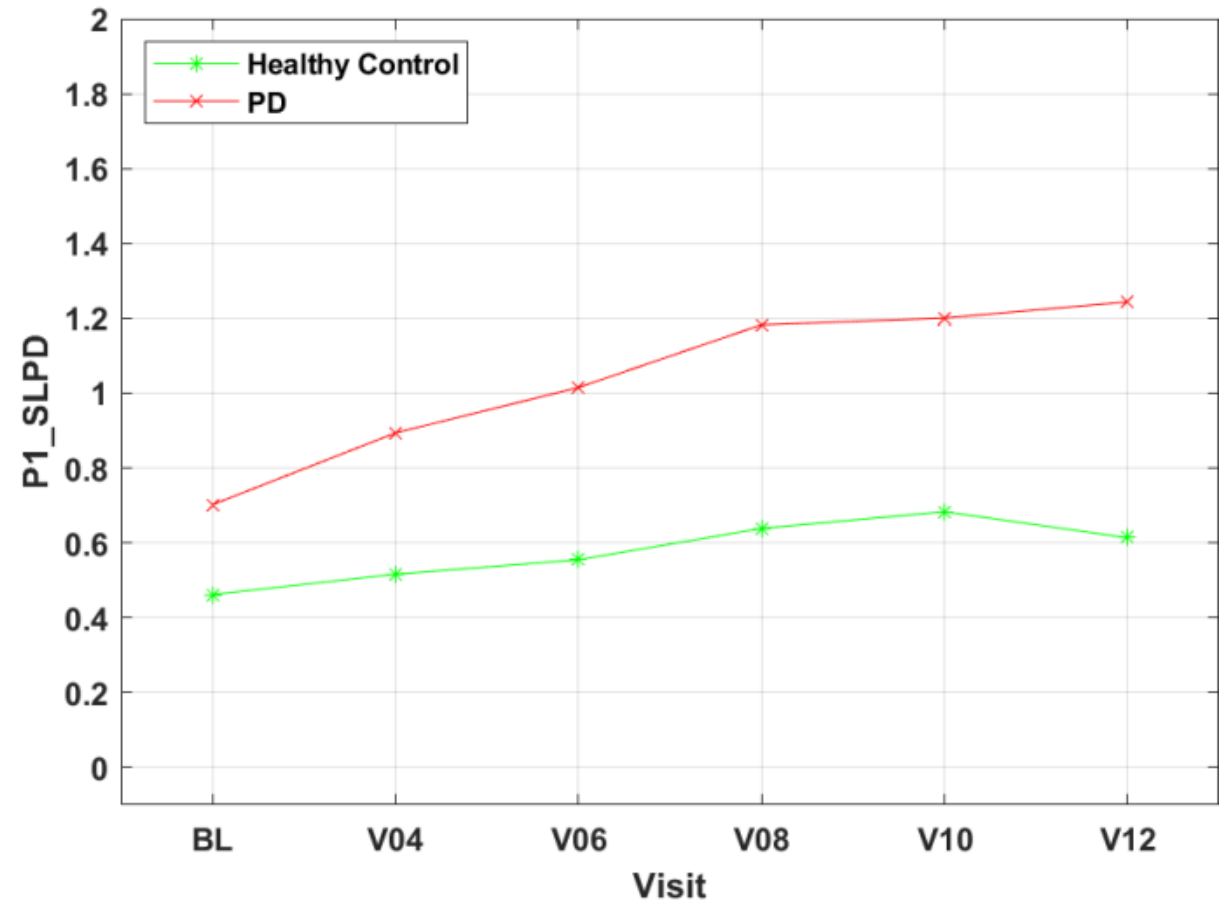

(h) Daytime sleepiness

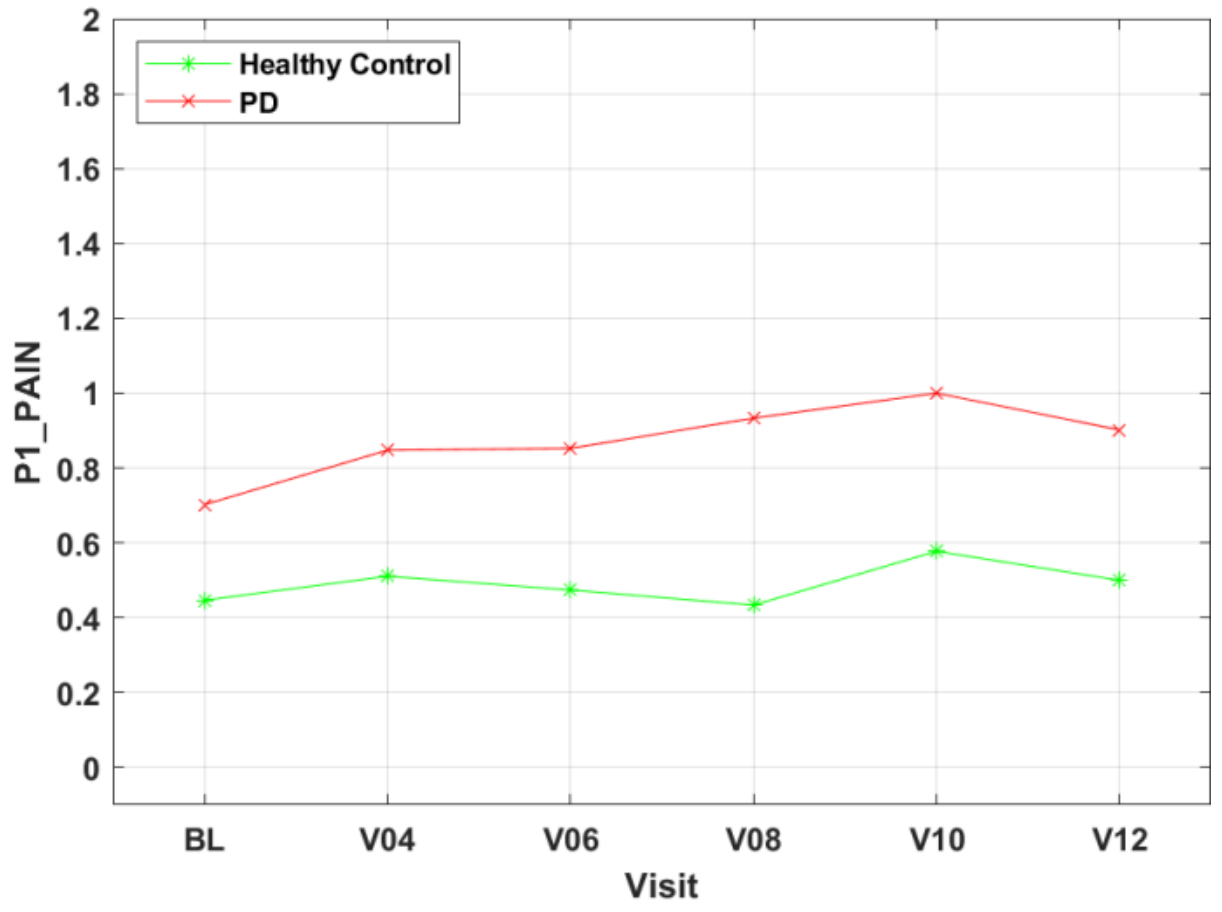

(i) Pain & other sensations

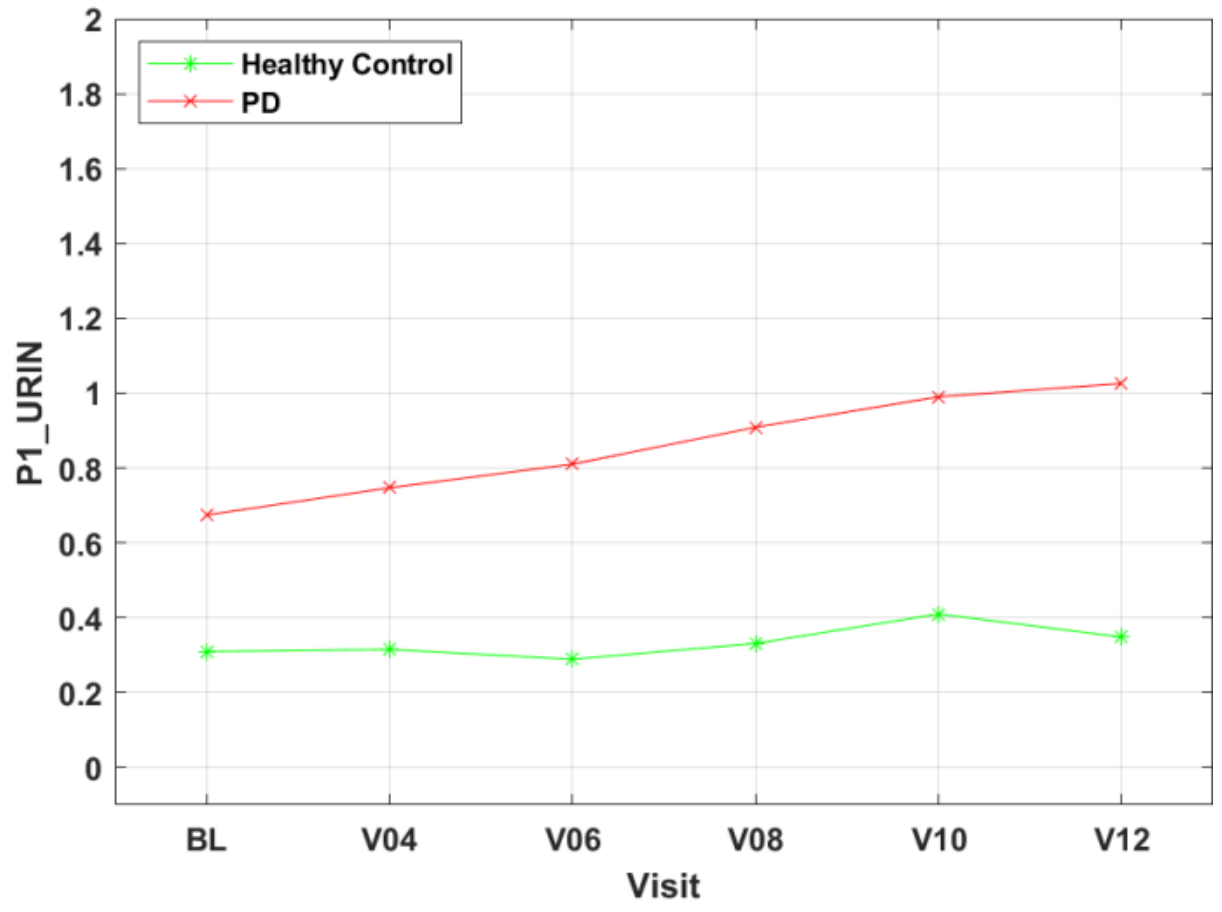

(j) Urinary problems

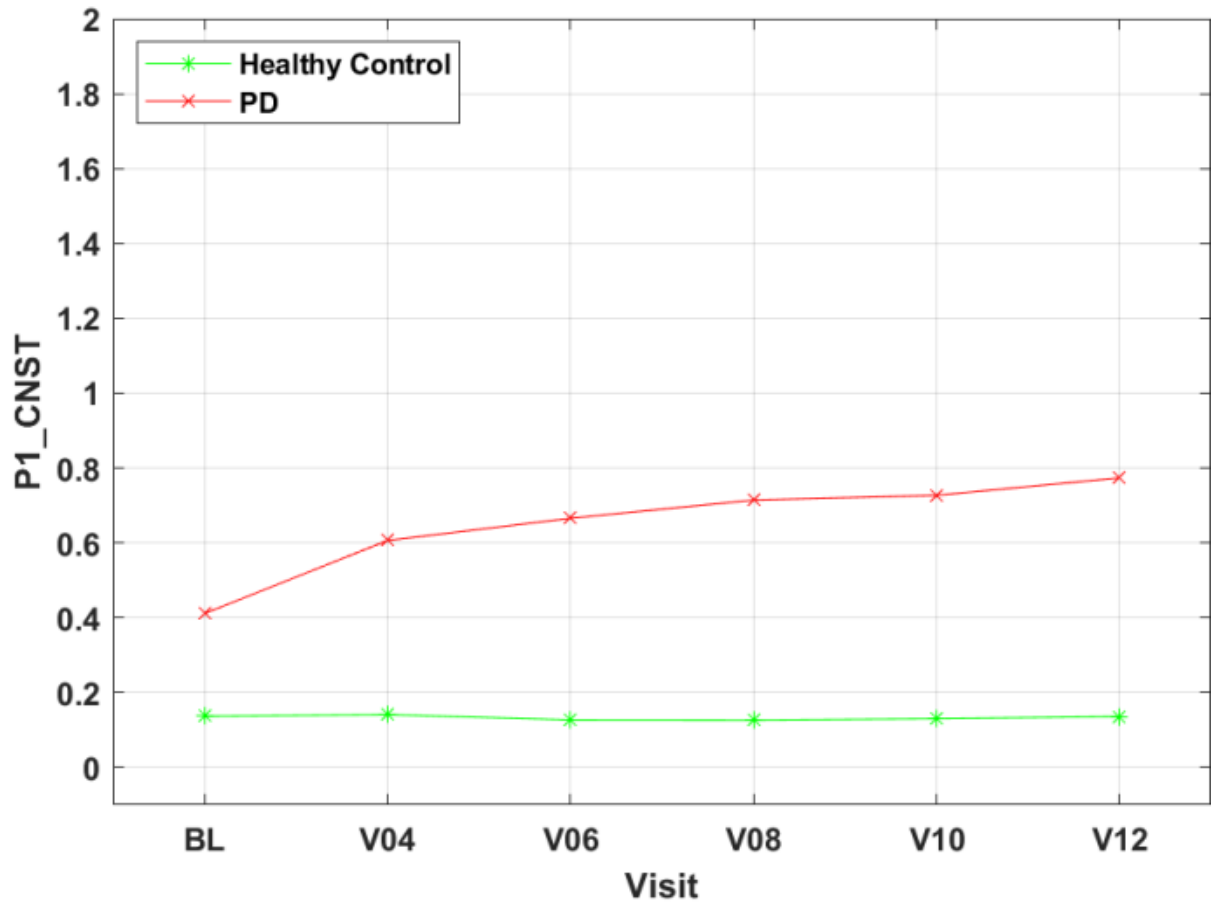

(k) Constipation

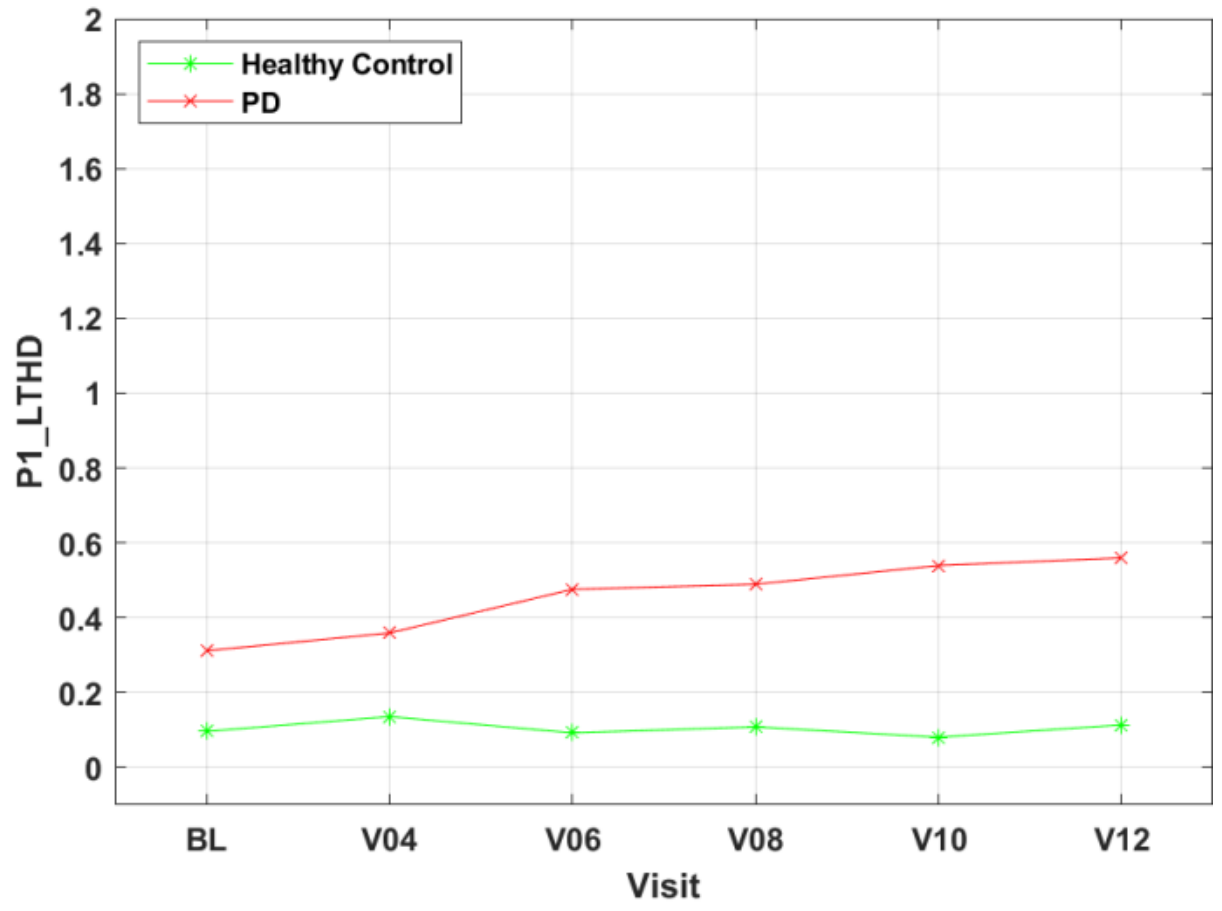

(l) Light headedness

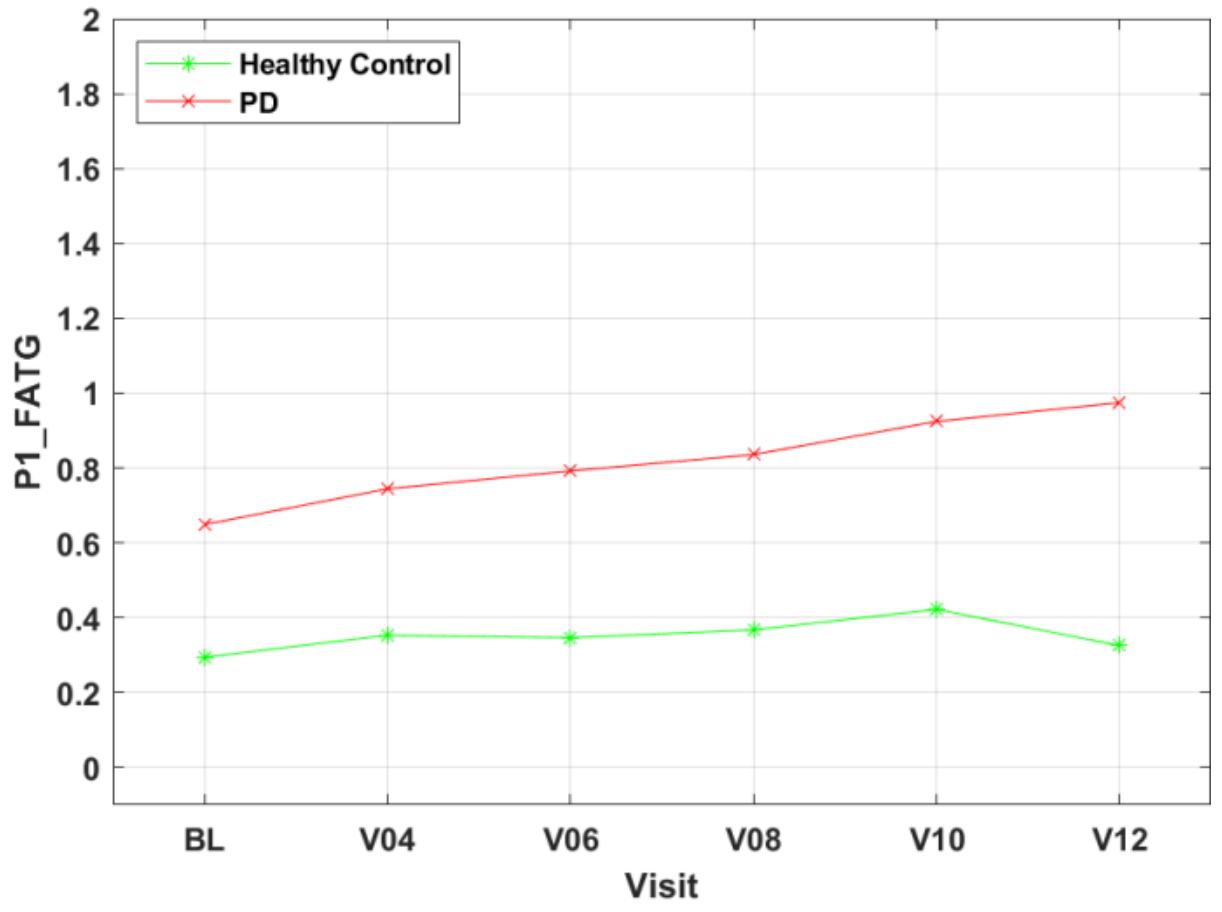

(m) Fatigue

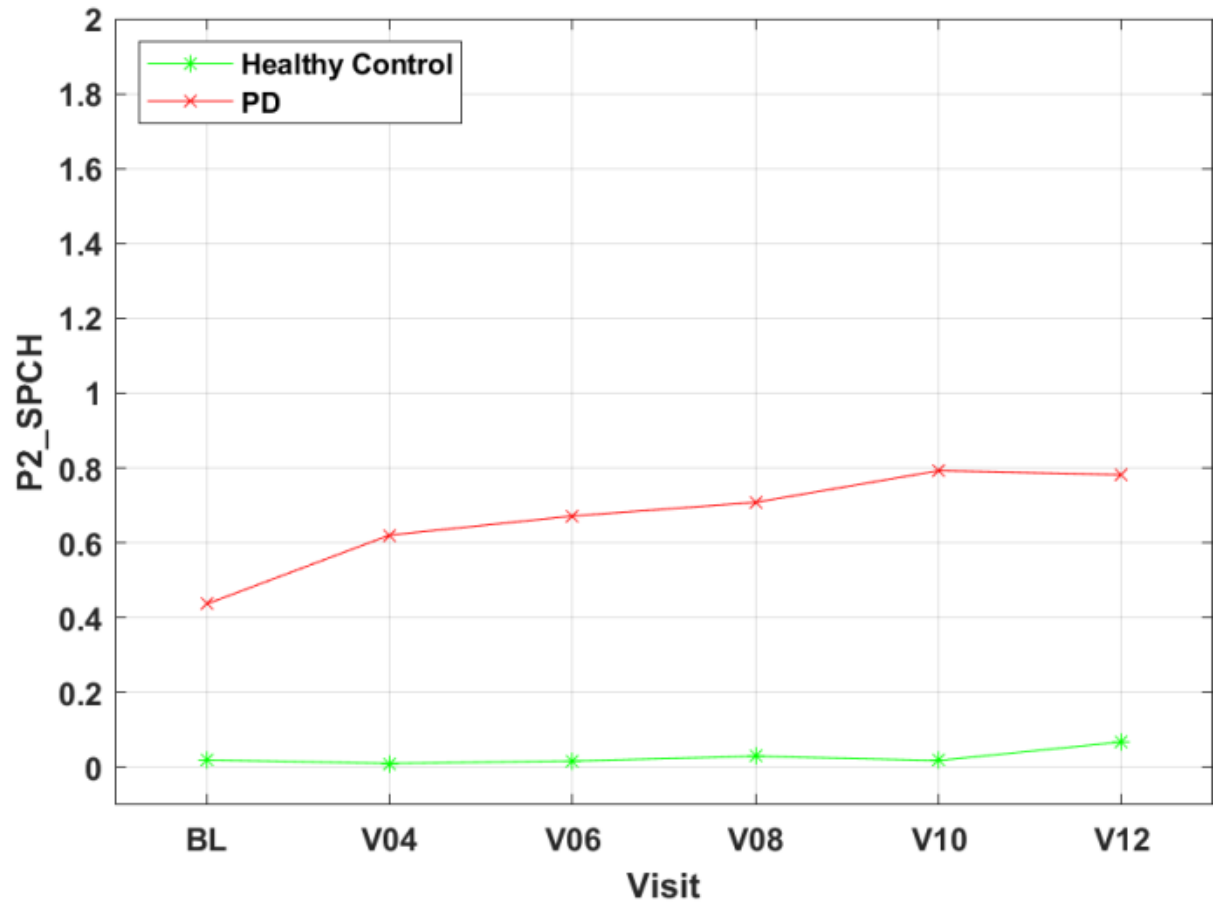

(n) Speech

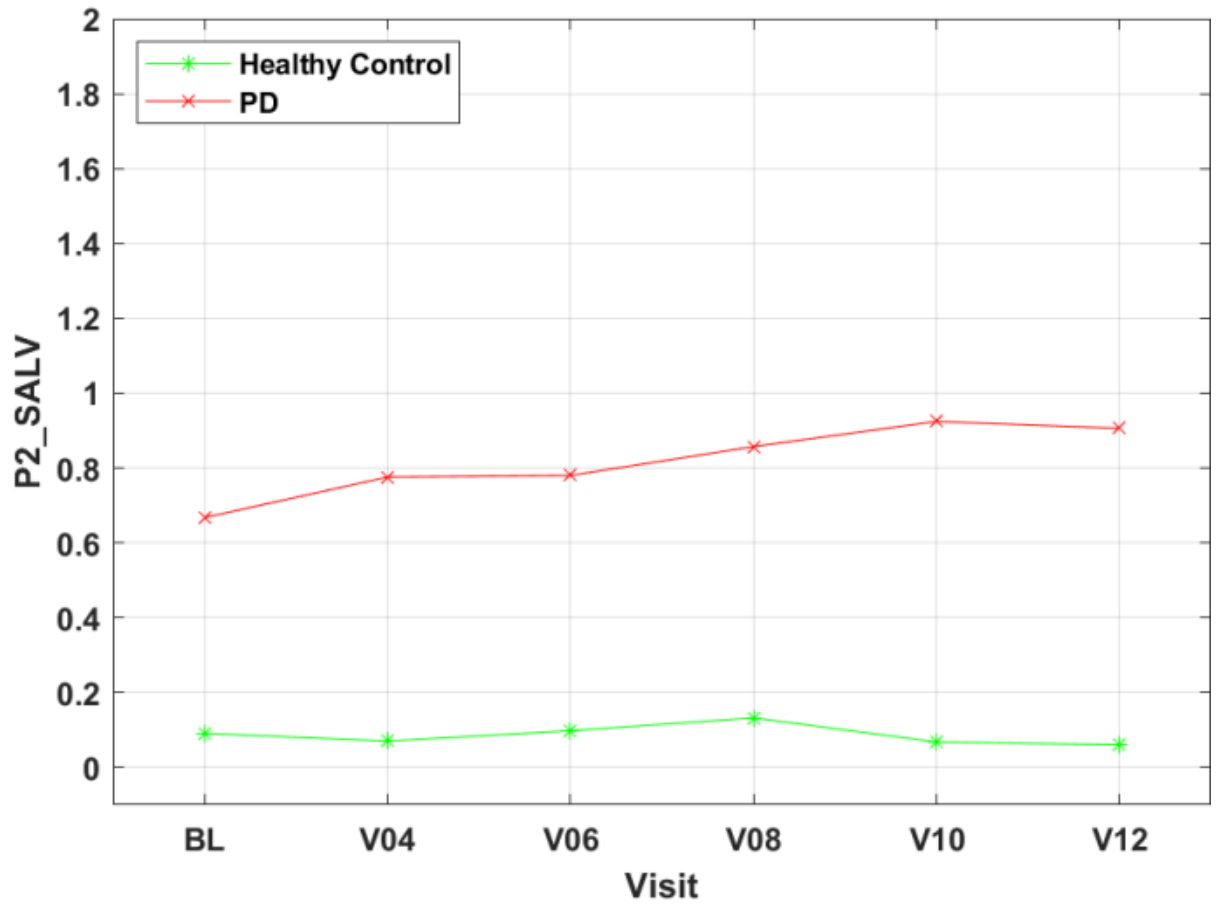

(o) Saliva & Drooling

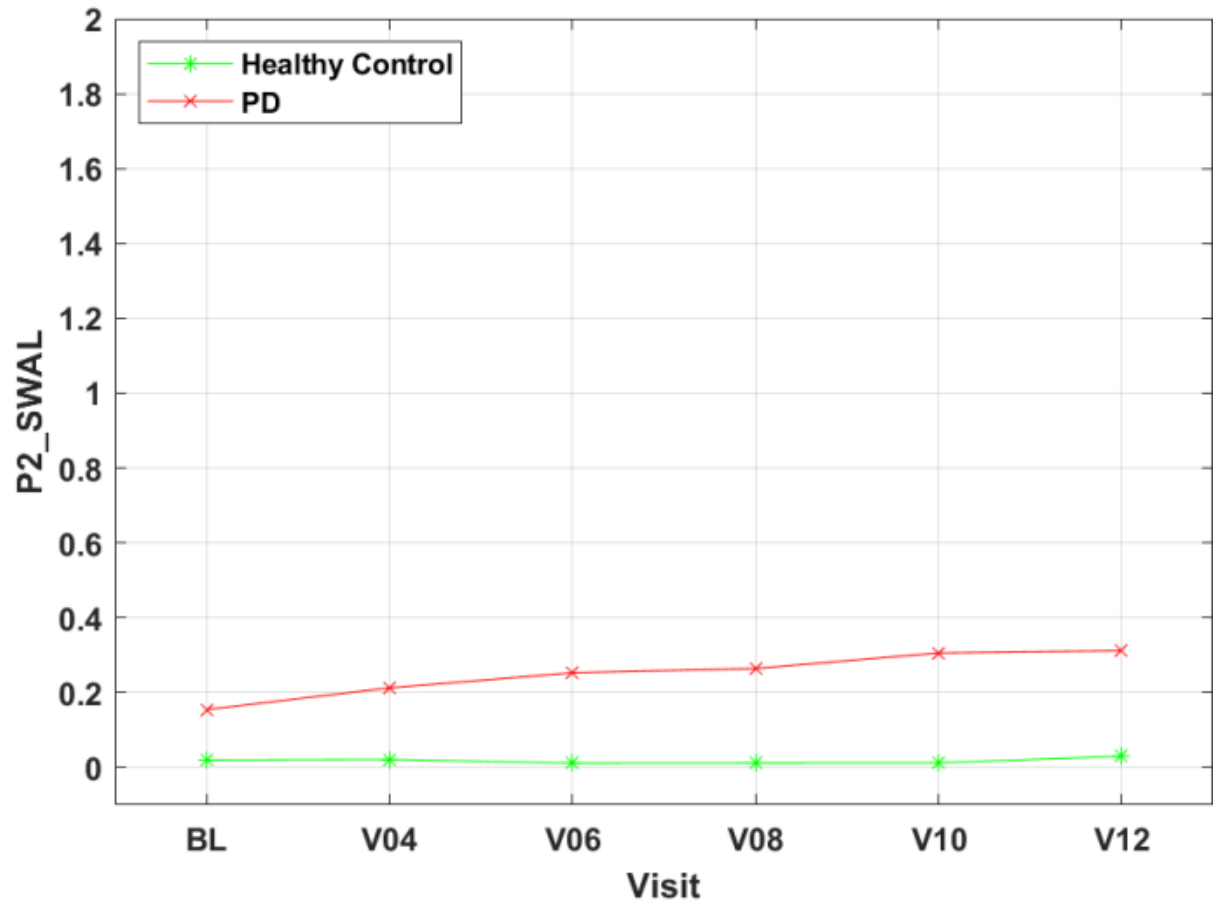

(p) Chewing & Swallowing

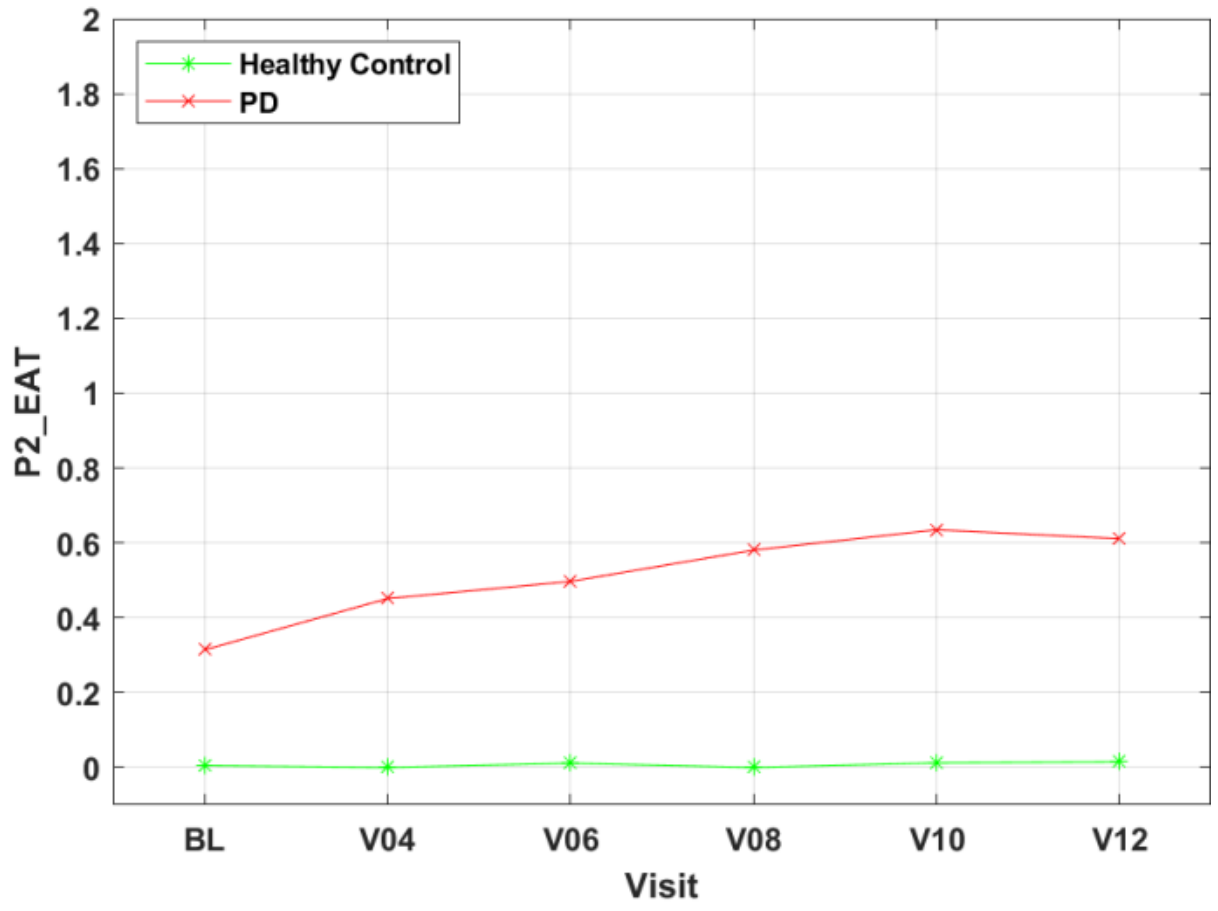

(q) Eating tasks

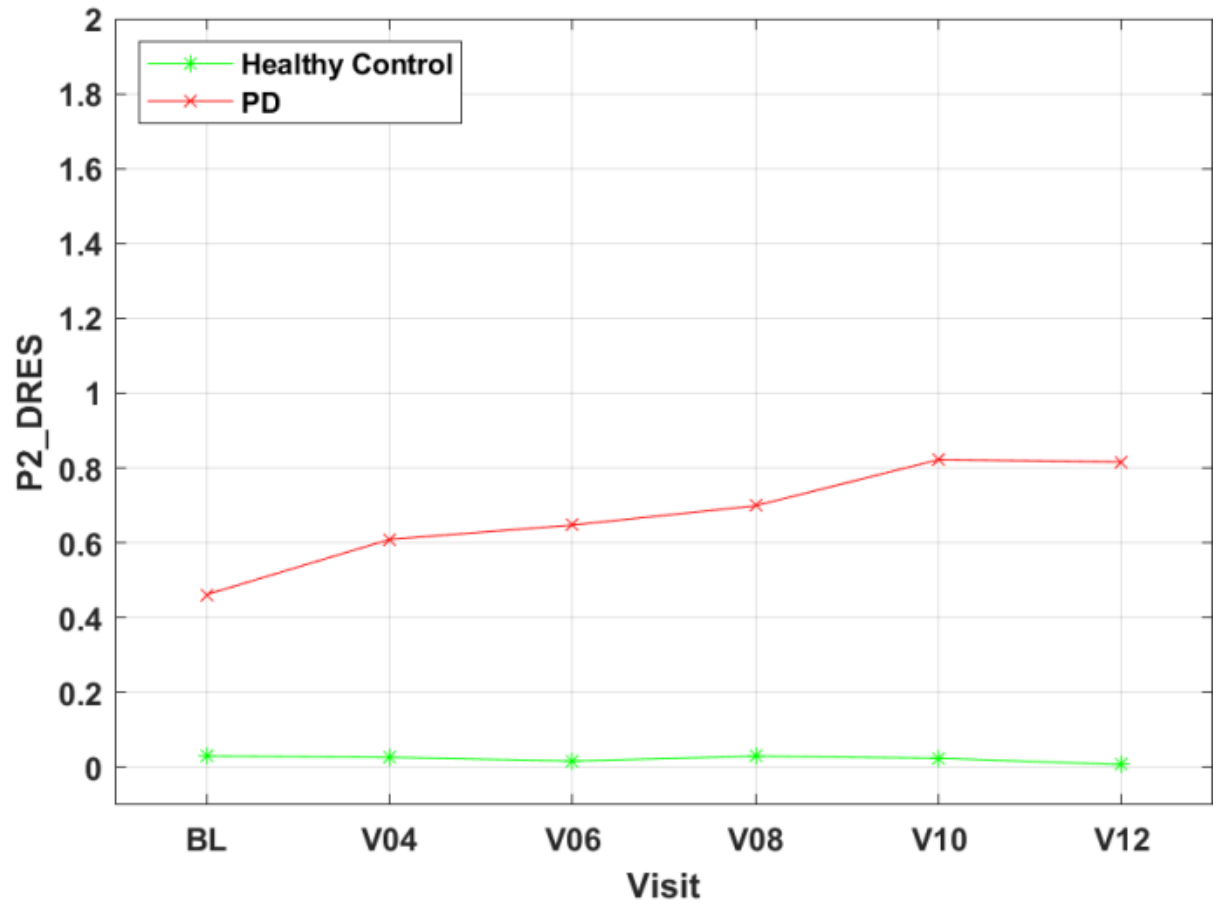

(r) Dressing

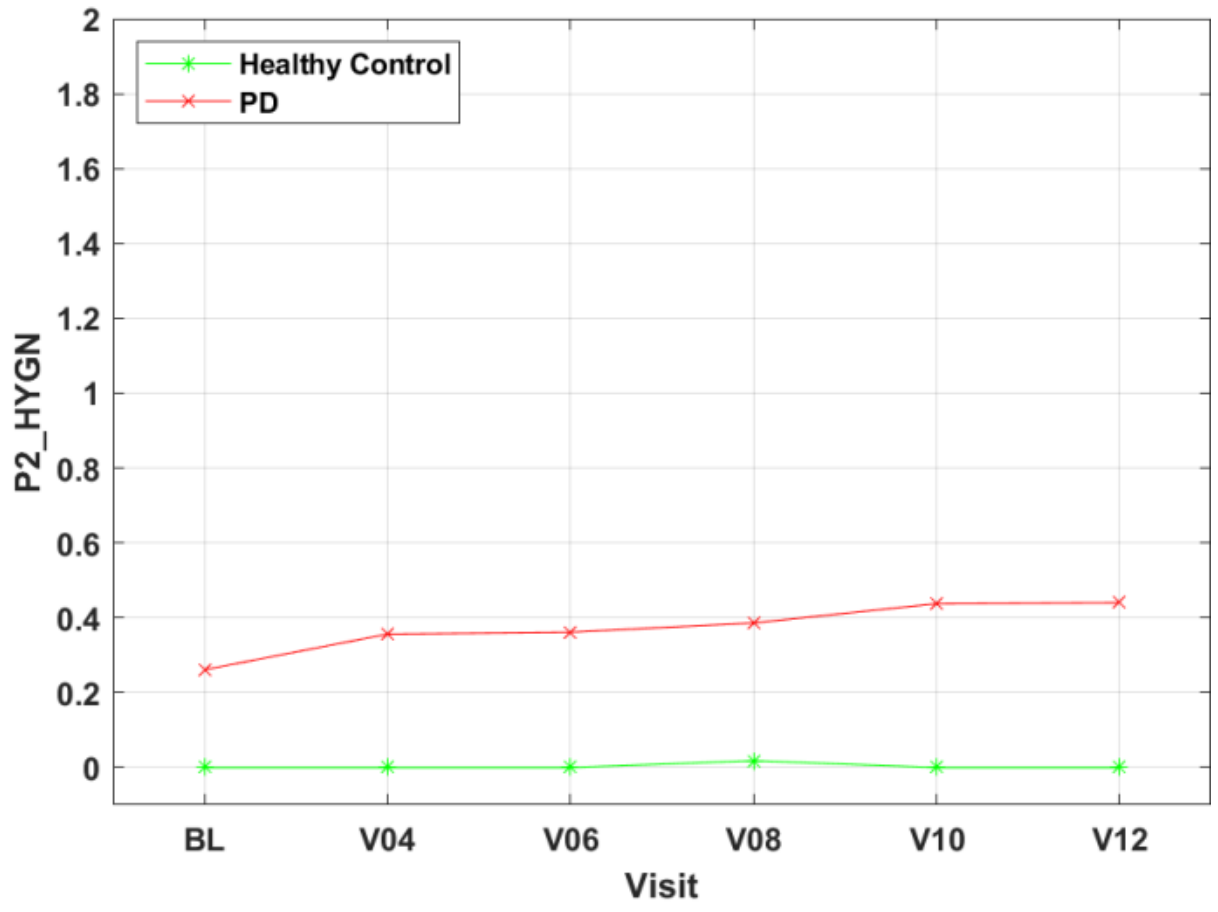

(s) Hygiene

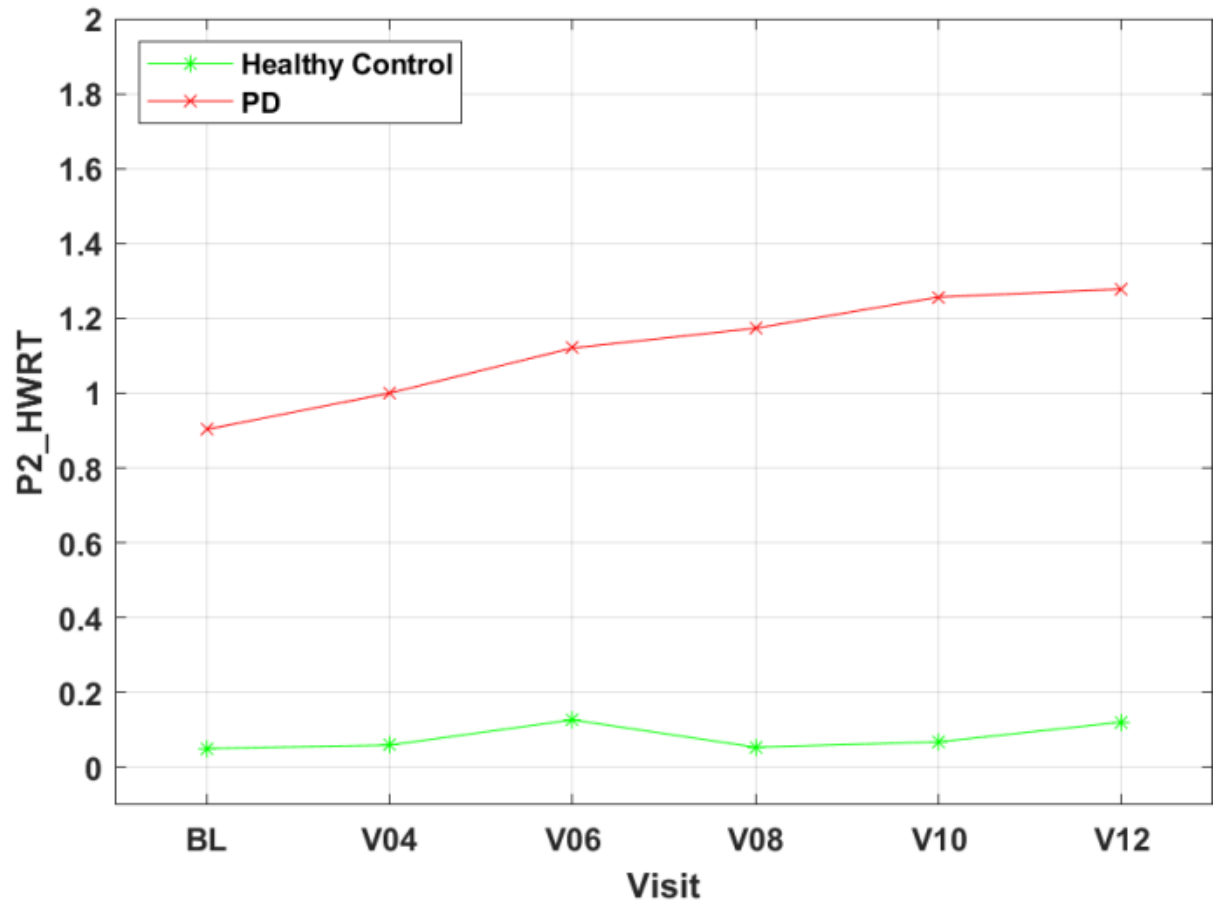

(t) Handwriting

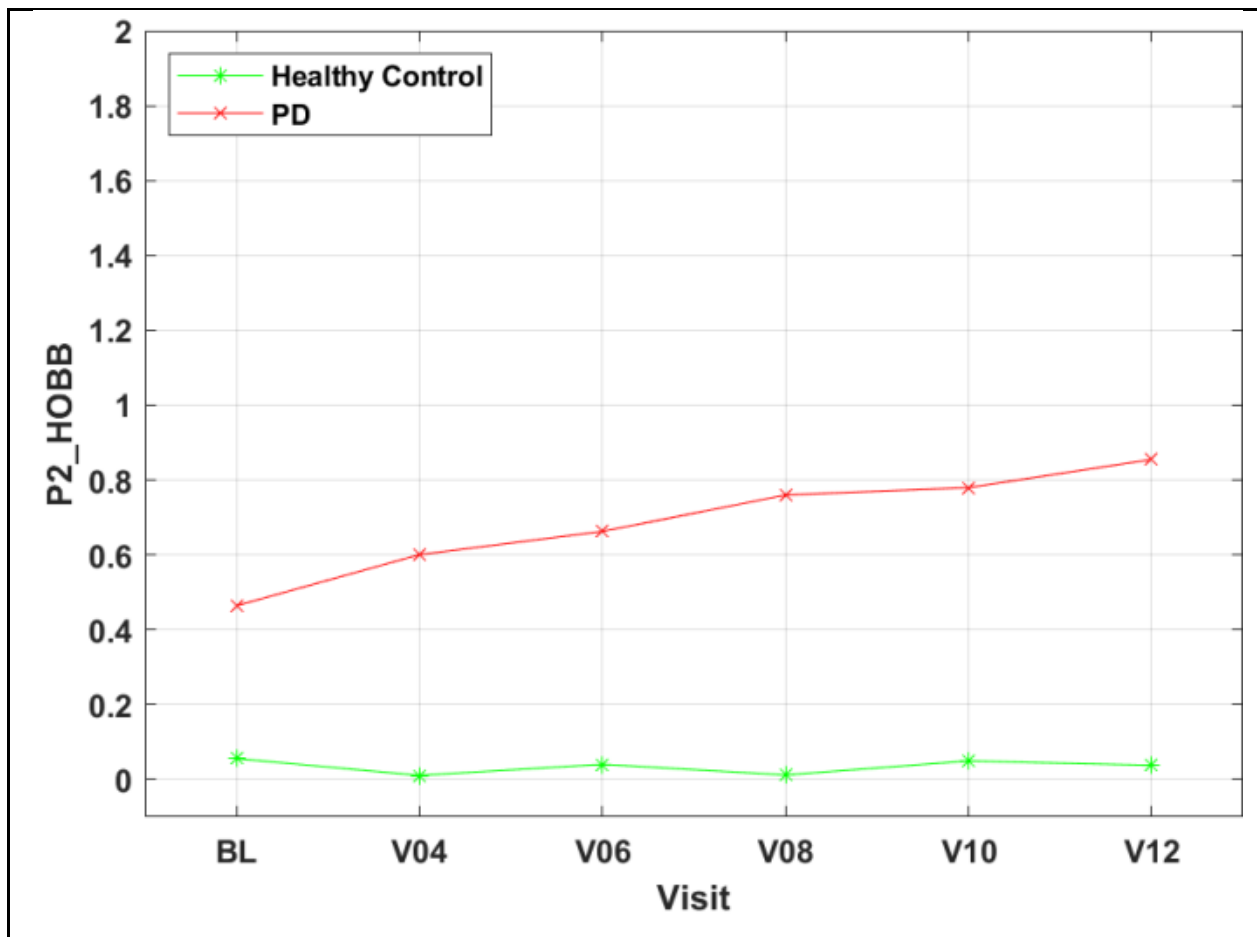

(u) Doing hobbies

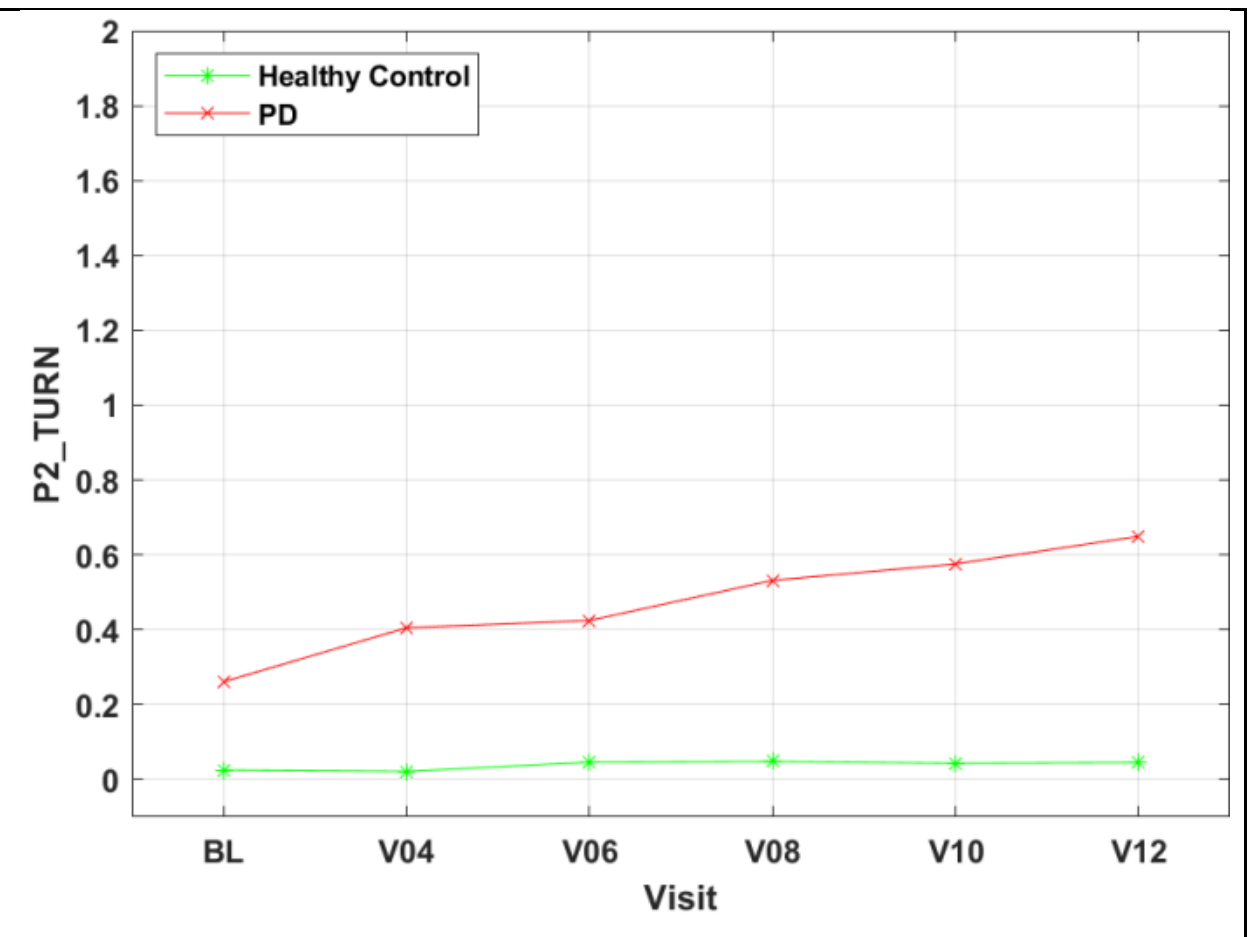

(v) Turning in bed

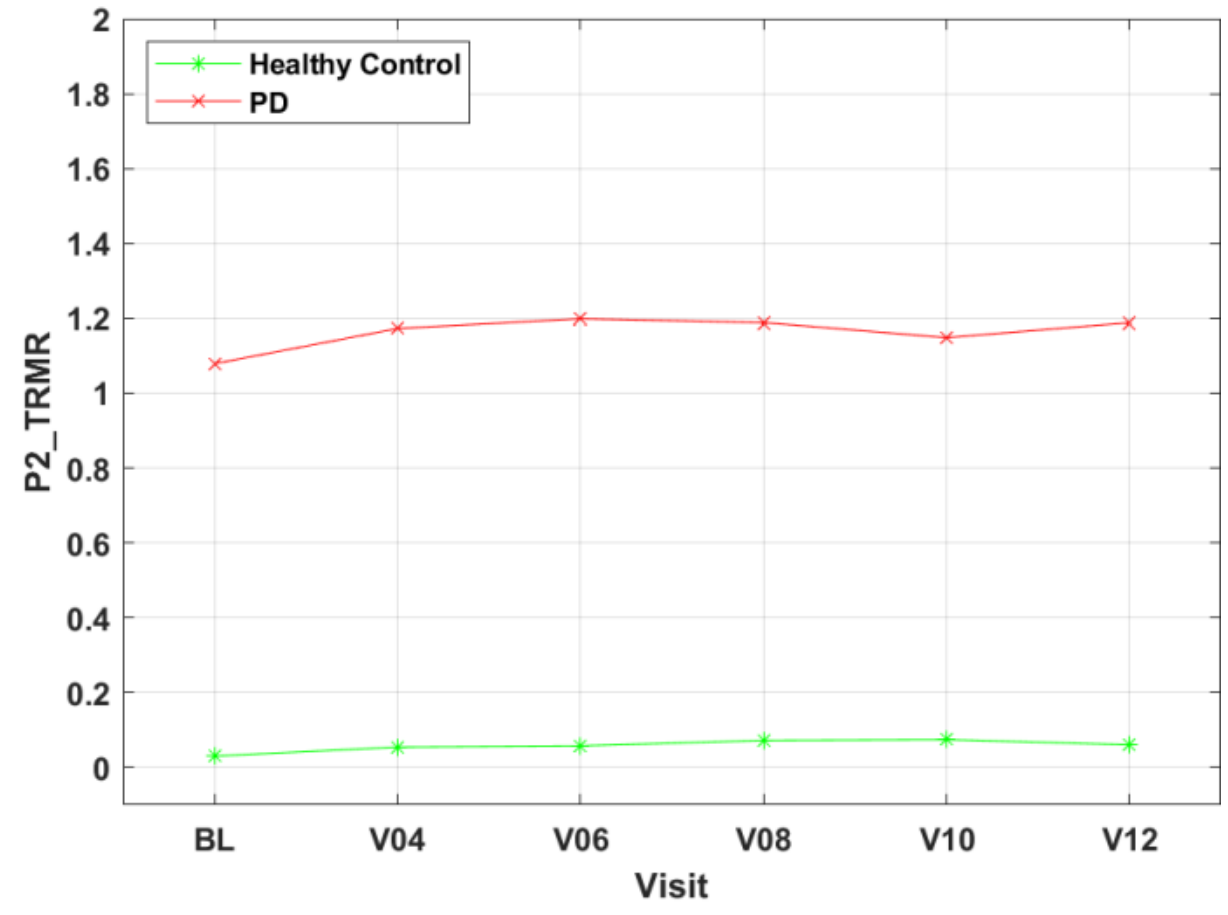

(w) Tremor

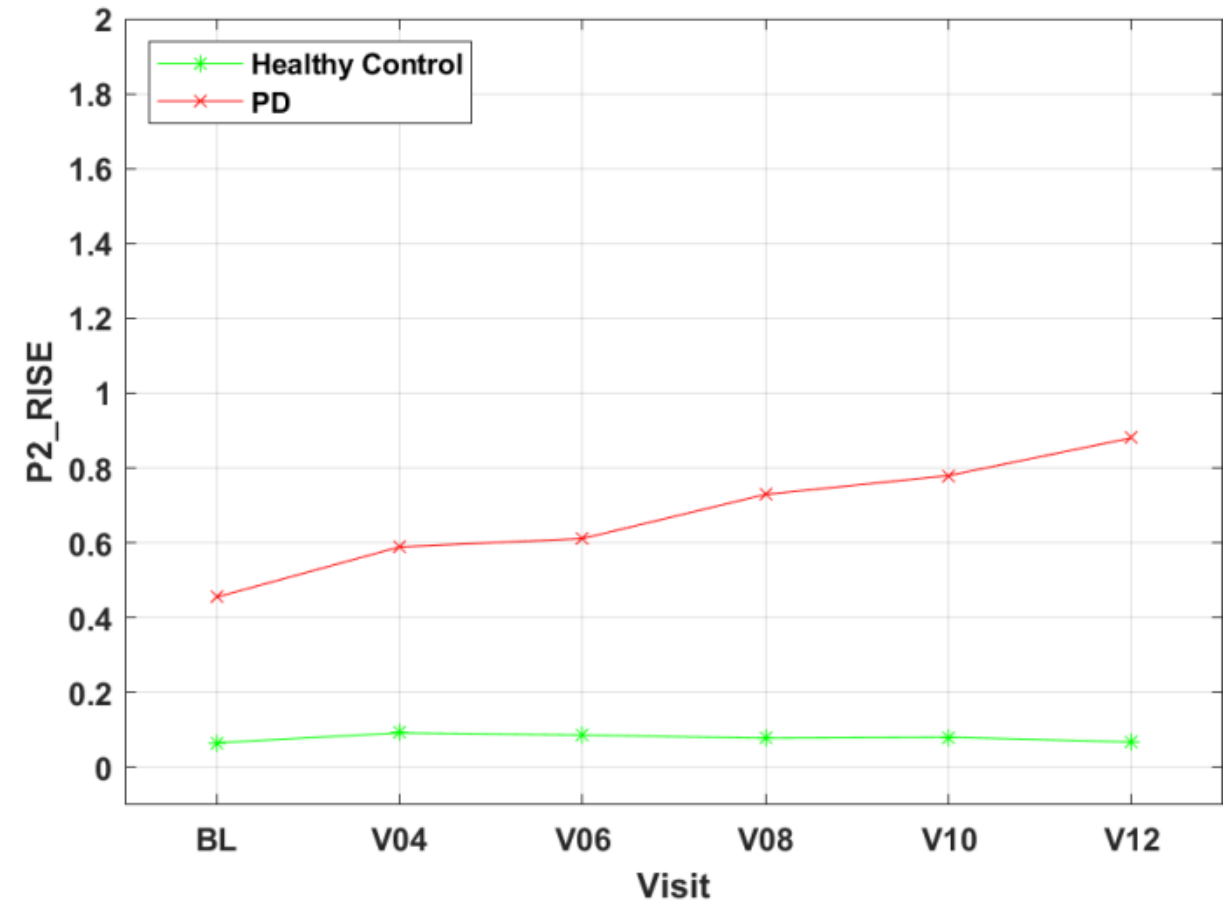

(x) Getting out of bed

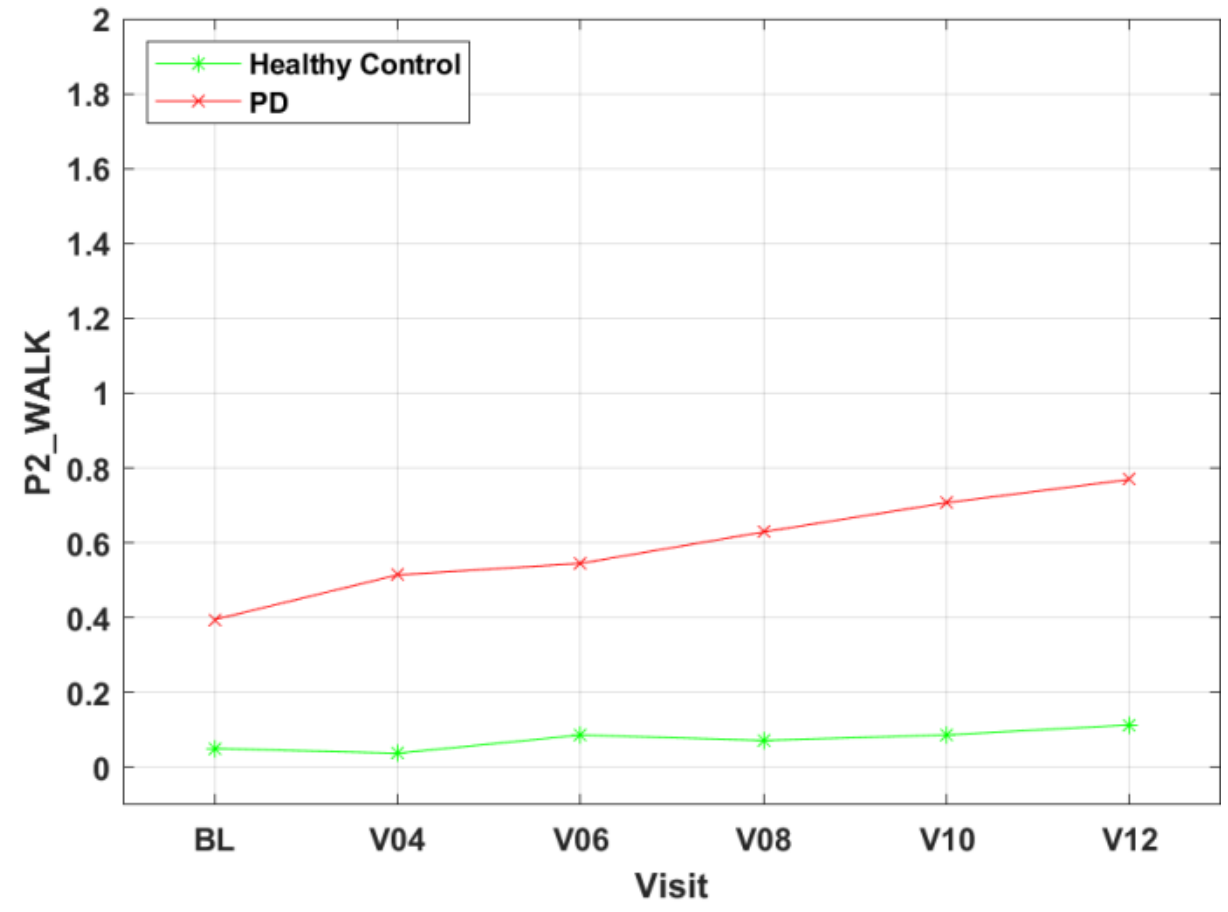

(y) Walking & Balance

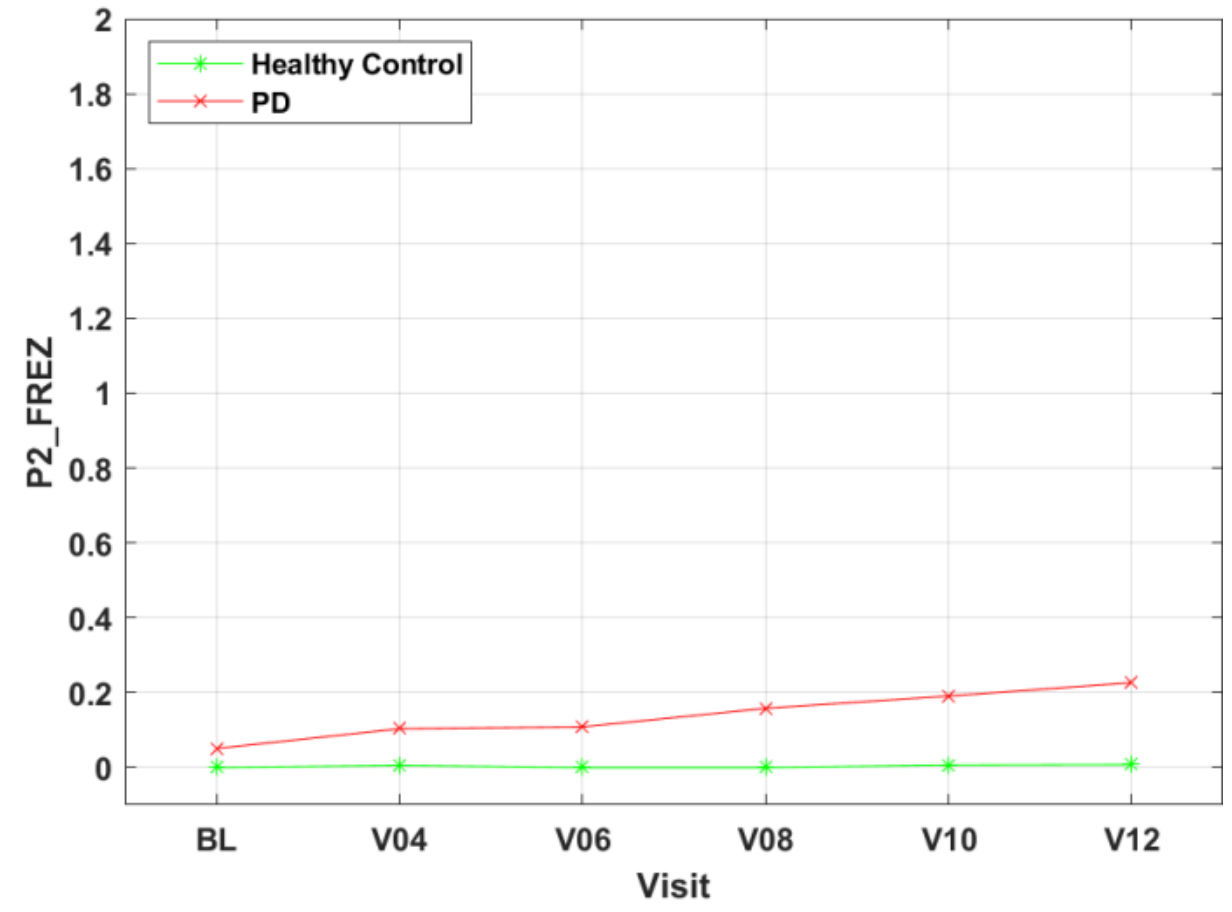

(z) Freezing

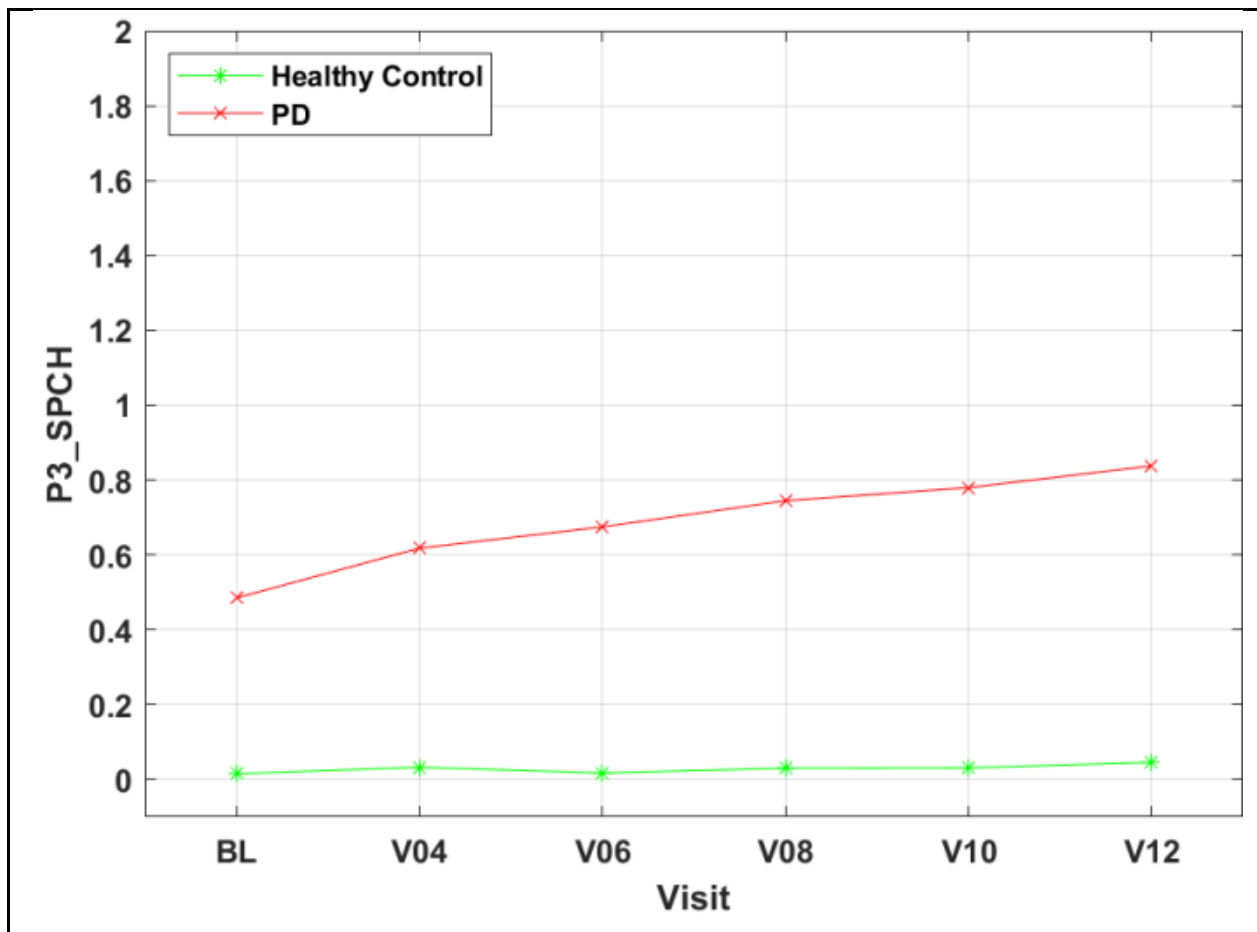

(aa) Speech

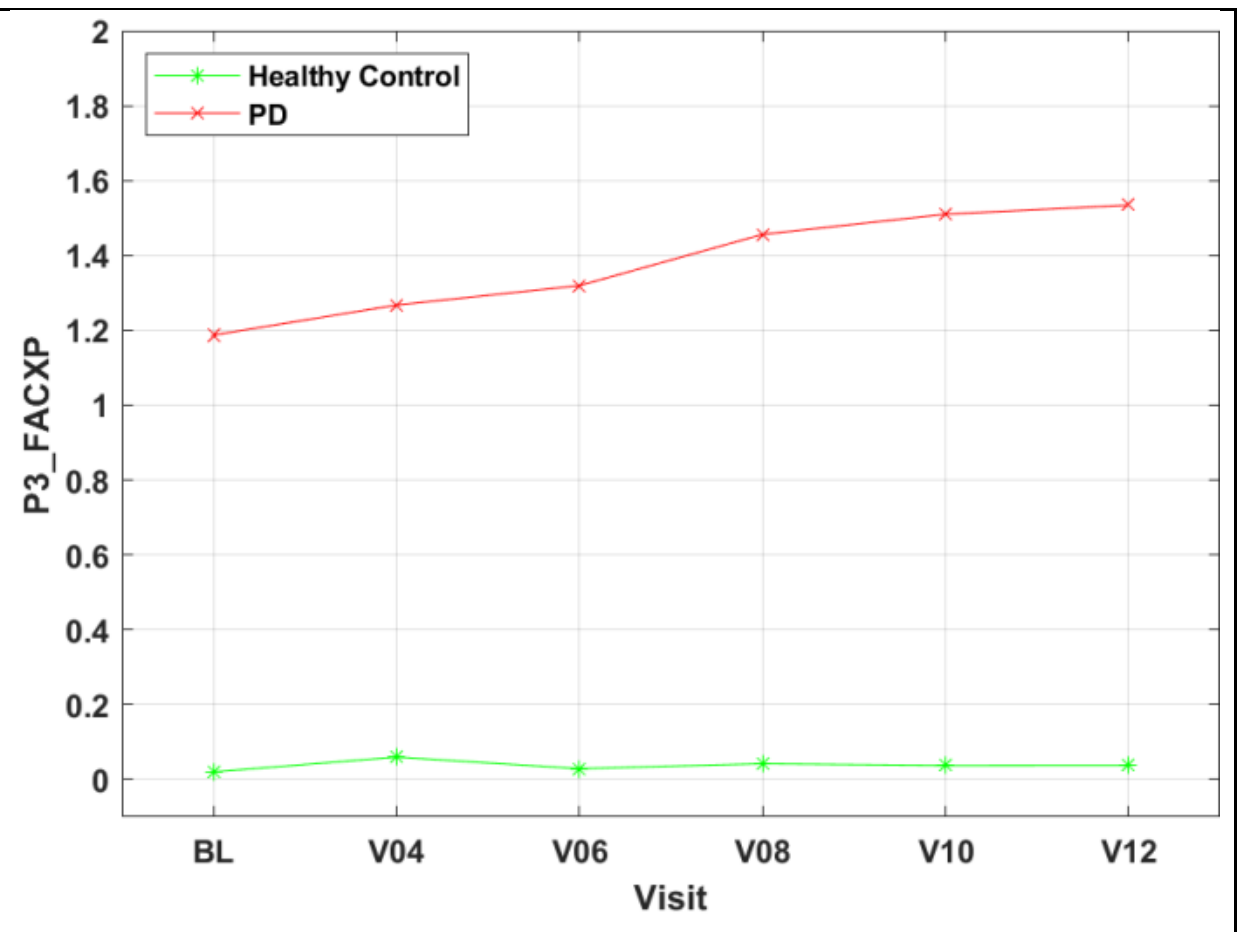

(ab) Facial expression

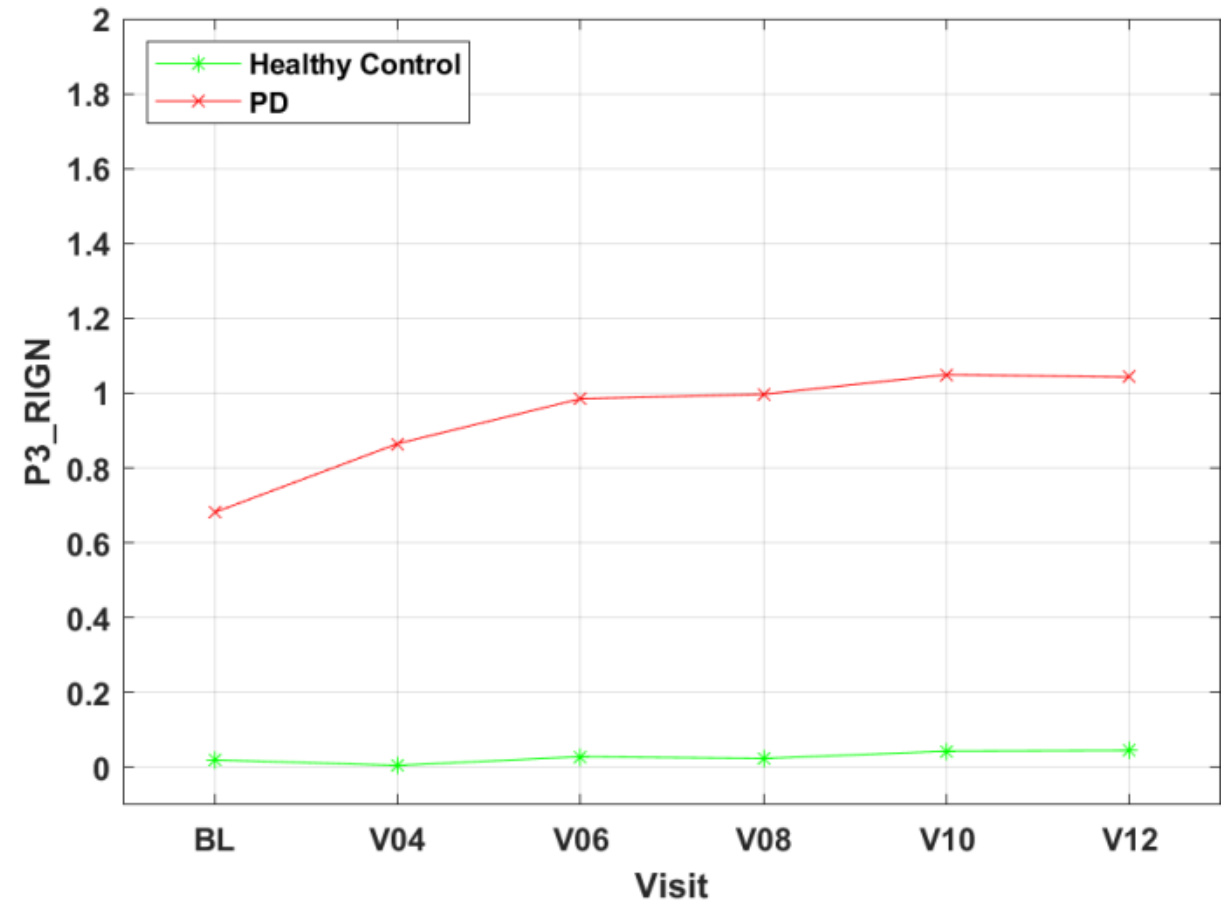

(ac) Rigidity neck

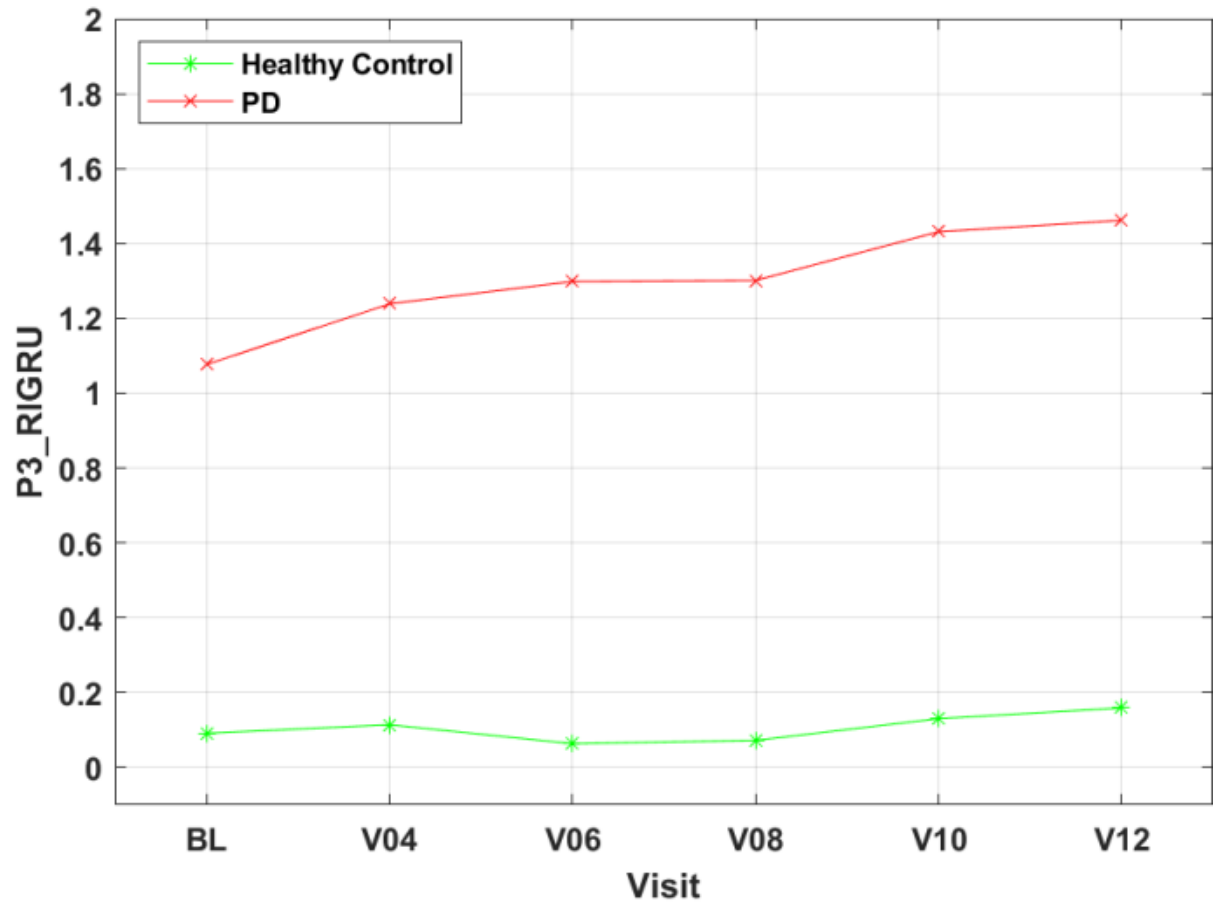

(ad) Rigidity Right Upper Extremity (RUE)

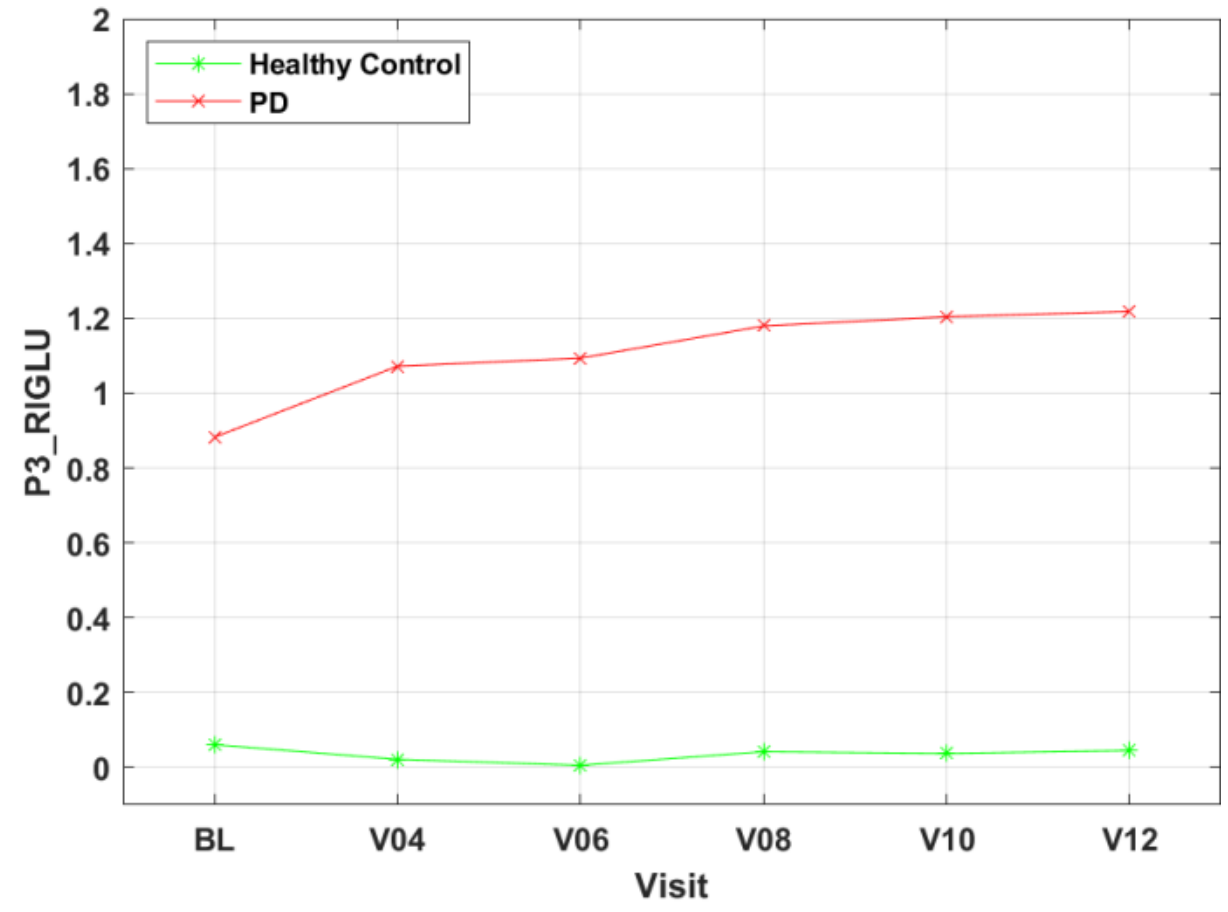

(ae) Rigidity Left Upper Extremity (LUE)

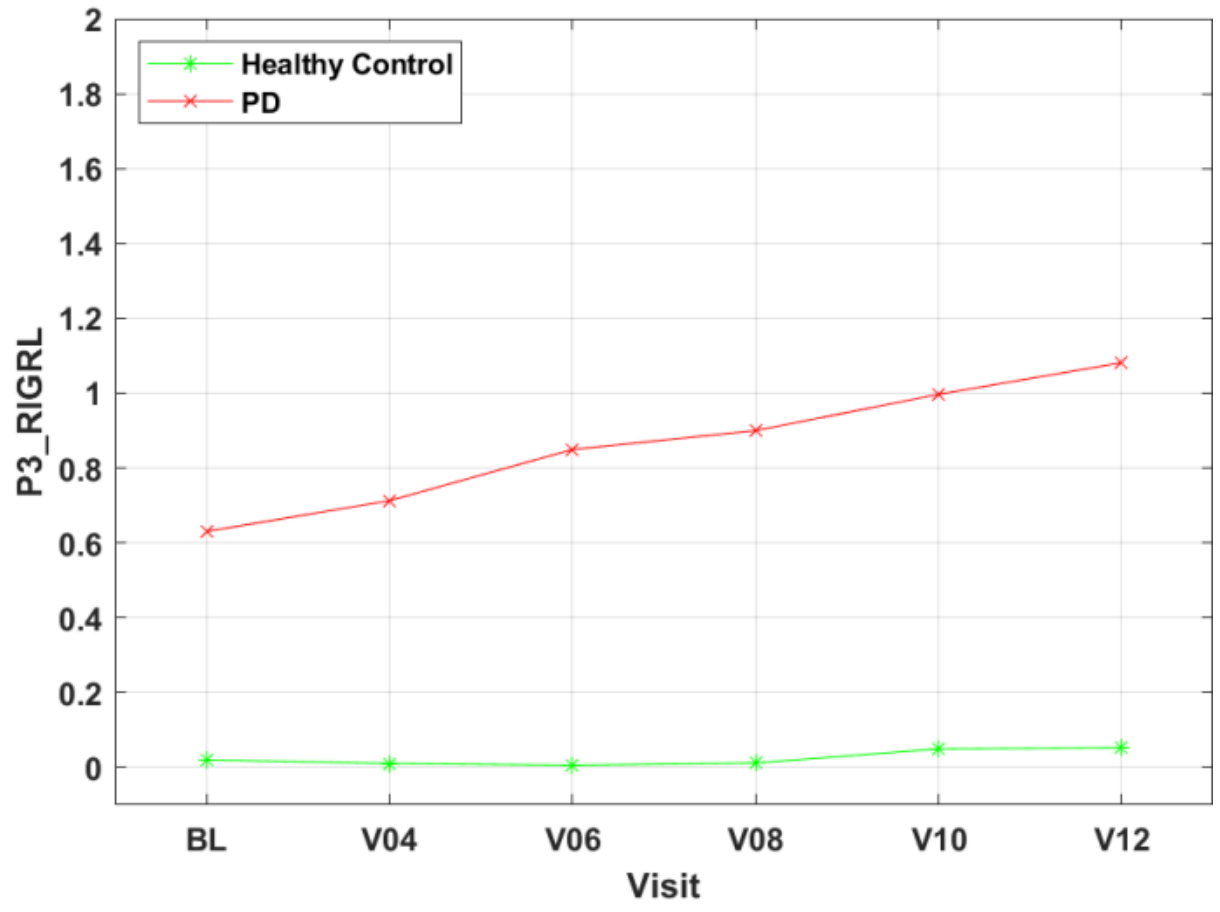

(af) Rigidity Right Lower Extremity (RLE)

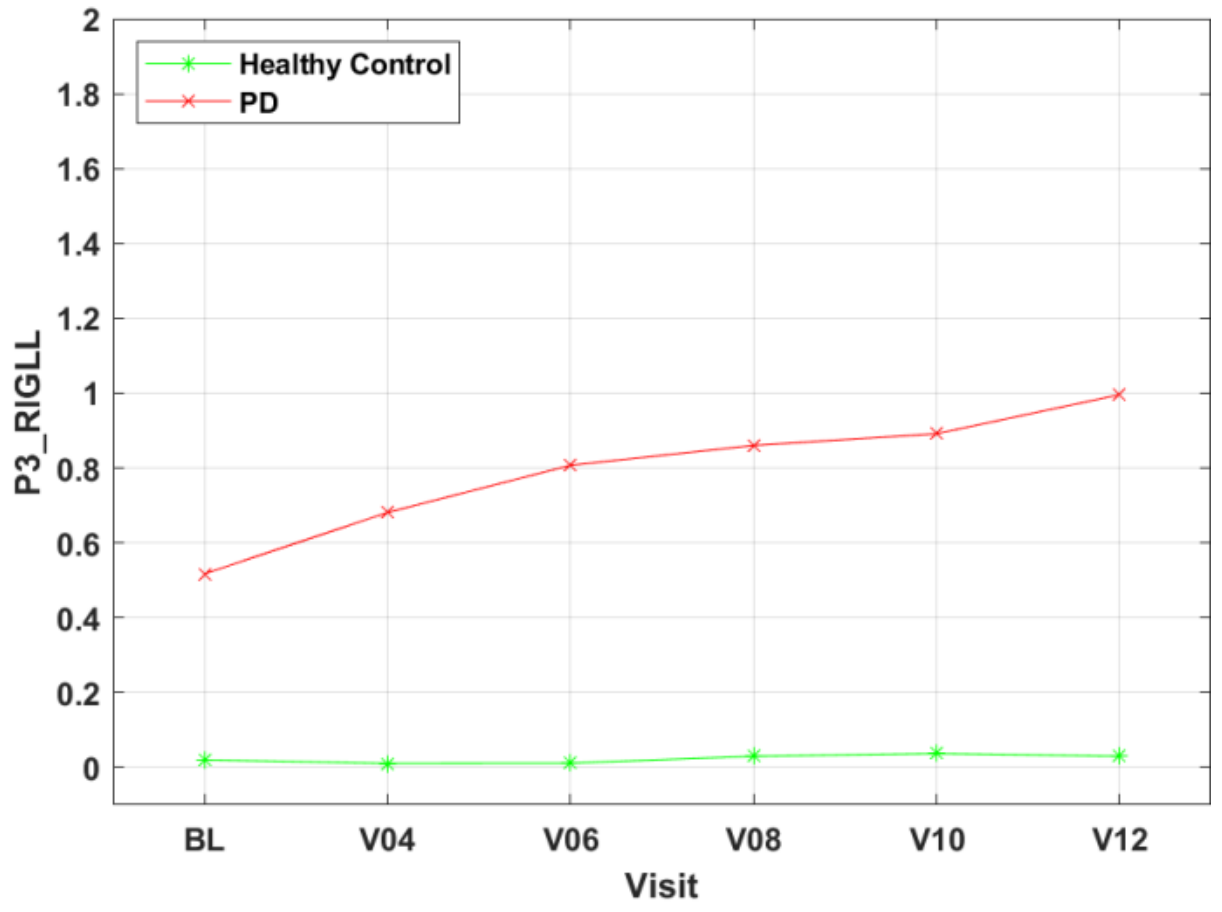

(ag) Rigidity Left Lower Extremity (LLE)

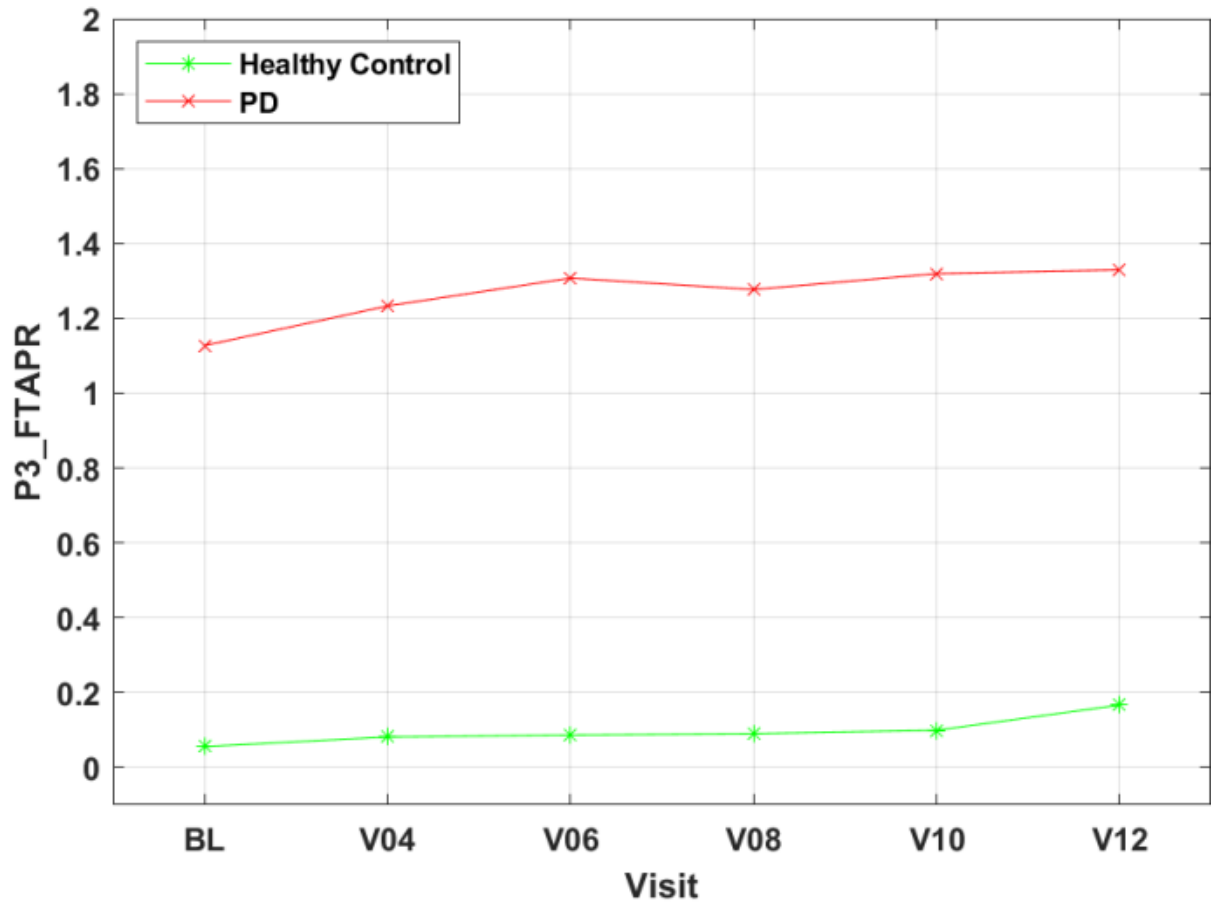

(ah) Finger tapping Right Hand (RH)

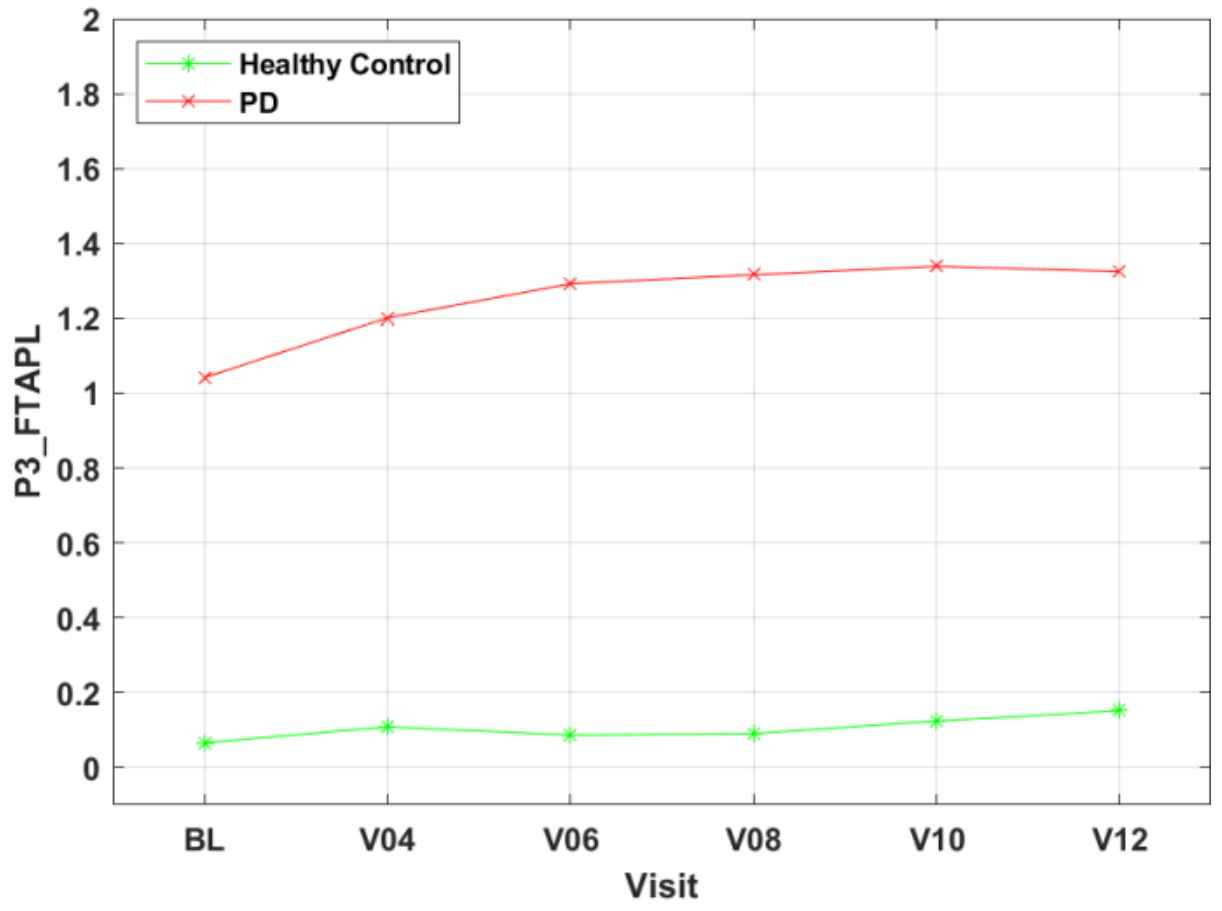

(ai) Finger tapping Left Hand (LH)

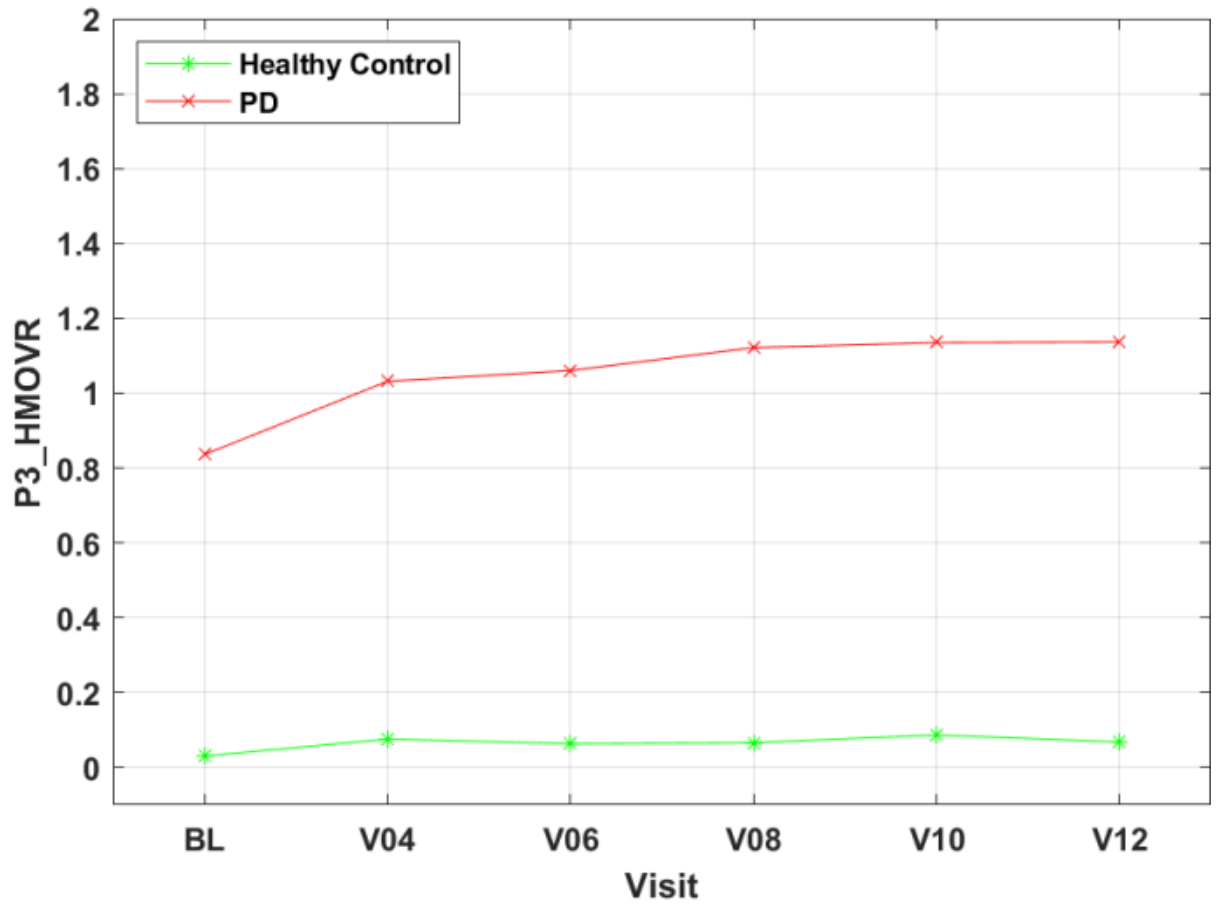

(aj) Hand movements RH

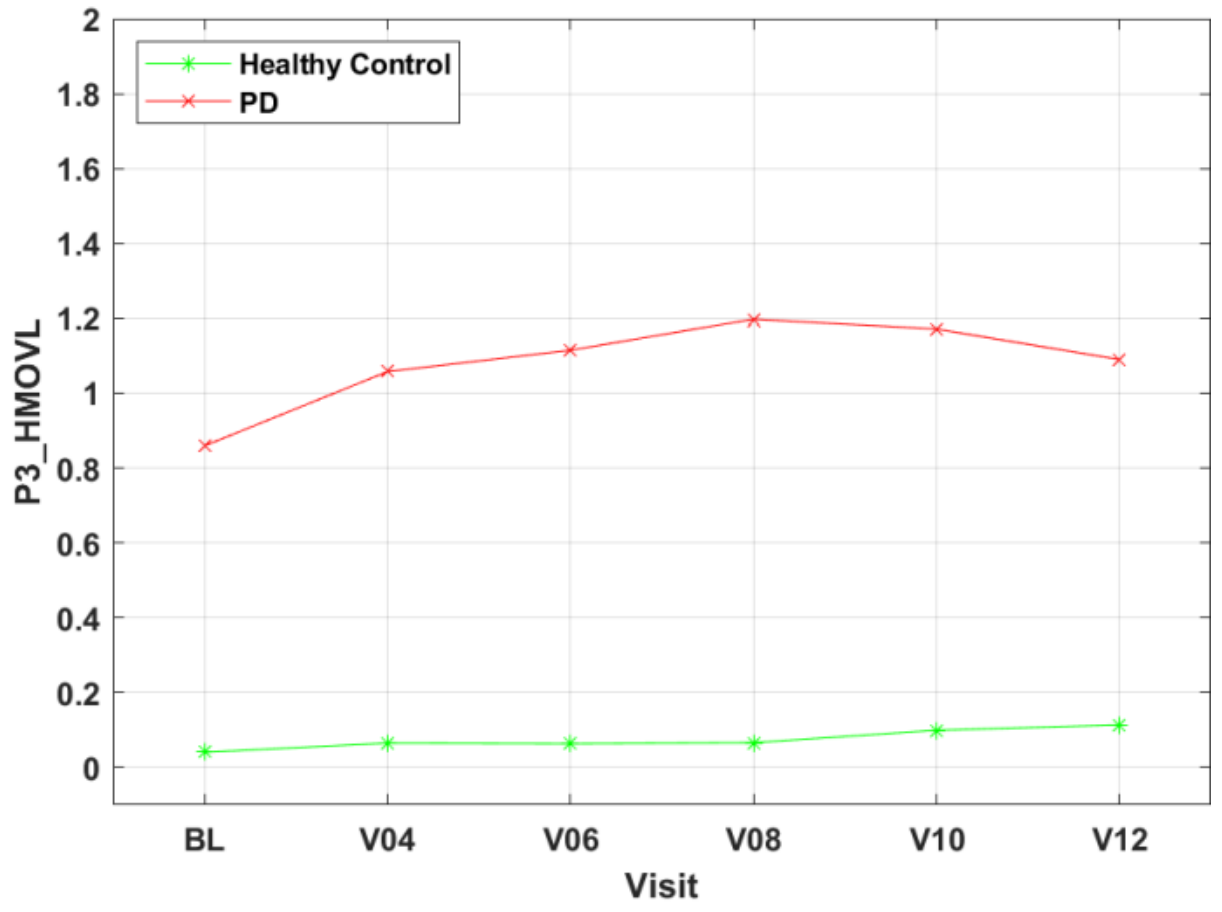

(ak) Hand movements LH

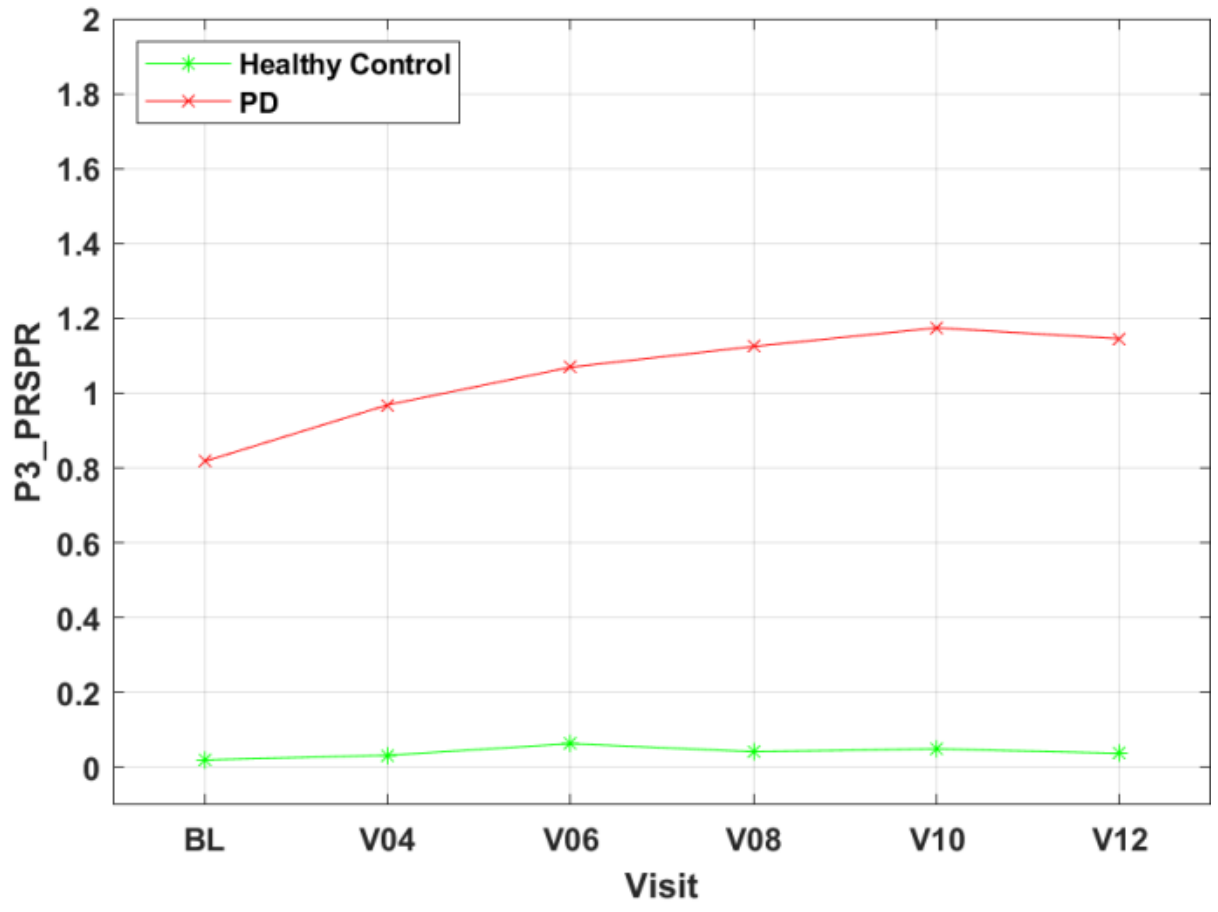

(al) Pronation-Supination (PS) movements RH

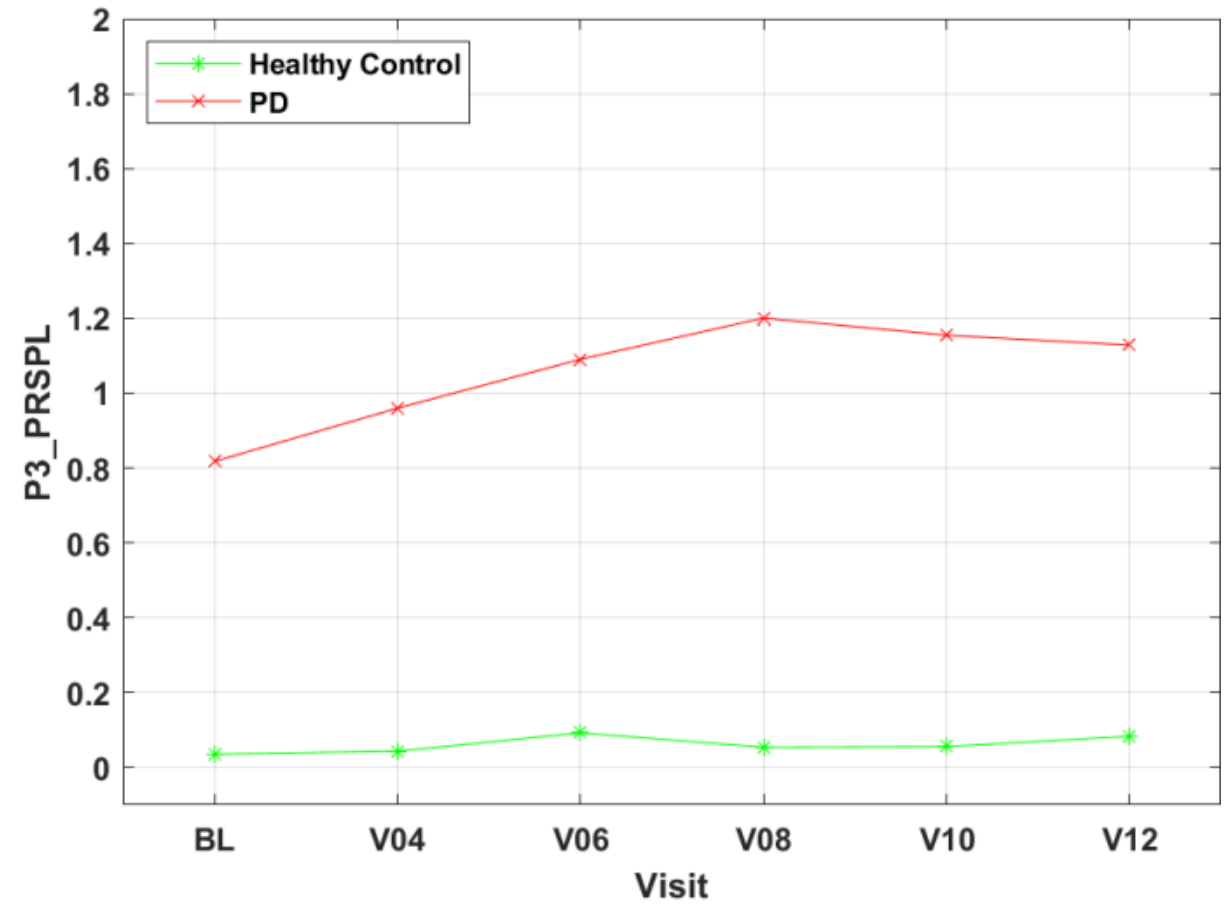

(am) PS movements LH

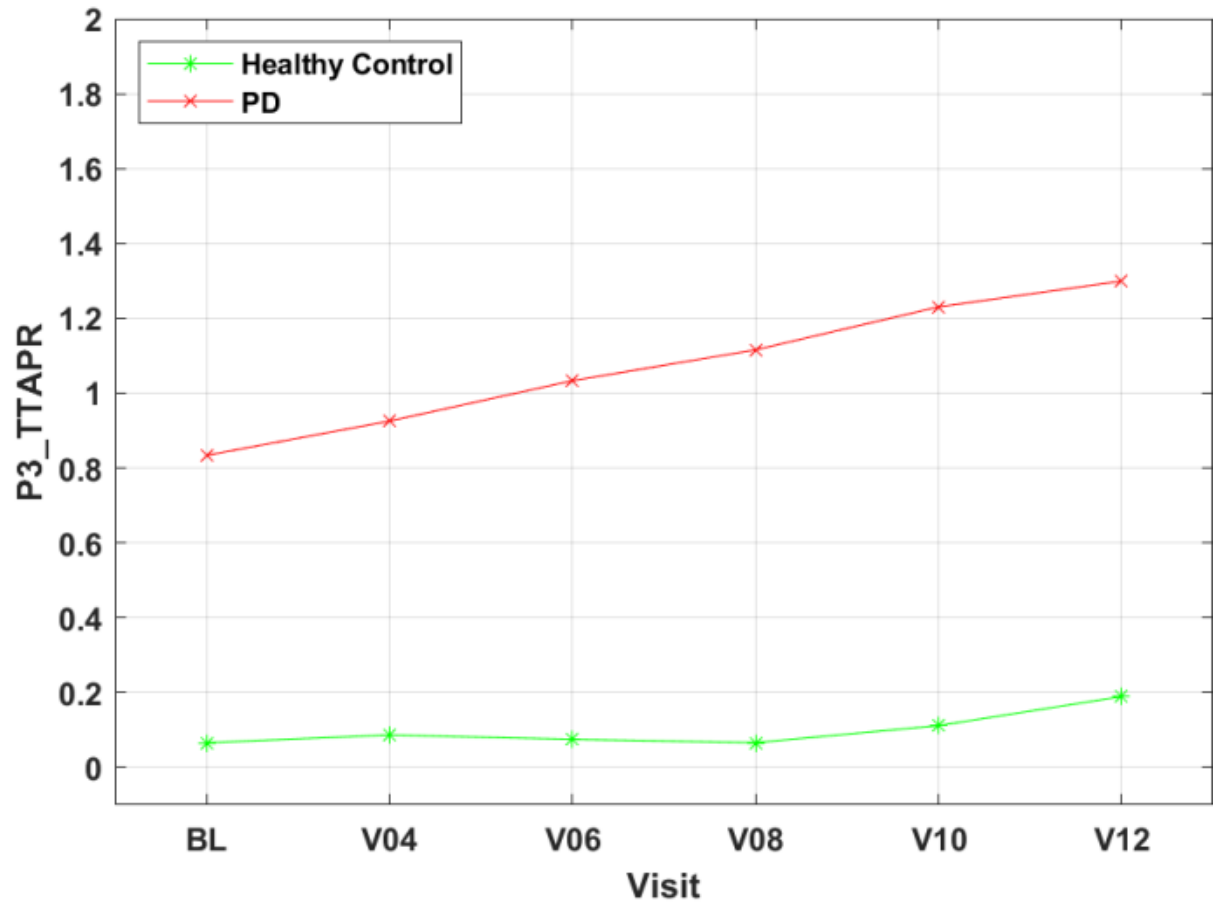

(an) Toe tapping Right Foot (RF)

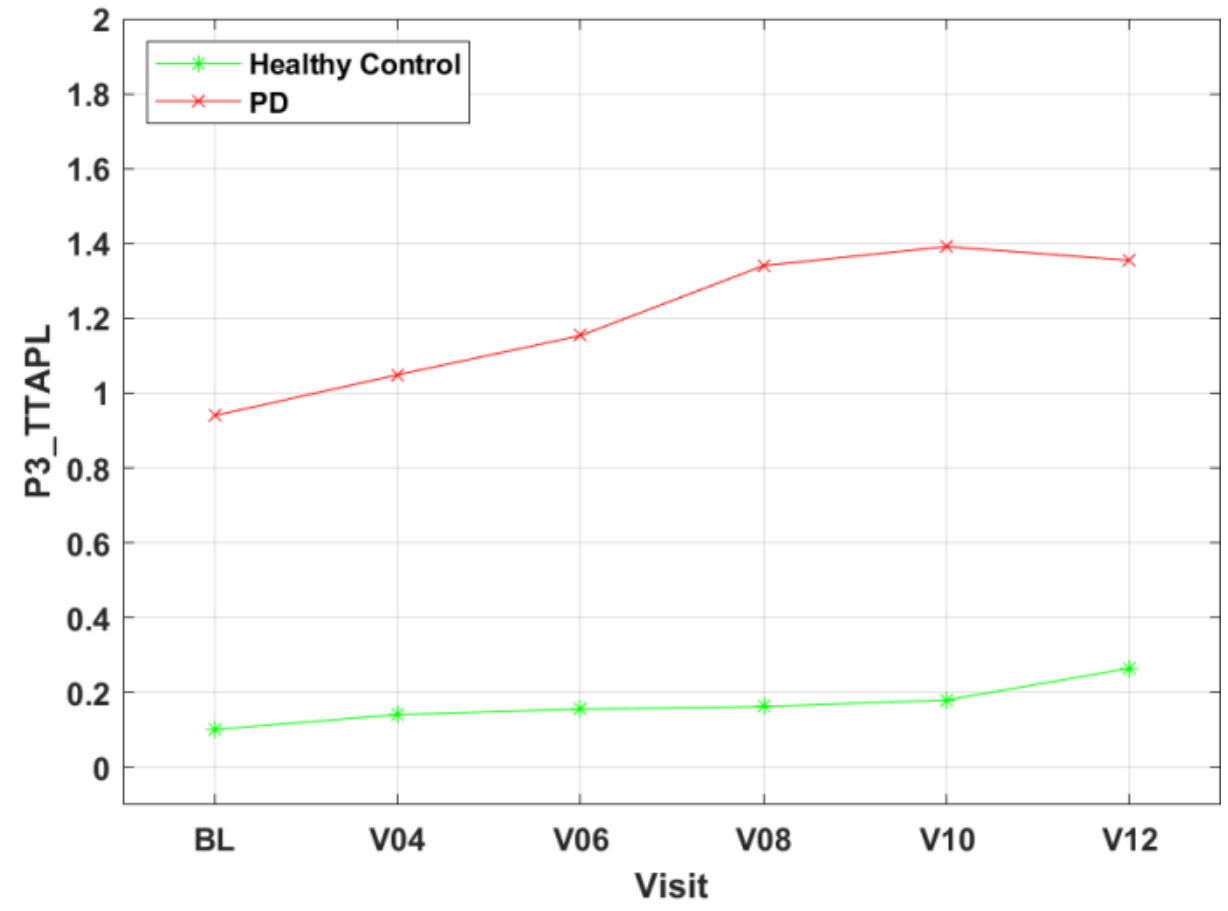

(ao) Toe tapping Left Foot (LF)

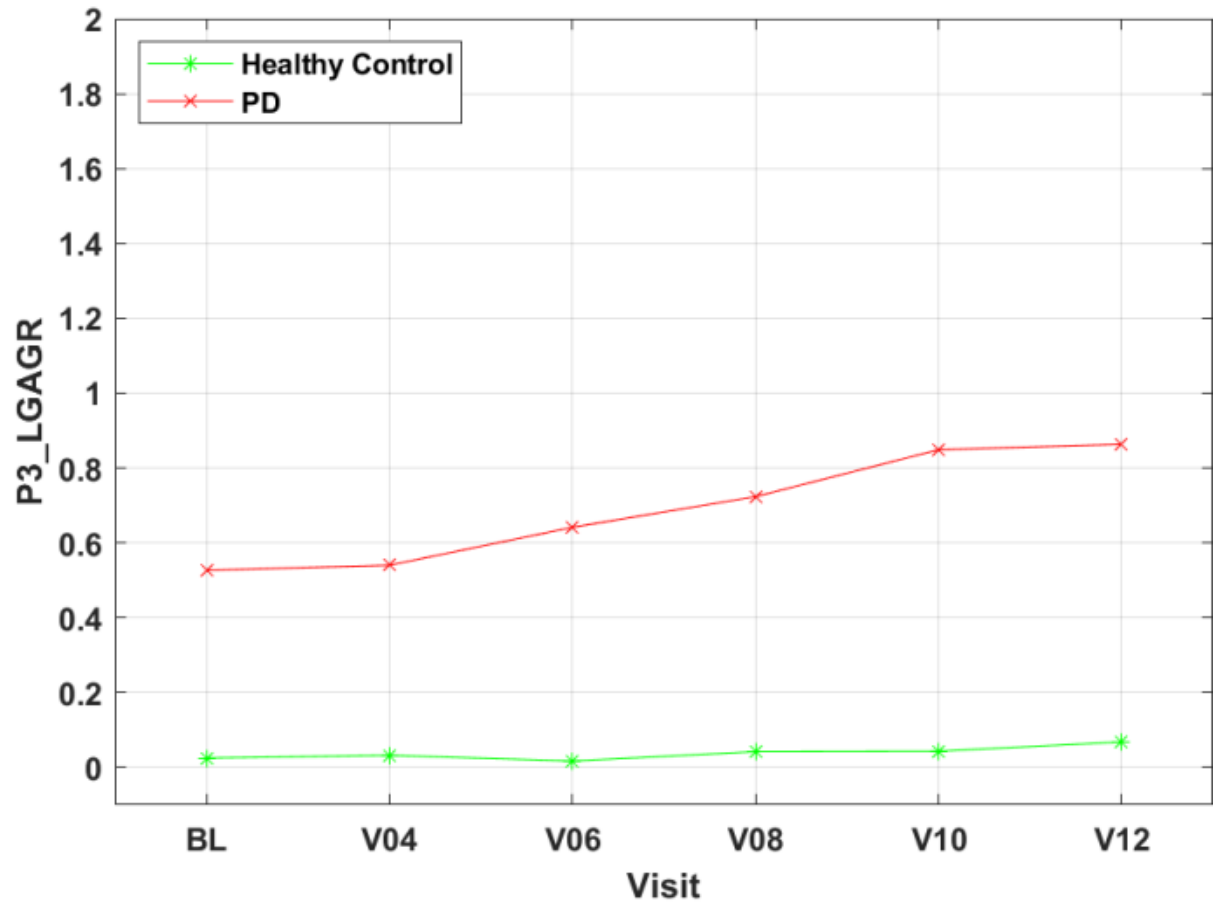

(ap) Leg Agility Right Leg (RL)

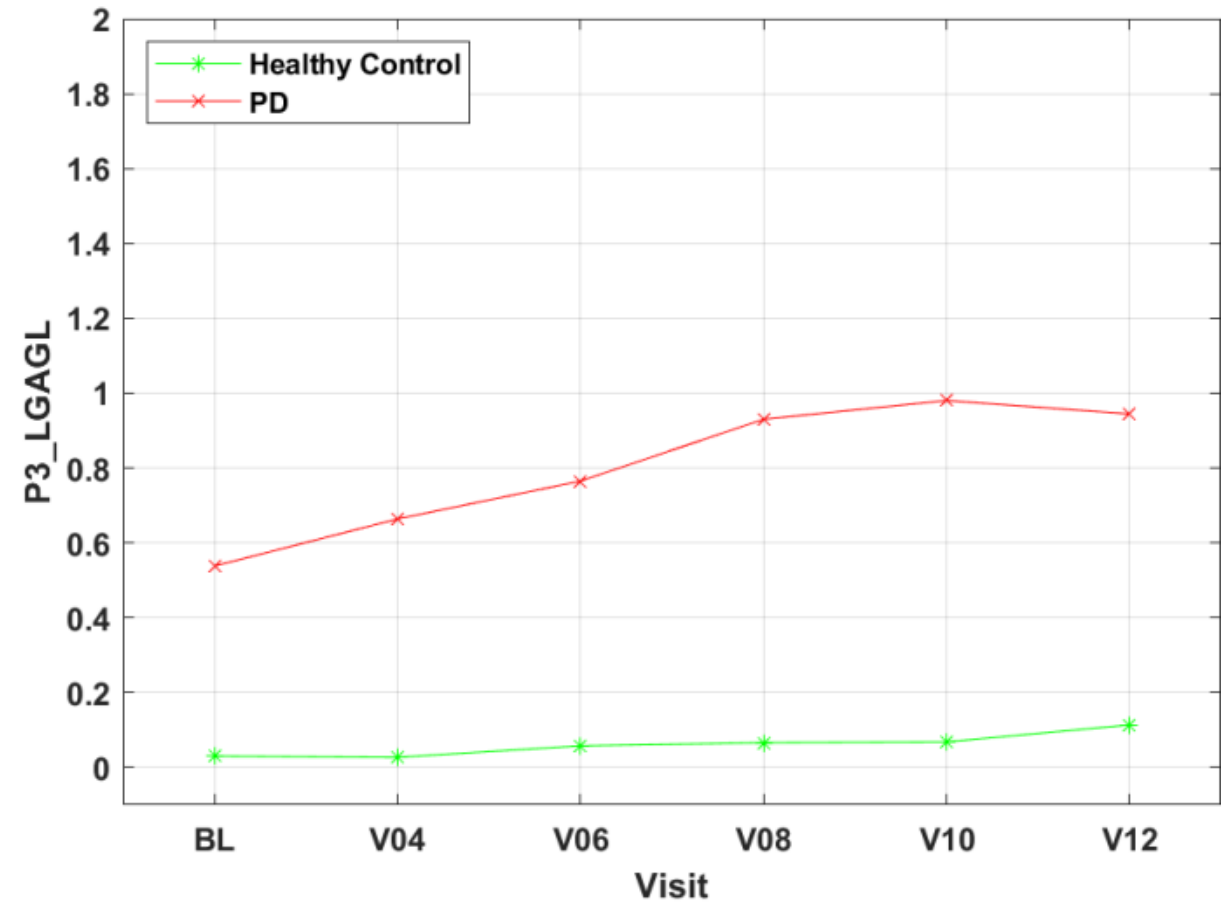

(aq) Leg Agility Left Leg (LL)

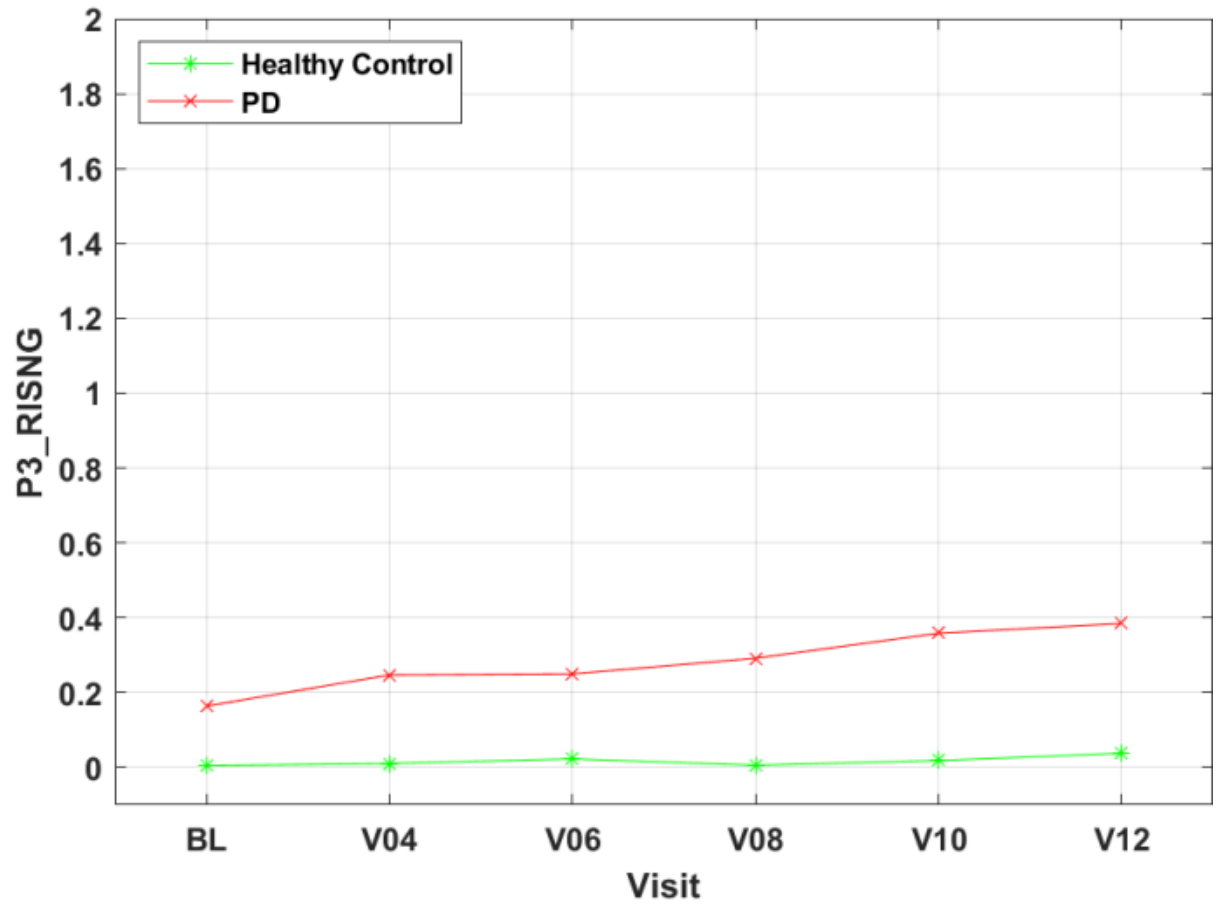

(ar) Arising from chair

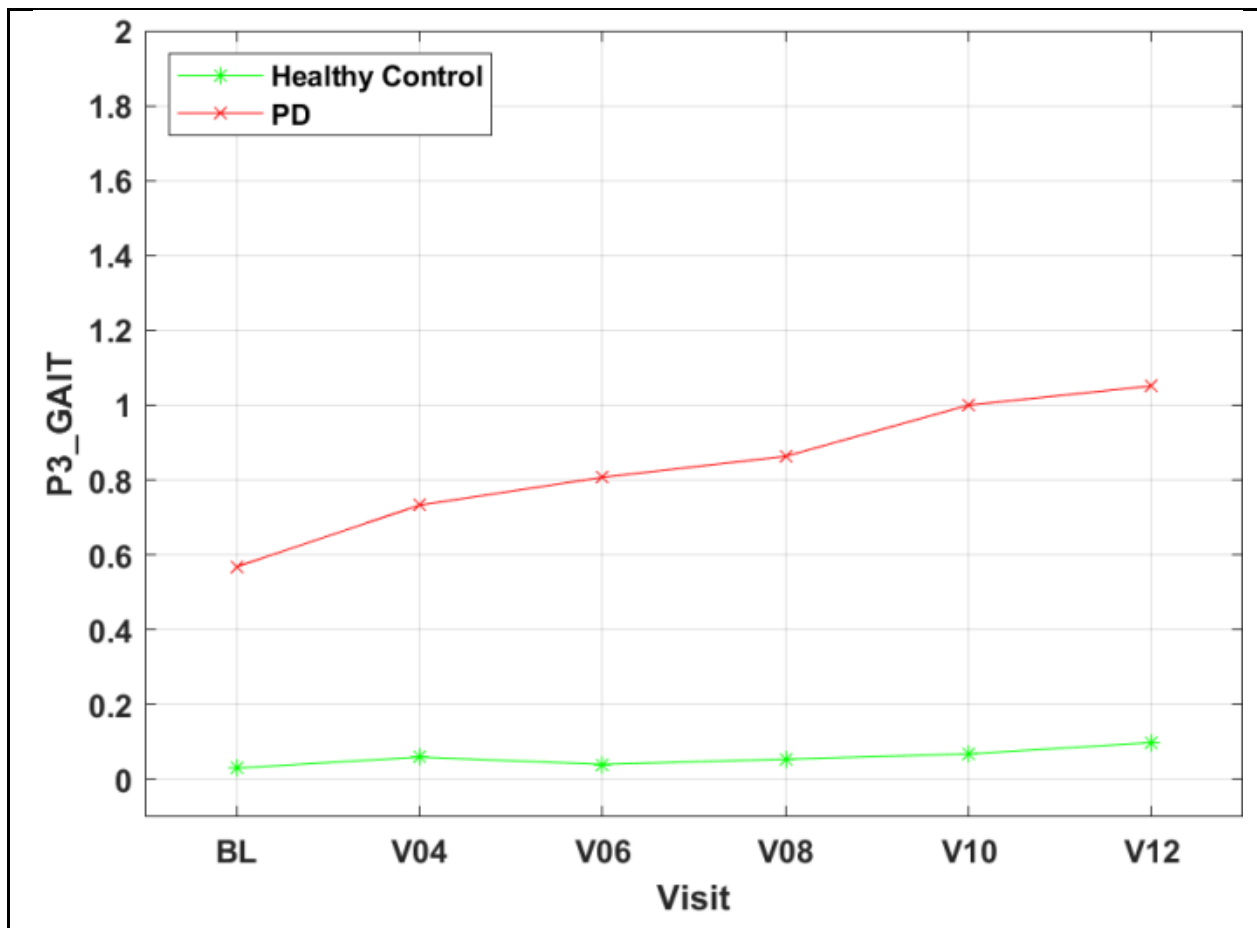

(as) Gait

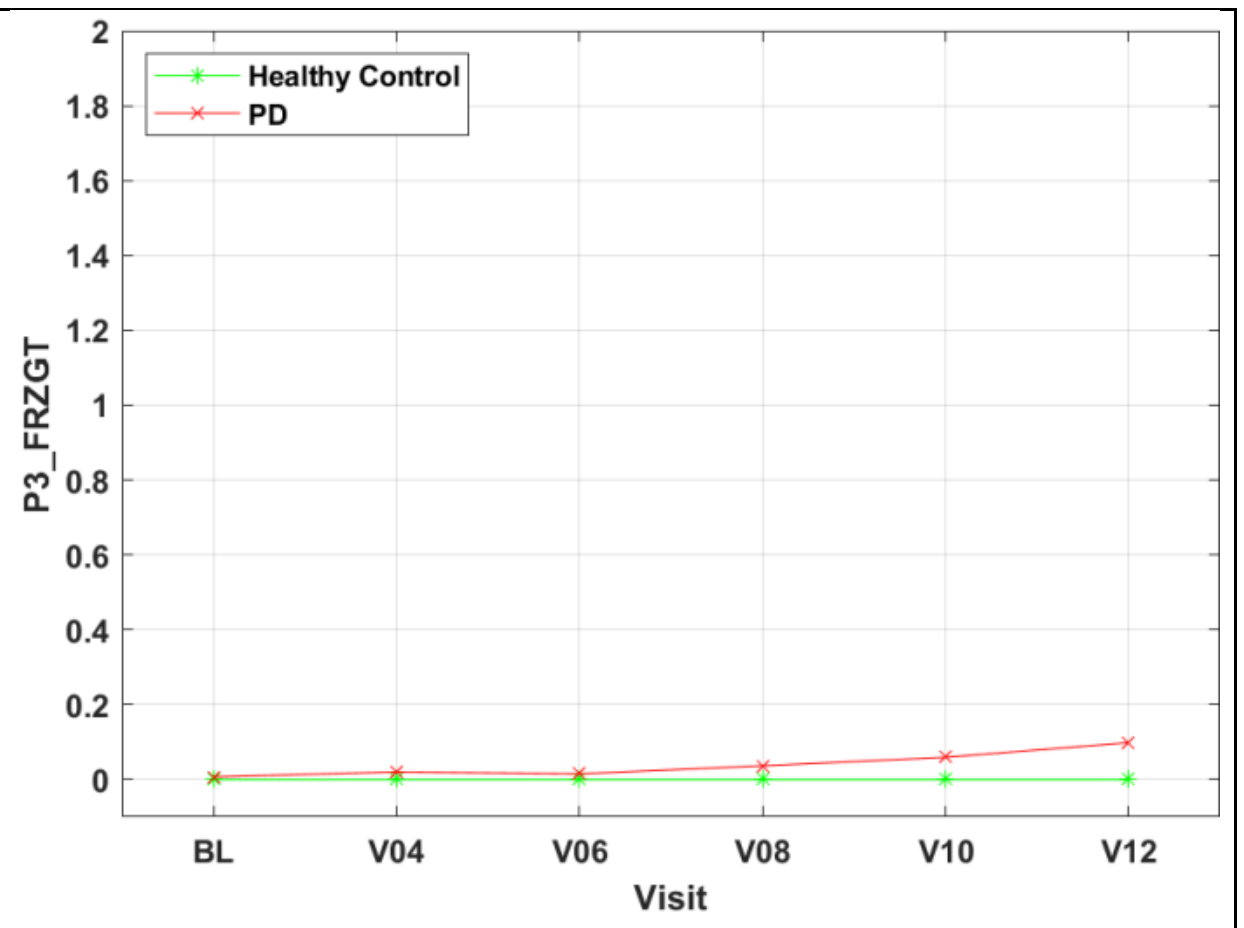

(at) Freezing of gait

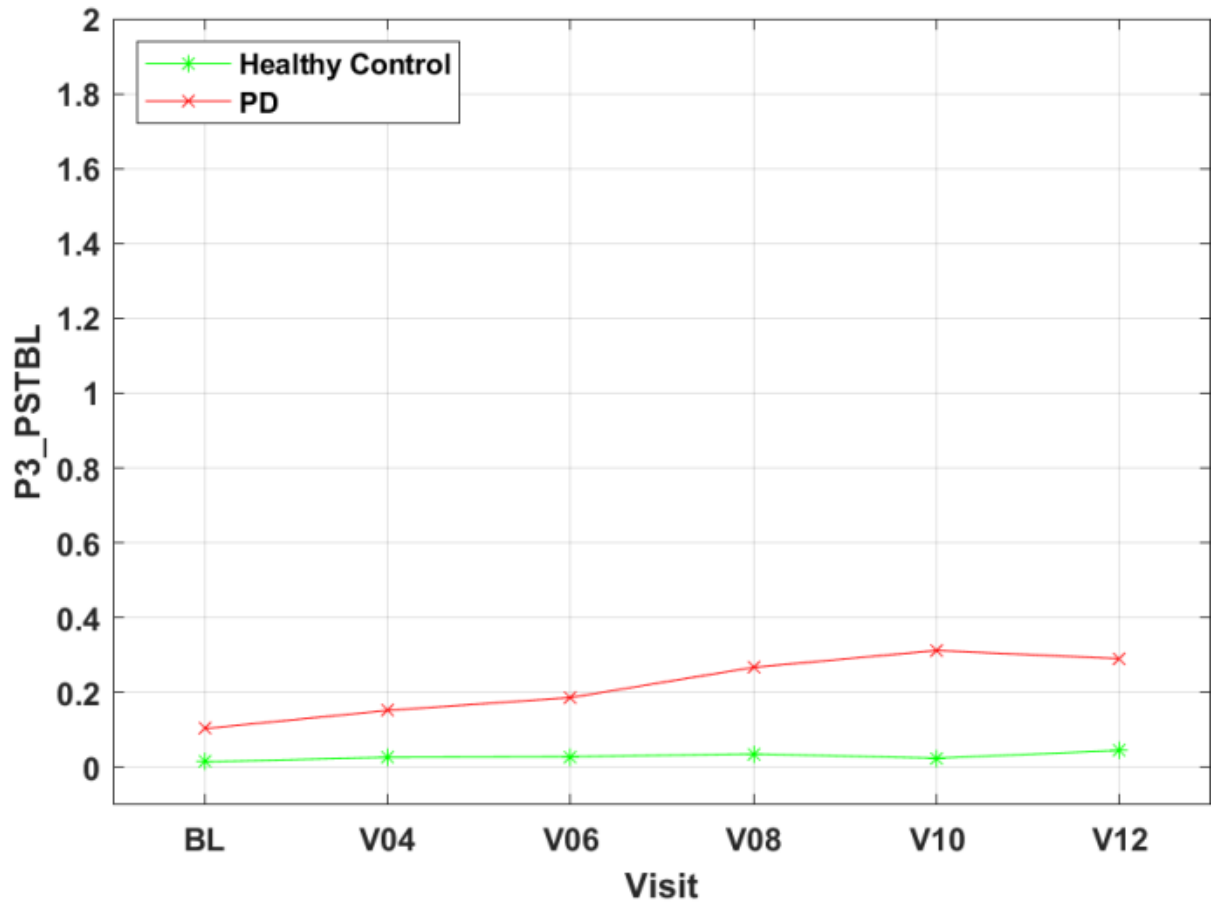

(au) Postural stability

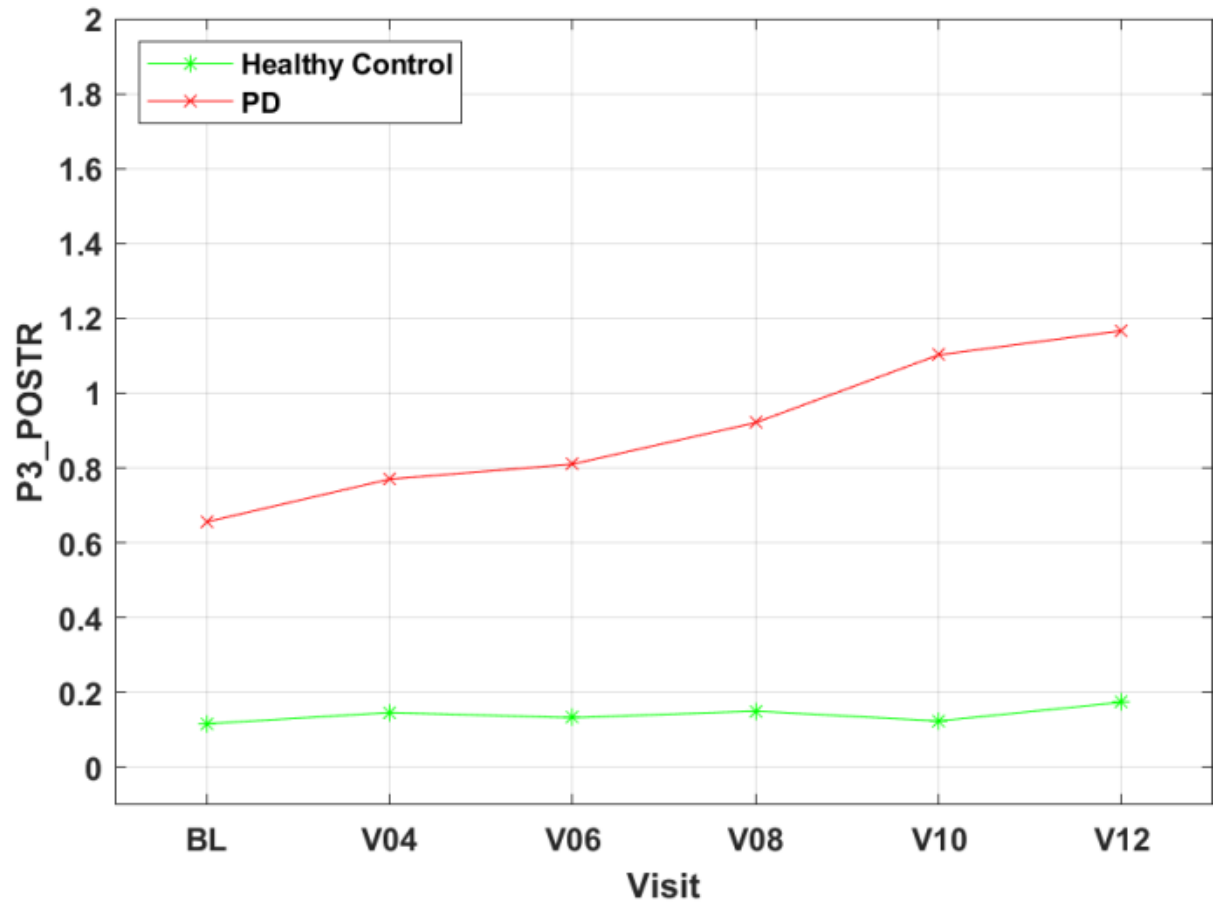

(av) Posture

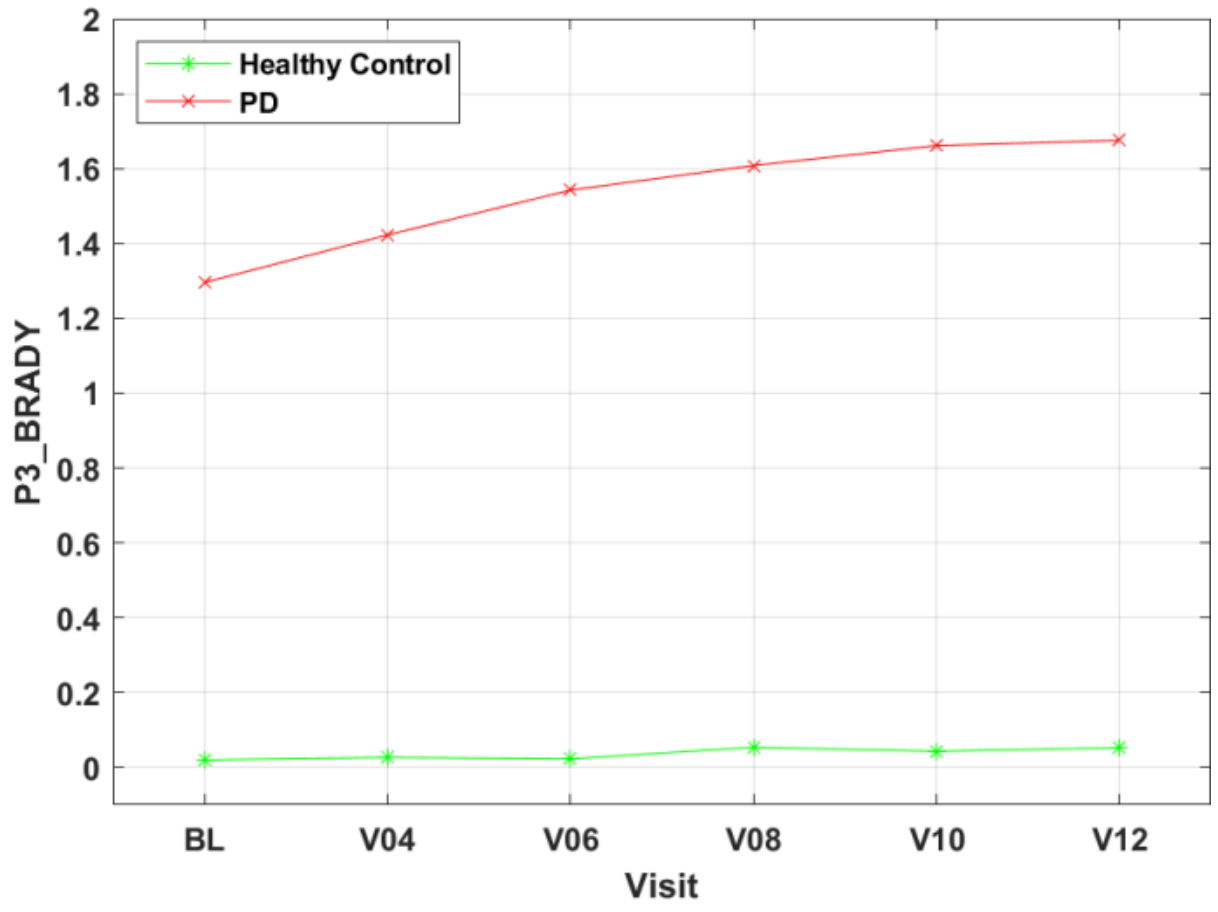

(aw) Global spontaneity of

Movement (bradykinesia)

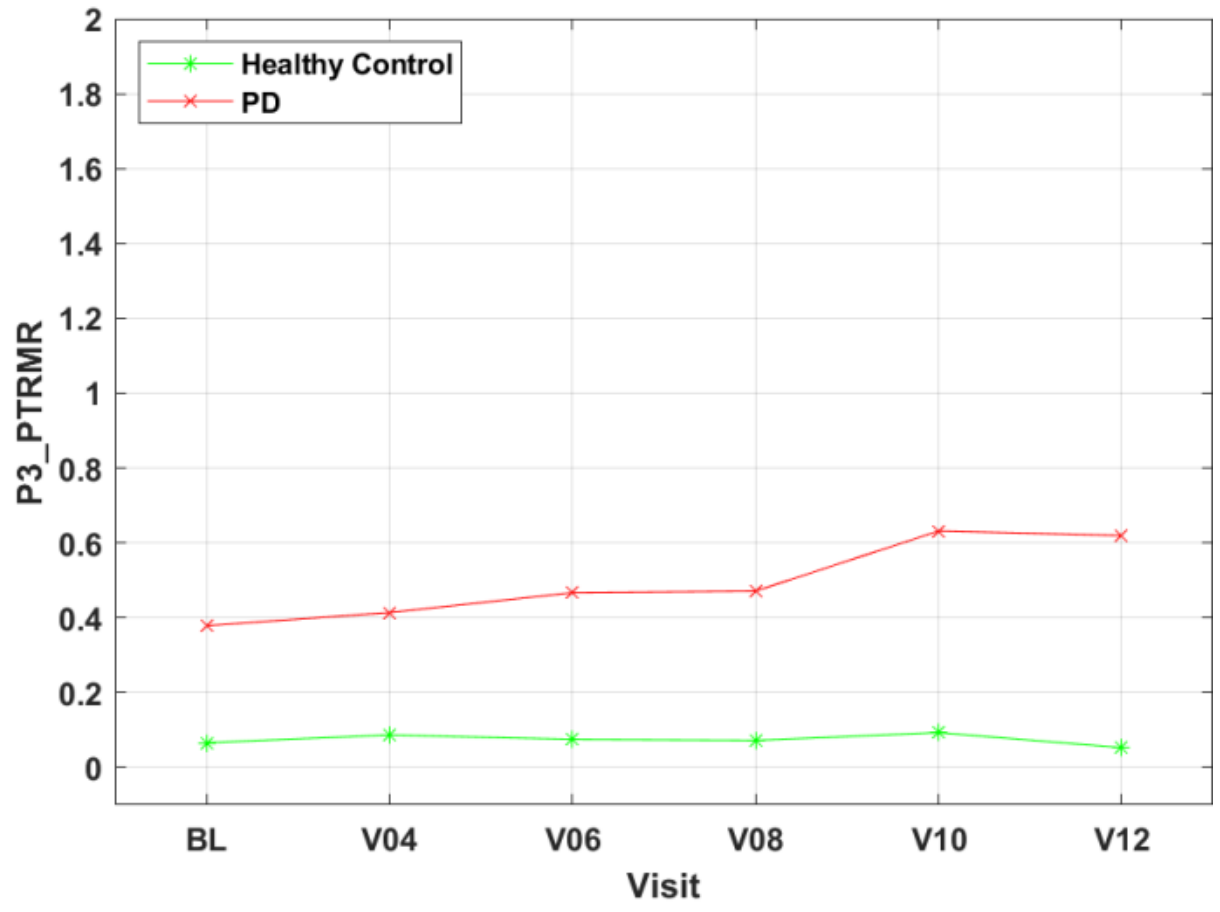

(ax) Postural tremor RH

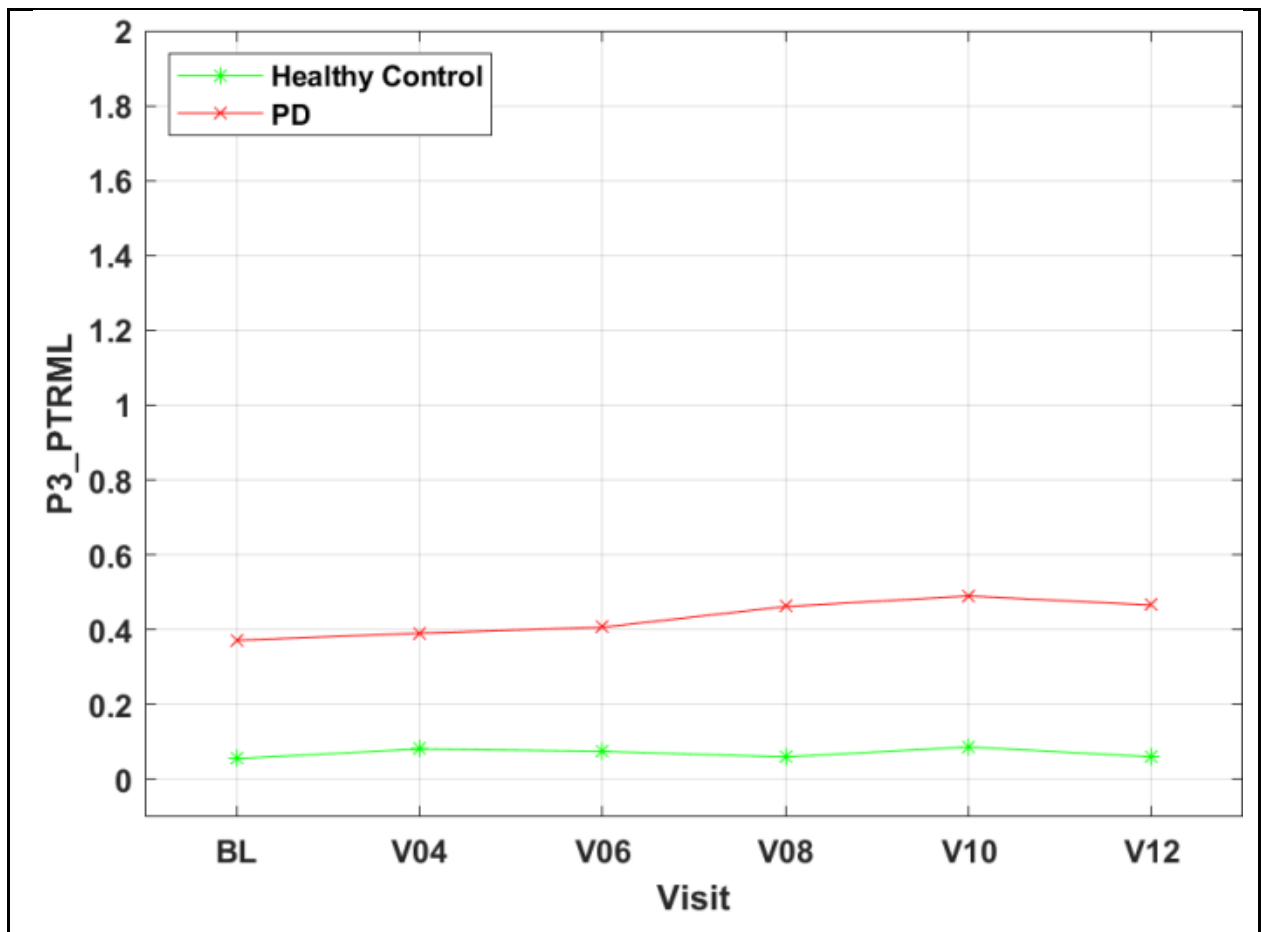

(ay) Postural tremor LH

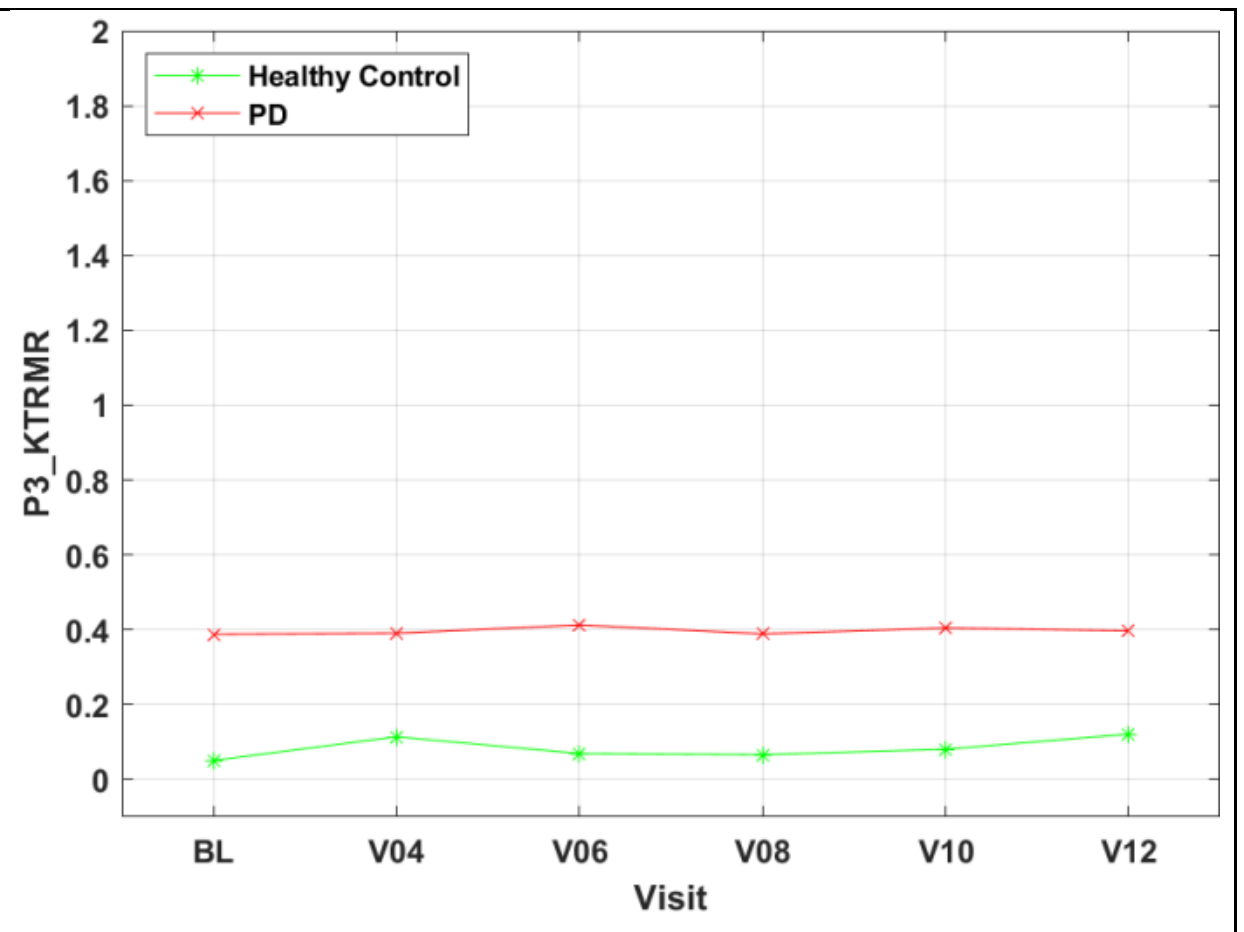

(az) Kinetic tremor RH

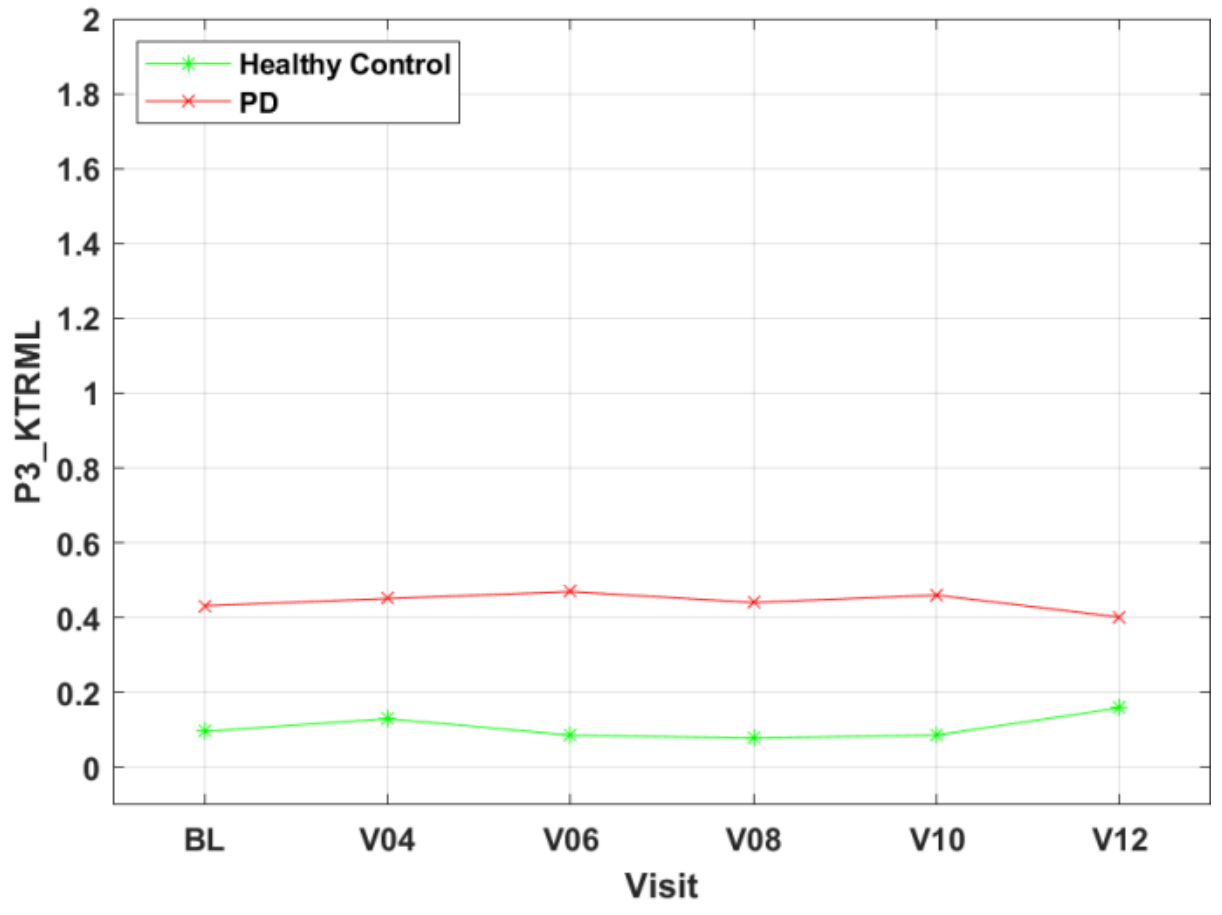

(ba) Kinetic tremor LH

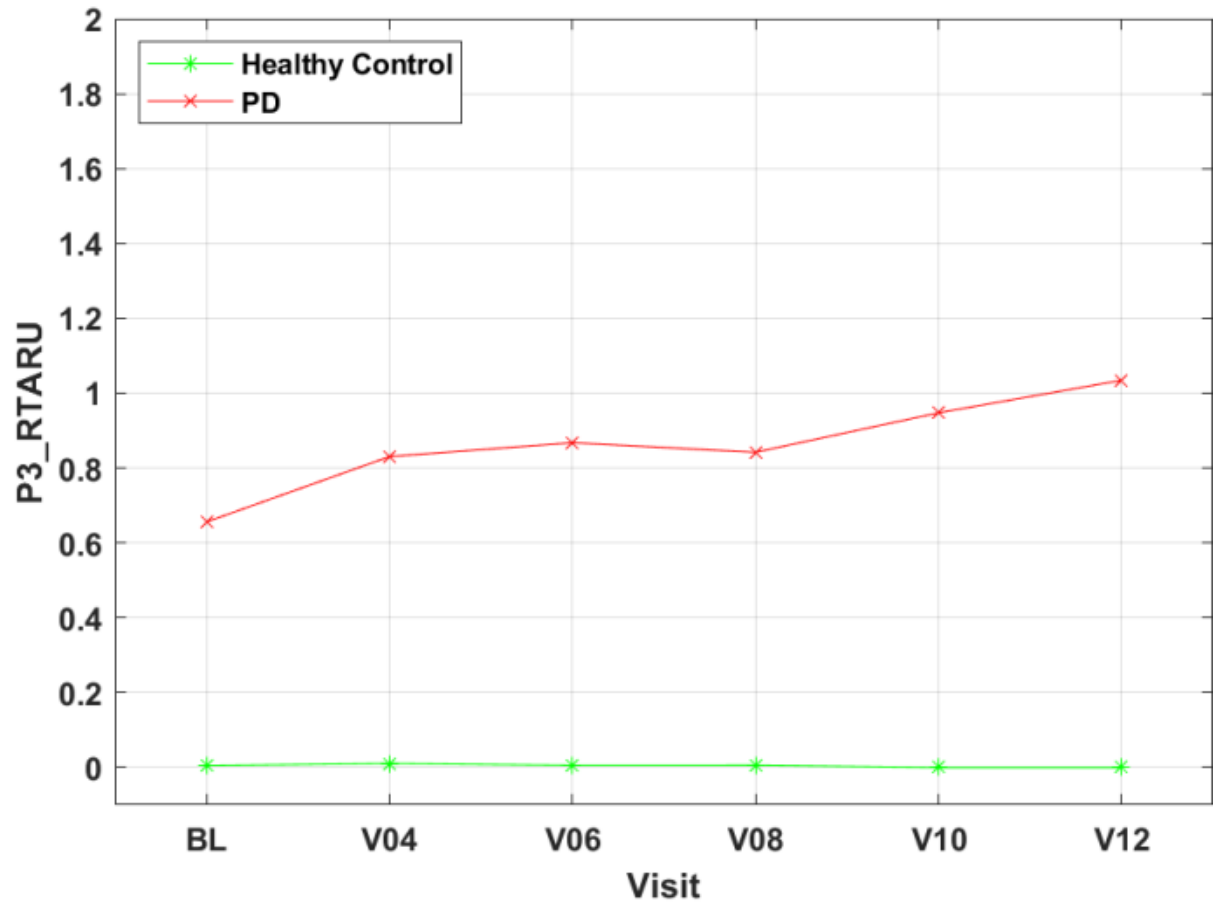

(bb) Rest Tremor Amplitude (RTA) RUE

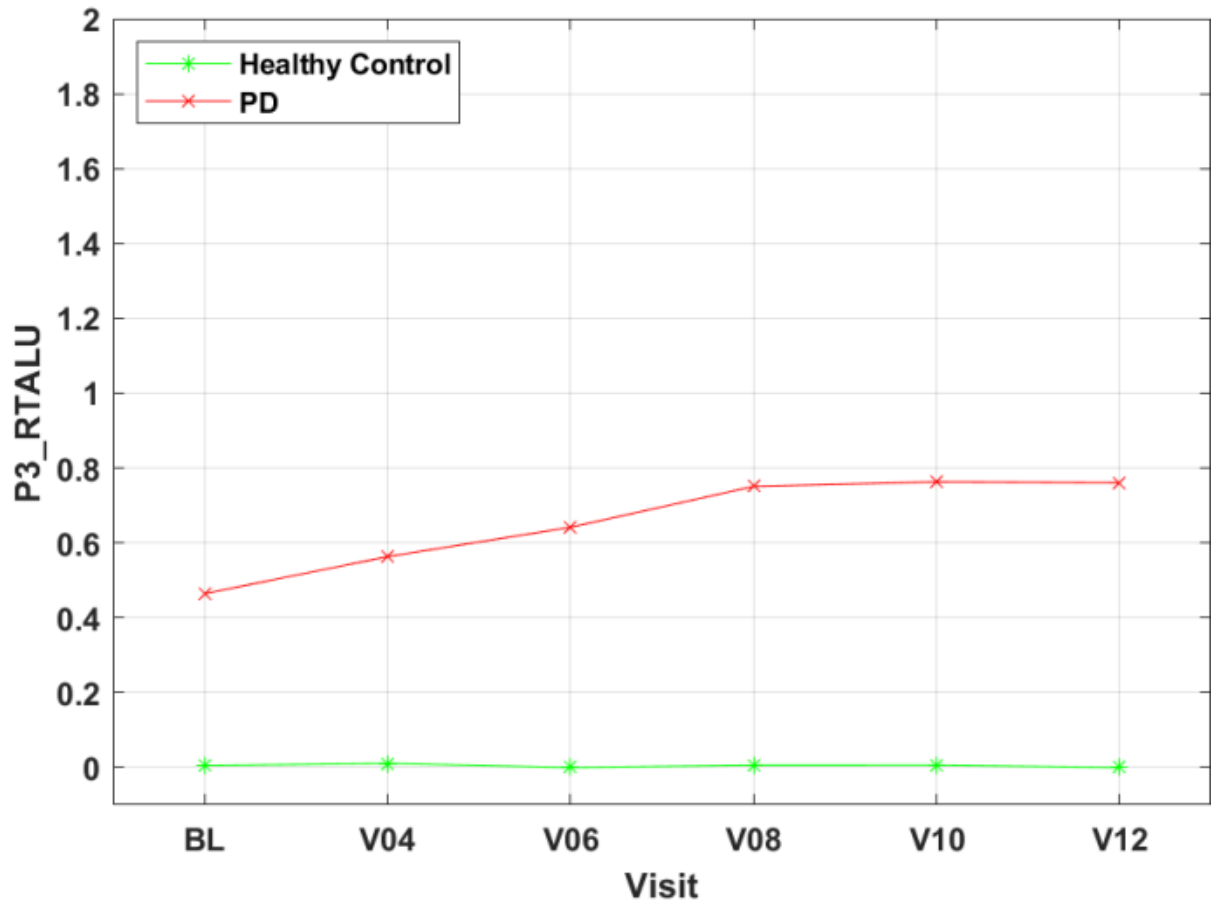

(bc) RTA LUE

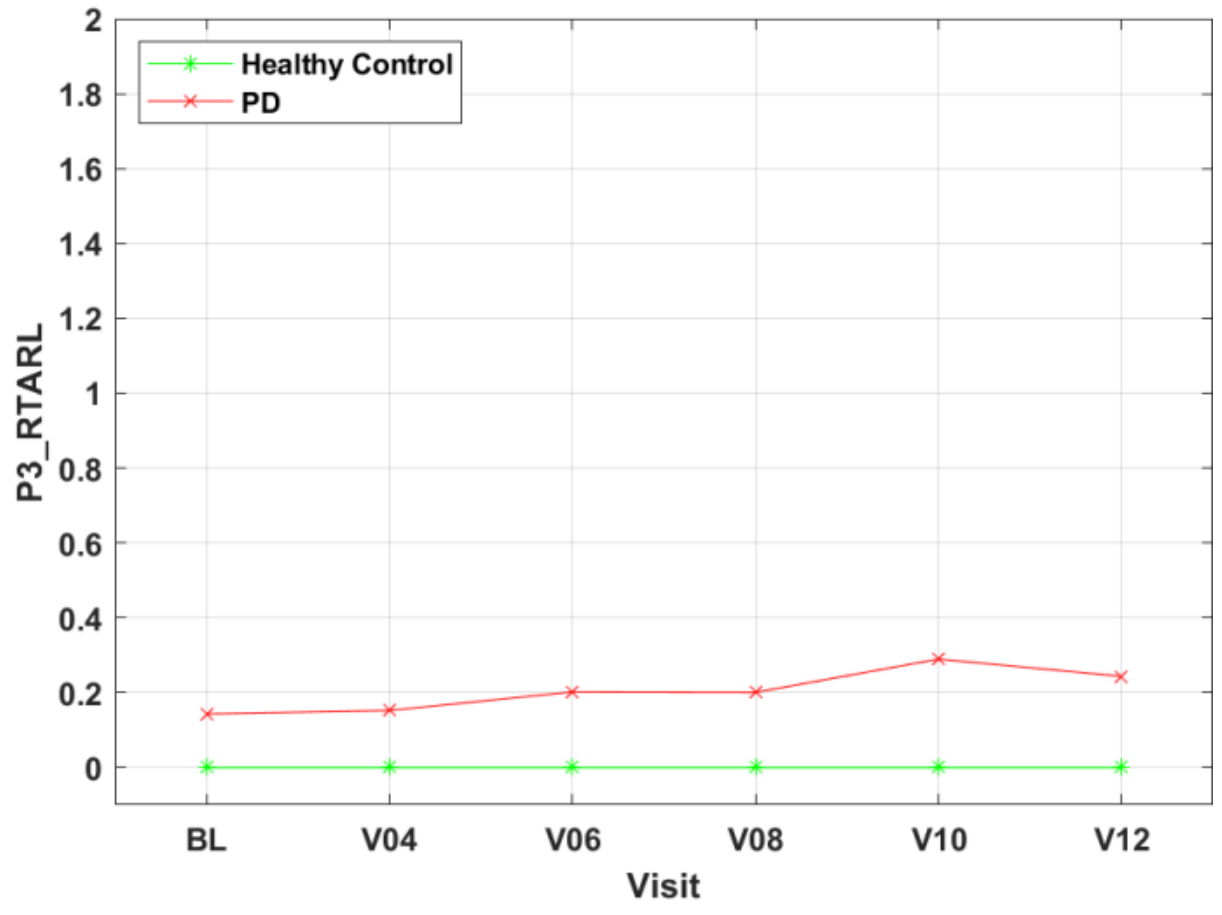

(bd) RTA RLE

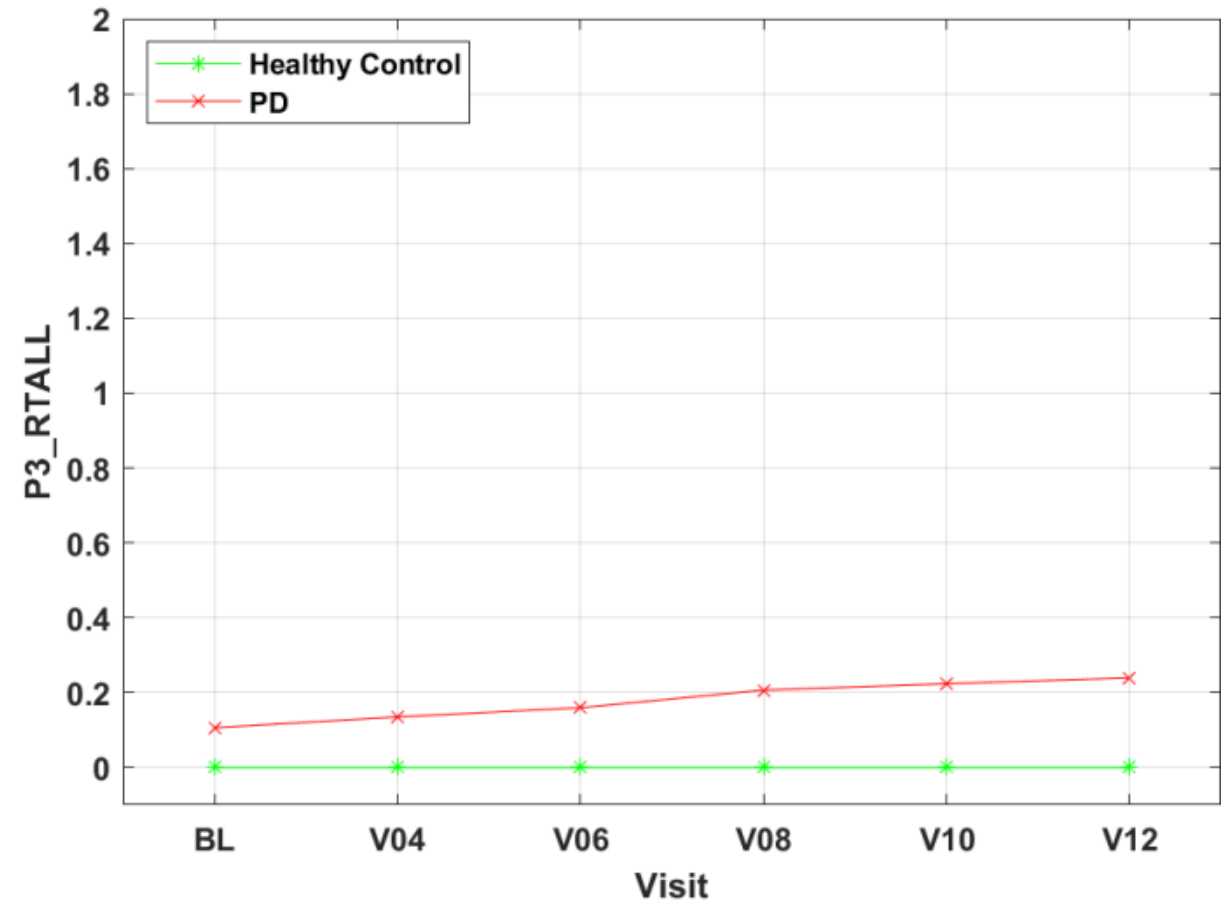

(be) RTA LLE

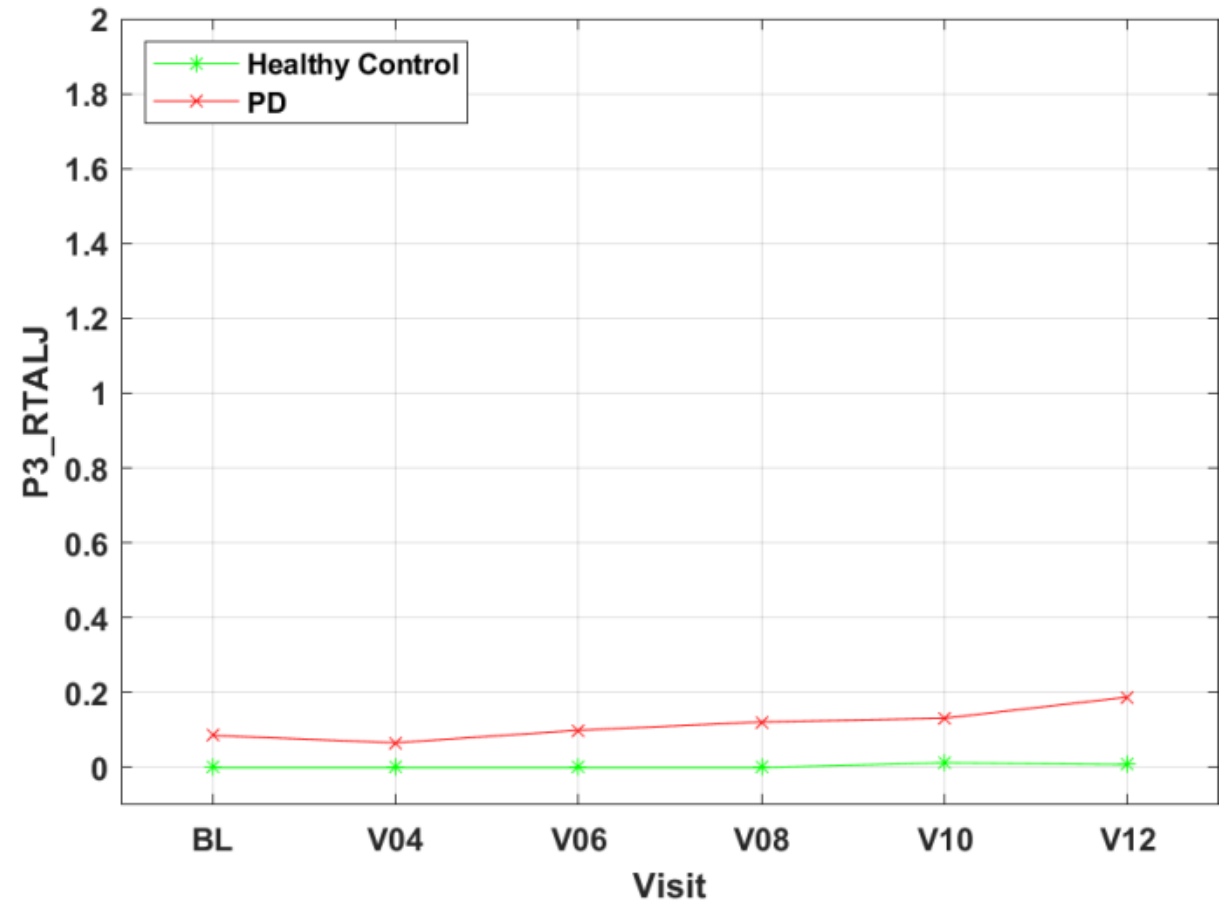

(bf) RTA Lip/jaw

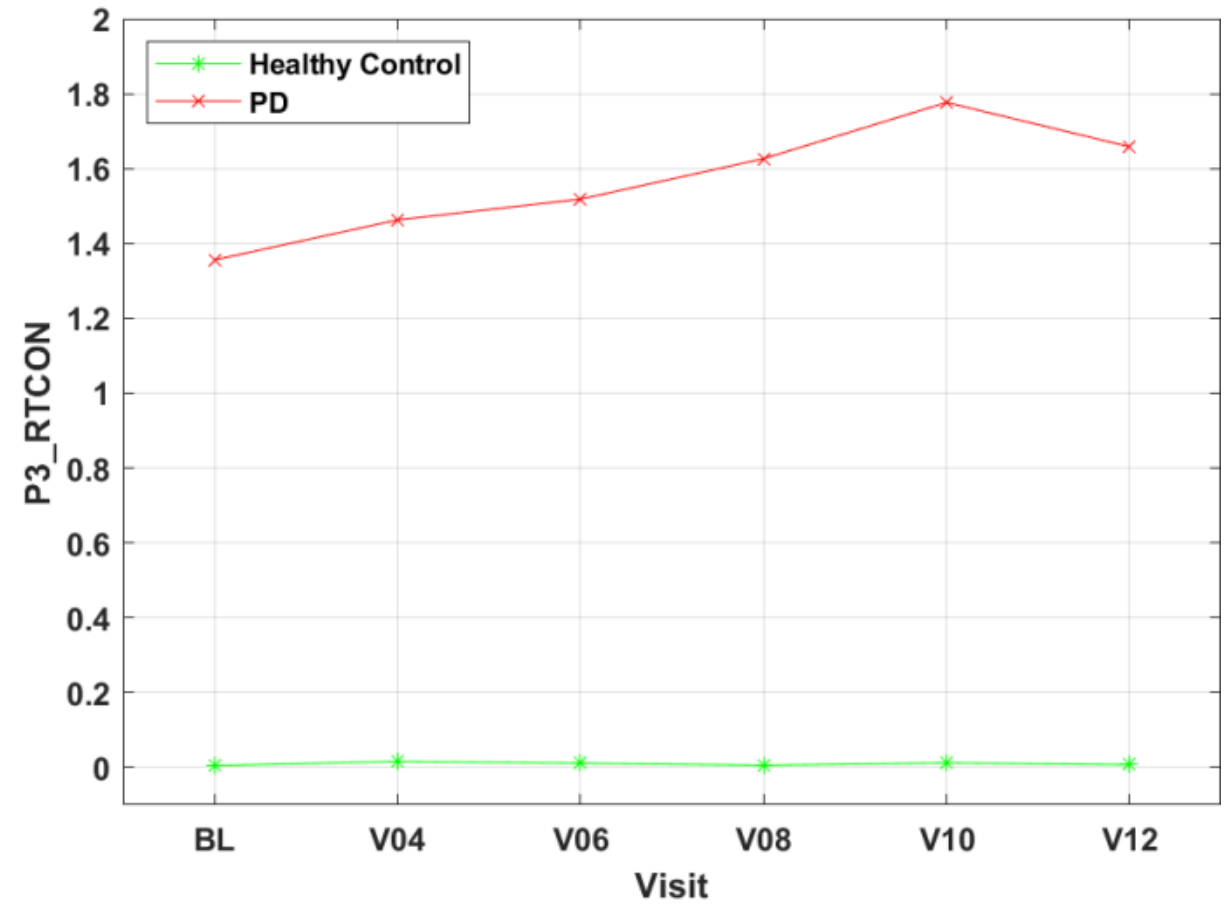

(bg) Constancy of Rest Tremor

Fig. C1. The plot of average severity of features for HC and PD for different visits. The visit at month 84 (V14) was not included in the plots as the number of observations for V14 were very low.